\documentclass[manuscript,screen]{acmart}

\usepackage{subfigure, graphicx, epsfig}
\usepackage{multirow}
\begin{document}

\title{Temporal Link Prediction: A Unified Framework, Taxonomy, and Review}

\author{Meng Qin}
\email{mengqin\_az@foxmail.com}
\author{Dit-Yan Yeung}
\email{dyyeung@cse.ust.hk}
\affiliation{%
  \institution{Department of Computer Science \& Engineering, Hong Kong University of Science \& Technology}
  \country{Hong Kong SAR}}

\renewcommand{\shortauthors}{Qin et al.}

\begin{abstract}
Dynamic graphs serve as a generic abstraction and description of the evolutionary behaviors of various complex systems (e.g., social networks and communication networks). Temporal link prediction (TLP) is a classic yet challenging inference task on dynamic graphs, which predicts possible future linkage based on historical topology. The predicted future topology can be used to support some advanced applications on real-world systems (e.g., resource pre-allocation) for better system performance. This survey provides a comprehensive review of existing TLP methods. Concretely, we first give the formal problem statements and preliminaries regarding data models, task settings, and learning paradigms that are commonly used in related research. A hierarchical fine-grained taxonomy is further introduced to categorize existing methods in terms of their data models, learning paradigms, and techniques. From a generic perspective, we propose a unified encoder-decoder framework to formulate all the methods reviewed, where different approaches only differ in terms of some components of the framework. Moreover, we envision serving the community with an open-source project \textit{OpenTLP}\footnote{We will open source and constantly update \textit{OpenTLP} at \href{https://github.com/KuroginQin/OpenTLP}{https://github.com/KuroginQin/OpenTLP}.} that refactors or implements some representative TLP methods using the proposed unified framework and summarizes other public resources.
As a conclusion, we finally discuss advanced topics in recent research and highlight possible future directions.
\end{abstract}

\maketitle

\section{Introduction}
For various complex systems (e.g., social networks, biology networks, and communication networks), graphs provide a generic abstraction to describe system entities and their relationships. For instance, one can abstract each entity as a node (vertex) and represent the relationship between a pair of entities as an edge (link) between the corresponding node pair.
Each edge can also be associated with a weight to encode additional information about the interactions between system entities (e.g., trust rating between users \cite{kumar2016edge,kumar2018rev2} and traffic between telecommunication devices \cite{borgnat2009seven,sivanathan2018classifying}).

Dynamic graphs, which can be represented as sequences of snapshots or time-induced edges, are widely used to describe behaviors of systems that change over time \cite{holme2012temporal}. Temporal link prediction (TLP) is a classic yet challenging inference task on dynamic graphs. It aims to predict possible linkage in specific future time steps based on the observed historical topology, playing an essential role in revealing the dynamic nature of systems and pre-allocating key resources (e.g., caches, CPU time, and communication channels) for better system performance \cite{lei2018adaptive,lei2019gcn}. The predicted future topology can also be used to support various advanced applications on real-world systems including (\romannumeral1) friend and next item recommendation in online social networks and media \cite{campana2017recommender,wang2020next}, (\romannumeral2) intrusion detection in enterprise Internet \cite{king2023euler}, (\romannumeral3) channel allocation in wireless internet-of-things networks \cite{gao2020edge}, (\romannumeral4) burst traffic detection and dynamic routing in optical networks \cite{vinchoff2020traffic,aibin2021short}, as well as (\romannumeral5) dynamics simulation and conformational analysis of molecules \cite{ashby2021geometric}.

As an extension of the conventional link prediction on static graphs \cite{martinez2016survey,kumar2020link}, TLP is a more difficult task due to the following challenges. First, it is hard to capture the spatial-temporal characteristics of a dynamic graph, including the topology structures (e.g., interactions between nodes) in each time step and evolving patterns across successive time steps (e.g., changes of node interactions), which are usually complex and non-linear \cite{cui2018survey}. Second, the behaviors of some real-world systems may vary rapidly. Most of them also have the requirements of real-time inference. It is challenging to achieve fine-grained representations and predictions of the rapid variation while satisfying real-time constraints with low complexities. Third, most existing inference techniques on dynamic graphs have simple problem statements (e.g., TLP on unweighted graphs with fixed known node sets \cite{li2014deep,nguyen2018continuous}). Some advanced settings from real-world systems (e.g., prediction of weighted links between previously unobserved nodes) are not fully studied in recent research.

To the best of our knowledge, there are a series of related surveys published in recent years with different focuses. Kazemi et al. \cite{kazemi2020representation}, Xue et al. \cite{xue2022dynamic}, and Barros et al. \cite{barros2021survey} gave overviews of existing dynamic network embedding (DNE) techniques, which learn a low-dimensional vector representation (a.k.a. embedding) for each node with the evolving patterns of graph topology and other side information (e.g., node attributes) preserved. The derived embedding can then be used to support various downstream tasks including TLP. Also from the perspective of DNE, Skarding et al. \cite{skarding2021foundations} reviewed recent techniques of graph neural networks (GNNs) for dynamic graphs. However, there remain gaps between the research on DNE and TLP. On the one hand, some classic TLP approaches \cite{hill2006building,sharan2008temporal} are not based on the DNE framework.
On the other hand, some DNE methods \cite{li2018deep,goyal2020dyngraph2vec,min2021stgsn} are \textit{task-dependent} with model architectures and objectives designed only for a specific setting of TLP. Moreover, most \textit{task-independent} DNE techniques can only support simple TLP settings based on some common but naive strategies (e.g., treating the prediction of unweighted links as binary edge classification \cite{nguyen2018continuous,xu2020inductive,sankar2020dysat}). The aforementioned surveys lack detailed discussions regarding whether and how a DNE method can be used to handle different settings of TLP. This survey covers both (\romannumeral1) classic TLP methods that do not rely on DNE and (\romannumeral2) representative DNE approaches that can support TLP. In particular, for DNE-based techniques, we focus on how they can support TLP and why they cannot tackle some specific settings, highlighting their limitations.

Haghani et al. \cite{haghani2017temporal} and Divakaran et al. \cite{divakaran2020temporal} reviewed representative TLP methods only based on the techniques they used but existing TLP techniques may differ in terms of multiple aspects (e.g., data models, learning paradigms, and task settings). We aim to give a finer-grained description of existing TLP approaches covering multiple aspects via a unified framework. The major contributions of this survey can be summarized as follows.
\begin{itemize}
    \item We propose a hierarchical taxonomy to categorize existing TLP methods in terms of the (\romannumeral1) data models, (\romannumeral2) learning paradigms, and (\romannumeral3) techniques used to handle the dynamic topology. Compared with existing surveys, the proposed taxonomy is a finer-grained description of existing approaches covering multiple aspects.
    \item From a generic perspective, we introduce a unified encoder-decoder framework to formulate all the TLP methods reviewed in this survey. In this framework, each method can be described by (\romannumeral1) an encoder, (\romannumeral2) a decoder, and (\romannumeral3) a loss function. It is expected that different methods only differ in terms of these three components.
    \item By using the proposed unified framework, we refactor and implement some representative TLP methods and serve the community with an open-source project \textit{OpenTLP} (\href{https://github.com/KuroginQin/OpenTLP}{https://github.com/KuroginQin/OpenTLP}). This project also summarizes some other public resources regarding TLP and will be constantly updated.
    \item In addition, some typical advanced research topics, future directions, quality evaluation criteria, applications, and datasets of TLP are also discussed in this survey.
\end{itemize}
In the rest of this survey, we give the problem statements and preliminaries regarding (\romannumeral1) data models of dynamic graphs, (\romannumeral2) commonly-used task settings of TLP, and (\romannumeral3) learning paradigms of recent research in Section~\ref{Sec:Prob}. The unified encoder-decoder framework is introduced in the same section. In Section~\ref{Sec:Meth}, we review representative TLP methods based on a fine-grained hierarchical taxonomy. Section~\ref{Sec:Topic-Future} summarizes advanced topics in recent research and highlights possible future directions. Finally, Section~\ref{Sec:Conc} concludes this survey. We leave additional descriptions regarding the quality evaluation, detailed techniques, advanced applications, and public datasets of TLP in supplementary materials.

\section{Problem Statements \& Preliminaries}\label{Sec:Prob}
In this survey, we consider the TLP on undirected homogeneous dynamic graphs, which covers the settings of most related research. Other inference tasks on directed heterogeneous graphs (e.g., knowledge graphs \cite{rossi2021knowledge,cai2022temporal}) are not included. For convenience, we summarize the major notations and abbreviations used in this survey in supplementary materials. In the rest of this section, we introduce the (\romannumeral1) data models of dynamic graphs as well as (\romannumeral2) task settings, (\romannumeral3) quality evaluation criteria, and (\romannumeral4) learning paradigms that are commonly used in related research. From a generic perspective, a unified encoder-decoder framework is proposed to formulate all the TLP methods reviewed in this survey.

\subsection{Data Models}\label{Sec:Data-Model}
Existing graph inference techniques usually adopt two data models to describe the dynamic topology, which are the (\romannumeral1) evenly-sampled snapshot sequence description (ESSD) and (\romannumeral2) unevenly-sampled edge sequence description (UESD).

\begin{figure}[t]
  \centering
  \includegraphics[width=0.70\linewidth, trim=18 18 18 18,clip]{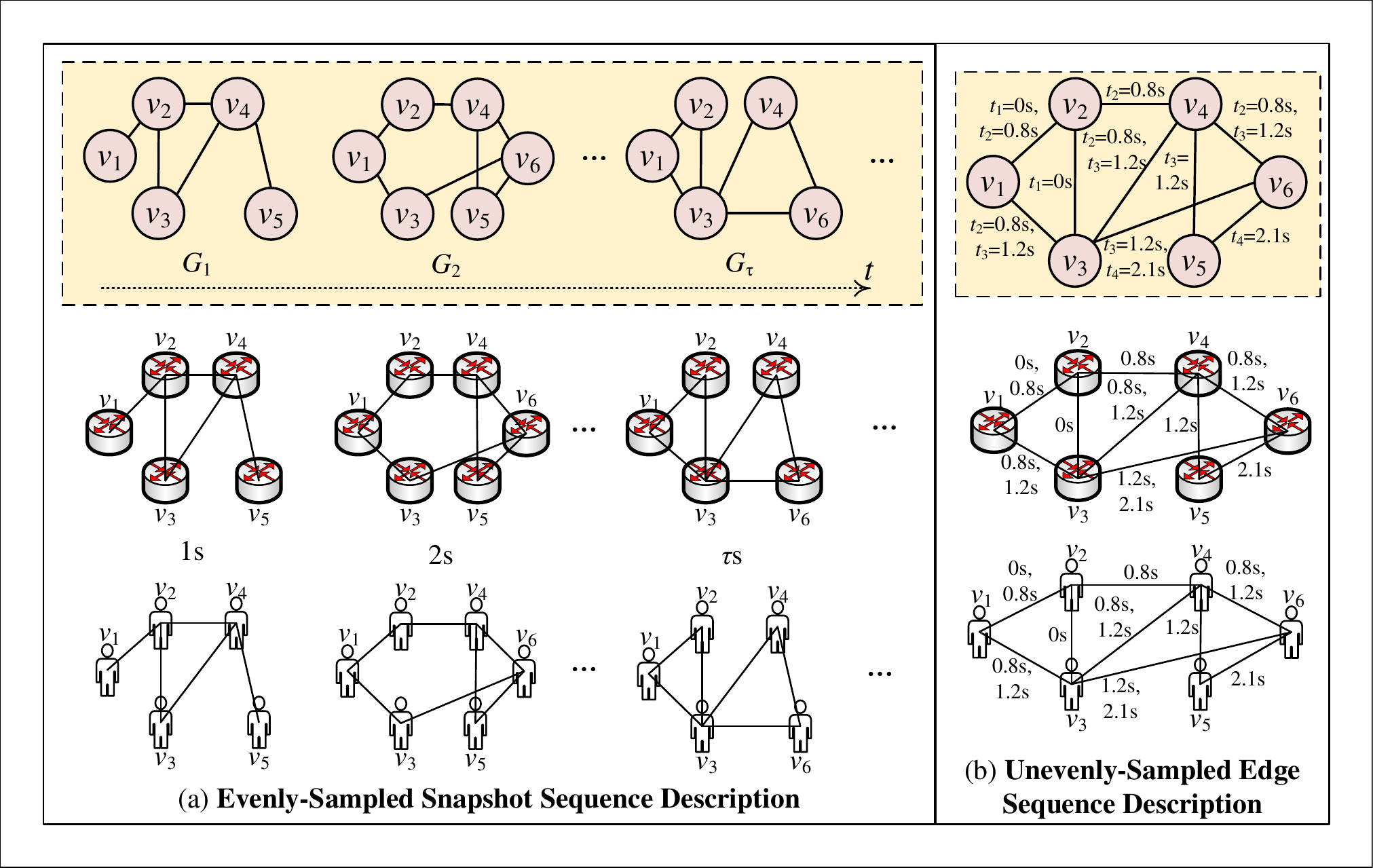}
  \vspace{-0.25cm}
  \caption{Examples of the (a) evenly-sampled snapshot sequence description (ESSD) and (b) unevenly-sampled edge sequence description (UESD) of a dynamic graph.}\label{fig:tmp_graph}
  \vspace{-0.3cm}
\end{figure}

\textbf{Definition 2.1} (\textbf{Evenly-Sampled Snapshot Sequence Description}, \textbf{ESSD}). A dynamic graph can be represented as a sequence of snapshots $G = ({G_1}, {G_2}, \cdots, {G_T})$ over a set of time steps $\{ 1, 2, \cdots, T\}$, where \textit{the time interval between successive snapshots is assumed to be regular}. Each snapshot can be described as a tuple $G_t = (V_t, E_t)$, where $V_t = \{ v_1^t, v_2^t, \cdots, v_{N_t}^t\}$ is the set of nodes (with $v_i^t$ denoting a node in $G_t$); $E_t = \{ ((v_i^t, v_i^t), w) | v_i^t,v_j^t \in V_t, w \in \Re^+\}$ is the set of edges with each edge $(v_i^t, v_j^t)$ associated with a weight $w$. For unweighted graphs, we omit $w$ and use $E_t = \{ (v_i^t, v_i^t)\}$ to denote the edge set. Some methods also assume that graph attributes are available, where each snapshot $G_t$ is associated with an attribute map $\mathcal{A}_t =\{ \varphi (v_1^t), \cdots ,\varphi (v_{N_t}^t)\}$ with $\varphi(v_i^t)$ mapping each node $v_i^t$ to its attributes.

In general, we can use an adjacency matrix ${\bf{A}}_t \in \Re^{N_t \times N_t}$ to describe the topology of each snapshot $G_t$. For unweighted graphs, $({\bf{A}}_t)_{ij} = ({\bf{A}}_t)_{ji} = 1$ when there is an edge between node pair $(v_i^t, v_j^t)$ and $({\bf{A}}_t)_{ij} = ({\bf{A}}_t)_{ji} = 0$ otherwise. For weighted graphs, $({\bf{A}}_t)_{ij} = ({\bf{A}}_t)_{ji} = w>0$ denotes the weight of edge $(v_i^t, v_j^t)$ while $({\bf{A}}_t)_{ij} = ({\bf{A}}_t)_{ji} = 0$ indicates that there is no edge between $(v_i^t, v_j^t)$. Graph attributes of snapshot $G_t$ can be described by an attribute matrix ${\bf{X}}_t \in \Re^{N_t \times M}$, where the $i$-th row $({\bf{X}}_t)_{i, :}$ denotes the attributes of node $v_i^t$. In the rest of this survey, $\{ {\bf{A}}_t, {\bf{X}}_t\}$ are used to describe the ESSD-based topology and attributes unless otherwise stated.

An example of ESSD for unweighted graphs is illustrated in Fig.~\ref{fig:tmp_graph} (a), where each snapshot $G_t$ describes the behavior of a system (e.g., data transmission in communication networks and user interactions in online social networks) at time step $t$ and successive snapshots have the same time interval (e.g., $1$ second).
When abstracting a real-world system via ESSD, one needs to manually select a fixed interval (or corresponding sampling rate) between successive time steps (i.e., snapshots). Accordingly, we can execute one prediction operation once it comes to a new time step.
To fully describe the system behavior, the time interval is usually set to be the minimum duration of interactions in a system, which may result in high space complexities and many redundant descriptions of topology in applications with rapid variations. Therefore, ESSD is usually adopted as a coarse-grained description of dynamic graphs.
Some other approaches use the UESD of dynamic graphs, which can describe system behaviors in a fine-grained manner.

\textbf{Definition 2.2} (\textbf{Unevenly-Sampled Edge Sequence Description}, \textbf{UESD}). A dynamic graph can also be represented as a tuple $G_{\Gamma}=(V_{\Gamma}, E_{\Gamma}, \Gamma)$, where $\Gamma  = \{ {t_1},{t_2}, \cdots \}$ is the set of time steps with $t_s \in \Re^+$ as the time of the $s$-th sampling and $t_1 < t_2 < \ldots < t_{|\Gamma|}$; $V_{\Gamma} = \{ v_1, v_2, \cdots, v_N\}$ is the set of nodes observed during time period $\Gamma$; $E_{\Gamma} = \{ ((v_i, v_j), w, t_e)| v_i,v_j\in V_\Gamma, w\in \Re^+, t_e\in \Gamma\}$ is the set of edges during $\Gamma$. In $E_{\Gamma}$, each edge $(v_i, v_j)$ is associated with a weight $w$ and a time step $t_e$, representing that there is an edge between node pair $(v_i, v_j)$ with weight $w$ at time step $t_e$. For unweighted graphs, we omit $w$ and use $E_{\Gamma} = \{ ((v_i, v_j), t_e)\}$ to denote the edge set. \textit{$t_e$ can be defined on the continuous domain}. \textit{The interval between edges observed in two successive time steps can also be irregular}. Some methods assume that graph attributes $\mathcal{A}_\Gamma = \{ \varphi ({v_i},t)|t \in \Gamma ,{v_i} \in {V_\Gamma }\}$ are available, where $\varphi ({v_i},t)$ maps a node $v_i$ to its attributes observed at time step $t$.

An example of UESD for unweighted graphs is demonstrated in Fig.~\ref{fig:tmp_graph} (b), where each edge $(v_i, v_j)$ is associated with one or more positive numbers, implying that the interaction (e.g., data transmission and message communication) between system entities $v_i$ and $v_j$ is observed at corresponding time steps.
When using UESD, we sample a corresponding edge once there is a new interaction in the system without manually specifying the sampling rate.
Hence, \textit{the interval between two successive time steps can be irregular}, which makes it space-efficient to be a fine-grained description of dynamic graphs without the redundant descriptions of topology in ESSD.
However, we still need to set a proper execution frequency of TLP for UESD.

Different from ESSD, UESD cannot use simple matrix representations of dynamic graphs (e.g., adjacency matrices). Hence, some mature matrix-based techniques (e.g., matrix factorization \cite{huang2012non,de2021survey}) cannot be applied to tackle the TLP with UESD. In contrast, most UESD-based methods rely on several continuous-time stochastic processes (e.g., 
Hawkes process \cite{gonzalez2016spatio,yuan2019multivariate}) to capture the evolution of graph topology. However, these stochastic processes are still inapplicable to some advanced tasks that conventional matrix representations can easily address (e.g., inference on weighted dynamic graphs) due to the lack of related studies.

Some literature uses terminologies different from the aforementioned definitions. We argue that some of these terminologies cannot precisely describe the two data models. For instance, \cite{nguyen2018continuous,zhang2021tg,xue2022dynamic} defined the first and second data models as \textit{discrete(-time)} and \textit{continuous(-time)} descriptions. Although the time index associated with each edge is defined on a continuous domain in the second data model, \textit{the edge sequence $E_{\Gamma}$ that describes the dynamic topology is still discrete}. Therefore, the term `\textit{continuous(-time)}' may be ambiguous in some cases. \cite{yu2018netwalk,ma2020streaming} described the second data model using the term `\textit{streaming}'. Nevertheless, `\textit{streaming}' is usually used to describe a process with the continual arrival of events, while \textit{each edge in $E_{\Gamma}$ is not required to be continually generated}. In fact, \textit{the major difference between the first and second data models is whether the interval between two successive time steps is irregular} (i.e., \textit{unevenly sampled}). In conclusion, we believe that ESSD and UESD can define the two data models more precisely.

\subsection{Task Settings}\label{Sec:TLP-Setting}
Existing TLP methods may have different hypotheses and settings regarding the variation of node sets and availability of attributes in a dynamic graph. We categorize task settings of TLP into two levels with different degrees of difficulty based on their assumptions regarding the variation of node sets.
In the rest of this survey, we use $\tau$ to represent the index of current time step. $L$ denotes the number of historical time steps or the historical time interval (a.k.a. window size) considered in a TLP method. $\Delta$ is the number of future time steps or the future time interval for prediction. For ESSD, $\mathcal{U}_s^d$ is adopted as a simplified notation of sequence $(\mathcal{U}_{s+1}, \mathcal{U}_{s+2}, \cdots, \mathcal{U}_d)$ w.r.t. a variable $\mathcal{U}$ (e.g., $G_{\tau-L}^{\tau}$ and ${\bf{A}}_{\tau}^{\tau+\Delta}$). For UESD, let $\Gamma(s, d) = \{ t | s < t \le d \}$ be the set of sampling time steps between $(s, d]$. $\mathcal{U}_{\Gamma(s, d)}$ denotes the sequence of a variable $\mathcal{U}$ associated with time steps in $\Gamma(s, d)$ (e.g., $G_{\Gamma(\tau-L, \tau)}$, $E_{\Gamma(\tau, \tau+\Delta)}$).

\textbf{Definition 2.3} (\textbf{TLP Level-1}). Level-1 assumes that \textit{the node set is known and fixed for all the time steps} in a dynamic graph. Namely, there is no addition or deletion of nodes as the topology evolves. For ESSD, level-1 takes snapshots $G_{\tau-L}^{\tau}$ w.r.t. previous $L$ time steps and attributes $\mathcal{A}_{\tau-L}^{\tau+\Delta}$ (if available) as inputs and then predicts the topology w.r.t. next $\Delta$ time steps, which can be formulated as
\begin{equation}
    \tilde G_{\tau}^{\tau  + \Delta } = {f_{{\mathop{\rm TLP}\nolimits} }}(G_{\tau  - L}^\tau , \mathcal{A}_{\tau-L}^{\tau + \Delta}),
\end{equation}
with $\tilde G_{\tau}^{\tau  + \Delta }$ as the prediction result. For UESD, given historical topology $G_{\Gamma(\tau-L, \tau)}$ and attributes ${\mathcal{A}}_{\Gamma(\tau-L, \tau+\Delta)}$ (if available), we aim to predict the topology w.r.t. next $\Delta$ time steps. It can be formulated as
\begin{equation}
    {\tilde G_{\Gamma (\tau,\tau  + \Delta )}} = {f_{{\mathop{\rm TLP}\nolimits} }}({G_{\Gamma (\tau  - L,\tau )}},{\mathcal{A}_{\Gamma (\tau  - L,\tau  + \Delta )}}),
\end{equation}
where ${\tilde G}_{\Gamma(\tau, \tau+\Delta)}$ denotes the prediction result.

\textbf{Definition 2.4} (\textbf{TLP Level-2}). Level-2 assumes that \textit{the node set can be non-fixed and can evolve over time}, allowing the deletion and addition of nodes. In this setting, \textit{the prediction includes not only the future topology induced by old (i.e., previously observed) nodes but also edges (\romannumeral1) between an old node and a new (i.e., previously unobserved) node or (\romannumeral2) between two new nodes}. For ESSD, level-2 takes historical snapshots $G_{\tau-L}^{\tau}$, node sets $V_{\tau}^{\tau  + \Delta }$ w.r.t. next $\Delta$ snapshots, and attributes $\mathcal{A}_{\tau-L}^{\tau+\Delta}$ (if available) as inputs and then derives the prediction result $\tilde G_{\tau}^{\tau  + \Delta }$ induced by $V_{\tau}^{\tau+\Delta}$ via
\begin{equation}\label{Eq:TLP-L2-1}
    \tilde G_{\tau}^{\tau  + \Delta } = {f_{{\mathop{\rm TLP}\nolimits} }}(G_{\tau  - L}^\tau ,V_{\tau}^{\tau  + \Delta }, \mathcal{A}_{\tau-L}^{\tau+\Delta}).
\end{equation}
For UESD, given the historical topology $G_{\Gamma(\tau-L, \tau)}$, future node set $V_{\Gamma(\tau, \tau+\Delta)}$, and attributes ${\mathcal{A}_{\Gamma (\tau  - L,\tau  + \Delta )}}$ (if available), we can formulate the TLP in level-2 as
\begin{equation}\label{Eq:TLP-L2-2}
    {{\tilde G}_{\Gamma (\tau,\tau  + \Delta )}} = {f_{{\mathop{\rm TLP}\nolimits} }}({G_{\Gamma (\tau  - L,\tau )}},{V_{\Gamma (\tau,\tau  + \Delta )}},{\mathcal{A}_{\Gamma (\tau  - L,\tau  + \Delta )}}),
\end{equation}
where ${{\tilde G}_{\Gamma (\tau,\tau  + \Delta )}}$ denotes the prediction result induced by $V_{\Gamma(\tau, \tau+\Delta)}$.

\begin{figure}[t]
  \centering
  \includegraphics[width=0.95\linewidth, trim=18 18 18 18,clip]{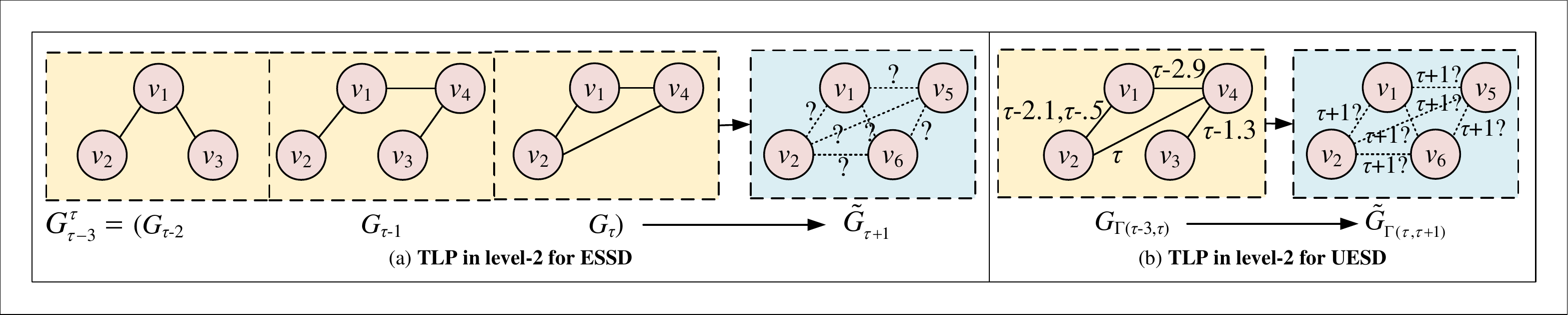}
  \vspace{-0.25cm}
  \caption{An example of the TLP in level-2 with $L=3$, $\Delta=1$, and $V_{\tau}^{\tau+\Delta}=\{ v_1, v_2, v_5, v_6\}$ for ESSD and UESD.}\label{fig:task_example}
  \vspace{-0.3cm}
\end{figure}

In general, level-2 is a more challenging setting, with level-1 as a special case of level-2. All the methods reviewed in this survey can deal with level-1 but only some of them can tackle level-2.
Fig.~\ref{fig:task_example} gives an example of level-2 with $L=3$, $\Delta=1$, and $V_{\tau}^{\tau+\Delta}=\{ v_1, v_2, v_5, v_6\}$.
Some literature \cite{trivedi2019dyrep,xu2020inductive,wang2021inductive} also divides the inference tasks on dynamic graphs into the \textit{transductive} and \textit{inductive} settings regarding the variation of node sets. For TLP, the \textit{transductive} setting only considers the prediction of edges induced by old (i.e., previously observed) nodes (e.g., $(v_1, v_2)$ at time step $(\tau+1)$ in Fig.~\ref{fig:task_example}). In contrast, the \textit{inductive} setting considers predicted edges (\romannumeral1) between an old node and a new node (e.g., $(v_1, v_6)$ at $(\tau+1)$) or (\romannumeral2) between two new nodes (e.g., $(v_5, v_6)$ at $(\tau+1)$). Some studies \cite{trivedi2019dyrep,xu2020inductive,wang2021inductive} separately treated and evaluated the three sources of prediction results, while level-2 defined in this survey covers all the results.

For ESSD, settings with $\Delta=1$ and $\Delta>1$ are defined as the \textit{one-step} and \textit{multi-step} prediction. Most related methods focus on the \textit{one-step} prediction. For a given current time step $\tau$, some approaches let $L = \tau$, where all historical snapshots are used for TLP and window size $L$ increases as the graph evolves. Other methods use a fixed setting of $L$ as $\tau$ increases, where only a fixed number of previous snapshots are utilized for TLP.

\subsection{Quality Evaluation}\label{Sec:Q-Eva}

Most existing TLP approaches focus on the prediction on unweighted graphs \textit{determining the existence of links between each pair of nodes in a future time step}, which can be considered as a binary edge classification problem. Hence, some classic metrics of binary classification can be used to measure the prediction quality, including \textit{accuracy}, \textit{F1-score}, \textit{receiver operating characteristic} (ROC) curve, and \textit{area under the ROC curve} (AUC).

TLP on weighted graphs is a more challenging case seldom considered in recent research, which \textit{should not only determine the existence of future links but also predict associated link weights}. Therefore, metrics of binary classification cannot be applied to measure the prediction quality. \textit{Root-mean-square error} (RMSE) and \textit{mean absolute error} (MAE) are widely-used metrics for the prediction of weighted graphs, which measure the reconstruction error between the prediction result and ground-truth. Qin et al. \cite{qin2023high} argued that \textit{conventional RMSE and MAE metrics cannot measure the ability of a TLP method to derive high-quality prediction results for weighted graphs} and proposed two new metrics of \textit{mean logarithmic scale difference} (MLSD) and \textit{mismatch rate} (MR) to narrow this research gap.

Due to space limit, we leave details of the aforementioned quality metrics in supplementary materials.

\subsection{Learning Paradigms}\label{Sec:Learn-Parad}
Existing TLP techniques usually follow three learning paradigms, which can be summarized as (\romannumeral1) \textit{direct inference} (DI), (\romannumeral2) \textit{online training and inference} (OTI), as well as (\romannumeral3) \textit{offline training and online generalization} (OTOG).

\textbf{Definition 2.5 (\textbf{Direct Inference}, \textbf{DI}).} Typical DI methods extract some manually designed or heuristic features from historical topology. The extracted features are directly used to infer the result of one prediction operation, in which there are no training procedures since DI methods do not have model parameters to be optimized. Once it comes to a new time step, one can repeat the direct inference procedure to derive a new prediction result.

\begin{figure}[t]
\begin{center}
 \includegraphics[width=0.95\linewidth, trim=18 18 18 18,clip]{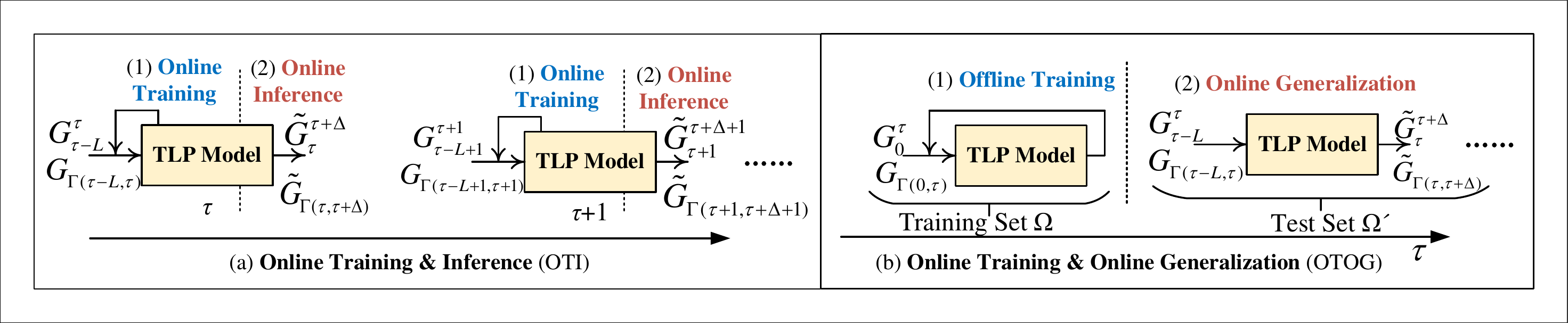}
\end{center}
\vspace{-0.25cm}
\caption{Illustrations of the TLP with (a) \textit{online training \& inference} (OTI) and (b) \textit{online training \& online generalization} (OTOG).}\label{fig:opt_scheme}
\vspace{-0.3cm}
\end{figure}

Most DI methods are easy to implement and have the potential to satisfy the real-time constraints of some systems because there are no time-consuming model optimization procedures. However, since DI approaches are still based on simple heuristics and intuitions, they may fail to capture complex and non-linear characteristics of dynamic topology. OTI and OTOG are more advanced paradigms that can automatically extract informative latent characteristics from the raw dynamic graphs via model optimization. Fig.~\ref{fig:opt_scheme} demonstrates examples of OTI and OTOG.

\textbf{Definition 2.6} (\textbf{Online Training \& Inference}, \textbf{OTI}). OTI methods \textit{are usually designed only for one prediction operation} including a training phase and an inference phase. For a given current time step $\tau$, we first optimize the TLP model (i.e., the training phase) according to the inputs of historical topology ($G_{\tau-L}^{\tau}$ or $G_{\Gamma(\tau-L, \tau)}$) and attributes ($\mathcal{A}_{\tau-L}^{\tau}$ or $\mathcal{A}_{\Gamma(\tau-L, \tau)}$ if available). Only after that, the TLP model can derive the prediction result $\tilde G_{\tau}^{\tau+\Delta}$ or $\tilde G_{\Gamma(\tau, \tau+\Delta)}$ (i.e., the inference phase).
When it comes to a new time step $(\tau+1)$, one should \textit{repeat the training and inference phases from scratch}, in order to derive the new prediction result $\tilde G_{\tau+1}^{\tau+\Delta+1}$ or $\tilde G_{\Gamma(\tau+1, \tau+\Delta+1)}$.

\textbf{Definition 2.7} (\textbf{Offline Training \& Online Generalization}, \textbf{OTOG}). In the OTOG paradigm, the sequence of snapshots or edges of a dynamic graph is divided into a training set ${\Omega}$ and a test set ${\Omega'}$, where snapshots or edges in ${\Omega}$ should occur before those in ${\Omega'}$. The TLP model is first trained on ${\Omega}$ in an offline way, with snapshots or edges in ${\Omega}$ as inputs and training ground-truth. After that, one can derive the prediction result of each new time step $\tau$ w.r.t. ${\Omega'}$ by directly generalizing the trained model to ${\Omega'}$ without additional optimization (i.e., with model parameters fixed).

In summary, OTI methods do not adopt the strategy that splits a dynamic graph into the training and test sets. Instead, they \textit{continually optimize the TLP model as time step $\tau$ increases in an online manner}, which can capture the latest evolving patterns. However, OTI approaches usually suffer from efficiency issues due to the high complexities of their online training and thus cannot be deployed to systems with real-time constraints.
In contrast, the OTOG paradigm includes offline training and online generalization. It is usually assumed that \textit{one has enough time to fully train a TLP model in an offline way}. \textit{The runtime of generating a prediction result only depends on the online generalization}. Since there is no additional optimization when generalizing the model to a test set, OTOG methods have the potential to satisfy the real-time constraints of systems. Nevertheless, they may also have the risk of failing to catch up with the latest variation of dynamic graphs, especially when there is a significant difference between the training and test sets.

\subsection{Unified Encoder-Decoder Framework}\label{Sec:Frame}

From a generic perspective, we introduce a unified encoder-decoder framework to formulate existing TLP methods, which includes (\romannumeral1) an \textit{encoder} ${\rm{Enc}}(\cdot)$, (\romannumeral2) a \textit{decoder} ${\rm{Dec}}(\cdot)$, and (\romannumeral3) a \textit{loss function} ${\mathcal{L}}(\cdot)$. It is expected that different TLP methods reviewed in this survey only differ in terms of these three components.

The \textit{encoder} ${\rm{Enc}}(\cdot)$ takes (\romannumeral1) historical topology and (\romannumeral2) attributes (if available) as inputs and then derives an \textit{intermediate representation} $R$ that captures the key properties of inputs. It can be formulated as
\begin{equation}\label{Eq:Enc-TLP}
    R = {\mathop{\rm Enc}\nolimits} (G_{\tau  - L}^\tau ,\mathcal{A}_{\tau  - L}^\tau ) {\rm{~~and~~}} R = {\mathop{\rm Enc}\nolimits} ({G_{\Gamma (\tau  - L,\tau )}},{\mathcal{A}_{\Gamma (\tau  - L,\tau )}})
\end{equation}
for ESSD and UESD.
The \textit{decoder} ${\rm{Dec}}(\cdot)$ further takes (\romannumeral1) \textit{intermediate representation} $R$, (\romannumeral2) node sets w.r.t. future time steps (if level-2 is considered), and (\romannumeral3) attributes (if available) as inputs and then derives the final prediction result, which can be described as
\begin{equation}
    \tilde G_{\tau}^{\tau  + \Delta } = {\mathop{\rm Dec}\nolimits} (R,V_{\tau}^{\tau  + \Delta }, \mathcal{A}_{\tau}^{\tau  + \Delta }) {\rm{~~and~~}} {{\tilde G}_{\Gamma (\tau,\tau  + \Delta )}} = {\mathop{\rm Dec}\nolimits} (R,{V_{\Gamma (\tau,\tau  + \Delta )}},{\mathcal{A}_{\Gamma (\tau,\tau  + \Delta )}})
\end{equation}
for ESSD and UESD.
Both the \textit{encoder} and \textit{decoder} may include a set of parameters that can be optimized (learned) via a \textit{loss function} ${\mathcal{L}}(\cdot)$ regarding the historical topology and attributes (if available). In the rest of this survey, $\delta$ is used to denote the set of learnable model parameters. The \textit{loss function} and its optimization objective can be formulated as
\begin{equation}
    {\min}_{\delta}~\mathcal{L}(G_{\tau  - L}^\tau, \mathcal{A}_{\tau  - L}^\tau ;\delta ) {\rm{~~and~~}} {\min}_{\delta}~\mathcal{L}(G_{\Gamma(\tau-L, \tau)} , \mathcal{A}_{\Gamma(\tau-L, \tau)} ;\delta )
\end{equation}
for ESSD and UESD.
In the rest of this survey, we omit $\mathcal{A}_{\tau  - L}^{\tau+\Delta}$ and $\mathcal{A}_{\Gamma(\tau-L, \tau+\Delta)}$ if attributes are not available. Furthermore, we omit $V_{\tau}^{\tau  + \Delta }$ or $V_{\Gamma(\tau, \tau + \Delta)}$ if the node set is assumed to be fixed (i.e., TLP in level-1).

In some cases, the intermediate representation $R$ given by $\rm{Enc}(\cdot)$ can be \textit{dynamic network embedding} (DNE) (a.k.a. \textit{dynamic graph representation learning}) \cite{kazemi2020representation,barros2021survey,xue2022dynamic} but is not limited to it.

\textbf{Definition 2.8} (\textbf{Dynamic Network Embedding}, \textbf{DNE}). Let $V$ denote the set of all possible nodes. Given the historical topology (described by $G_{\tau-L}^{\tau}$ or $G_{\Gamma(\tau-L, \tau)}$) and attributes (described by $\mathcal{A}_{\tau-L}^{\tau}$ or ${\mathcal{A}}_{\Gamma(\tau-L, \tau)}$ if available), DNE aims to learn a function ${f_{\rm{DNE}}}:V \mapsto {\Re ^d}$ that maps each node $v_i \in V$ to a low-dimensional vector representation ${\bf{z}}_i \in \Re^{d}$ (a.k.a. node embedding) with $d \ll |V|$ as the embedding dimensionality.
The derived embedding $\{{\bf{z}}_i\}$ is expected to preserve the evolving patterns of dynamic topology and attributes (if available). For example, nodes $(v_i, v_j)$ with an edge at a time step close to $\tau$ are more likely to have similar representations ${\bf{z}}_i$ and ${\bf{z}}_j$ with close distance or high similarity. The derived $\{{\bf{z}}_i\}$ can be used to support various downstream tasks on future topology including TLP.

Although there is a close relationship between DNE and TLP, there remain gaps between the research on these two tasks. First, some classic TLP methods (e.g., \textit{neighbor similarity} \cite{liben2003link} and \textit{graph summarization} \cite{hill2006building,sharan2008temporal} described in Section~\ref{Sec:SNAP-DI}) are not based on DNE. This survey covers DNE techniques that can support TLP but is not limited to them.

\begin{figure}[t]
\begin{center}
 \includegraphics[width=0.55\linewidth, trim=18 18 18 18,clip]{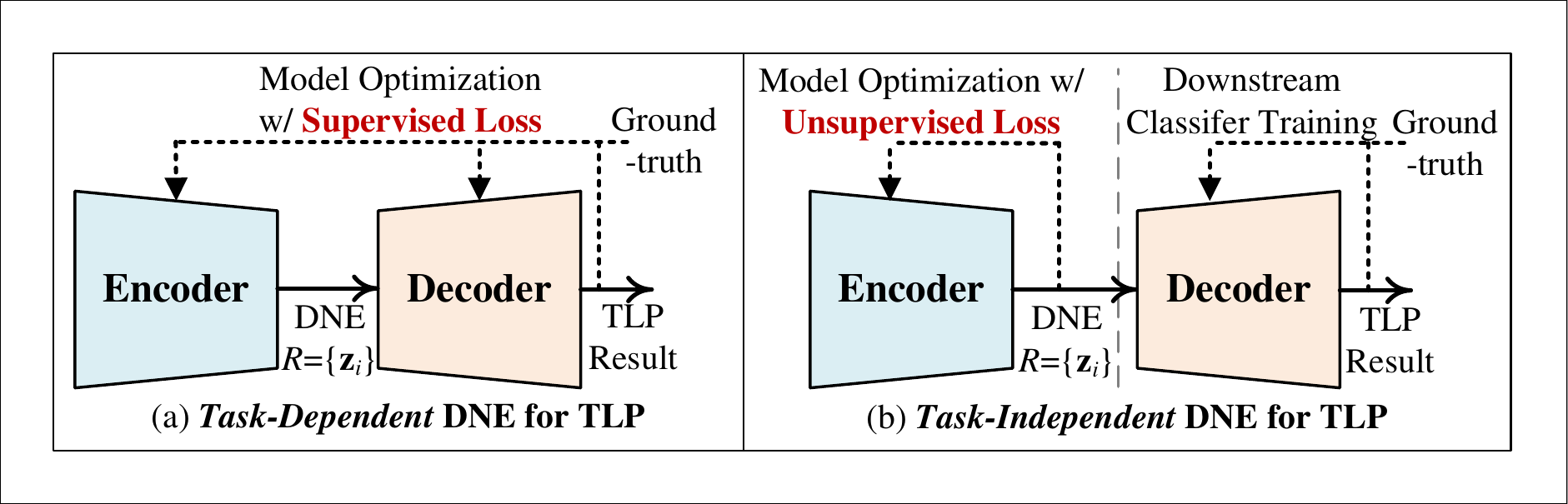}
\end{center}
\vspace{-0.25cm}
\caption{Illustrations of (a) \textit{task-dependent} and (b) \textit{task-independent} DNE for TLP in terms of the unified encoder-decoder framework.}\label{fig:DNE4TLP}
\vspace{-0.3cm}
\end{figure}

Furthermore, as shown in Fig.~\ref{fig:DNE4TLP} (a), some TLP approaches follow a \textit{task-dependent} DNE framework with specific \textit{encoders} and \textit{decoders} designed only for TLP. In most cases, supervised \textit{losses} regarding TLP are applied to optimize the DNE models (e.g., by minimizing the error between the topology reconstructed by $R$ and ground-truth) in an end-to-end manner. Therefore, these DNE-based approaches are optimized only for TLP.

Other DNE techniques are \textit{task-independent} as illustrated in Fig.~\ref{fig:DNE4TLP} (b), which optimize the model via unsupervised \textit{losses} to derive embedding regardless of downstream tasks. For TLP, these methods rely on some commonly-used but naive designs of \textit{decoders} to derive prediction results without specifying their own decoders. In these designs of \textit{decoders}, an auxiliary edge embedding ${\bf{e}}_{ij}$ is first derived for each pair of nodes $(v_i, v_j)$ using corresponding node embedding $({\bf{z}}_i, {\bf{z}}_j)$. A downstream binary classifier (e.g., logistic regression) is then trained with $\{ {\bf{e}}_{ij} \}$ as inputs and finally outputs the probability that an edge $(v_i, v_j)$ appears in a future time step. Table~\ref{tab:DNE-TLP} summarizes some commonly-used strategies to derive $\{ {\bf{e}}_{ij} \}$.
However, these strategies may still fail to handle some advanced applications of TLP (e.g., prediction of weighted links) due to the binary output of the downstream classifier.

In summary, existing DNE-based methods may have different designs of \textit{encoders}, \textit{decoders}, and \textit{loss functions} for TLP. Not all the designs can tackle some advanced settings (e.g., prediction of weighted links and TLP in level-2). In addition to the description of model architectures, this survey also highlights their limitations to TLP w.r.t. each component in the unified encoder-decoder framework, which is not covered in existing survey papers regarding DNE.

\begin{table}[t]\footnotesize
\centering
\caption{Commonly-used strategies of \textit{task-independent} DNE methods to derive the auxiliary edge embedding ${\bf{e}}_{ij}$ using corresponding node embedding $({\bf{z}}_i, {\bf{z}}_j)$, where a downstream binary classifier (e.g., logistic regression) can be applied to support TLP.}
\label{tab:DNE-TLP} 
\vspace{-0.25cm}
\begin{tabular}{ll|ll|ll}
\hline
\textbf{Strategies} & \textbf{Definitions} & \textbf{Strategies} & \textbf{Definitions} & \textbf{Strategies} & \textbf{Definitions} \\ \hline
Concatenation & ${{\bf{e}}_{ij}} = [{{\bf{z}}_i} || {{\bf{z}}_j}]$ & Hadamard Product & ${{\bf{e}}_{ij}} = {{\bf{z}}_i} \odot {{\bf{z}}_j}$ & Weighted-$l_2$ Norm & ${{\bf{e}}_{ij}} = |{{\bf{z}}_i} - {{\bf{z}}_j}|^2$ \\
Average & ${{\bf{e}}_{ij}} = ({{\bf{z}}_i} + {{\bf{z}}_j})/2$ & Weighted-$l_1$ Norm & ${{\bf{e}}_{ij}} = |{{\bf{z}}_i} - {{\bf{z}}_j}{|}$ &  &  \\ \hline
\end{tabular}
\vspace{-0.30cm}
\end{table}

\section{Review of Temporal Link Prediction Methods}\label{Sec:Meth}

\subsection{Overview of the Hierarchical Fine-Grained Taxonomy}

\begin{figure}[t]
\begin{center}
 \includegraphics[width=0.6\linewidth, trim=18 18 18 18,clip]{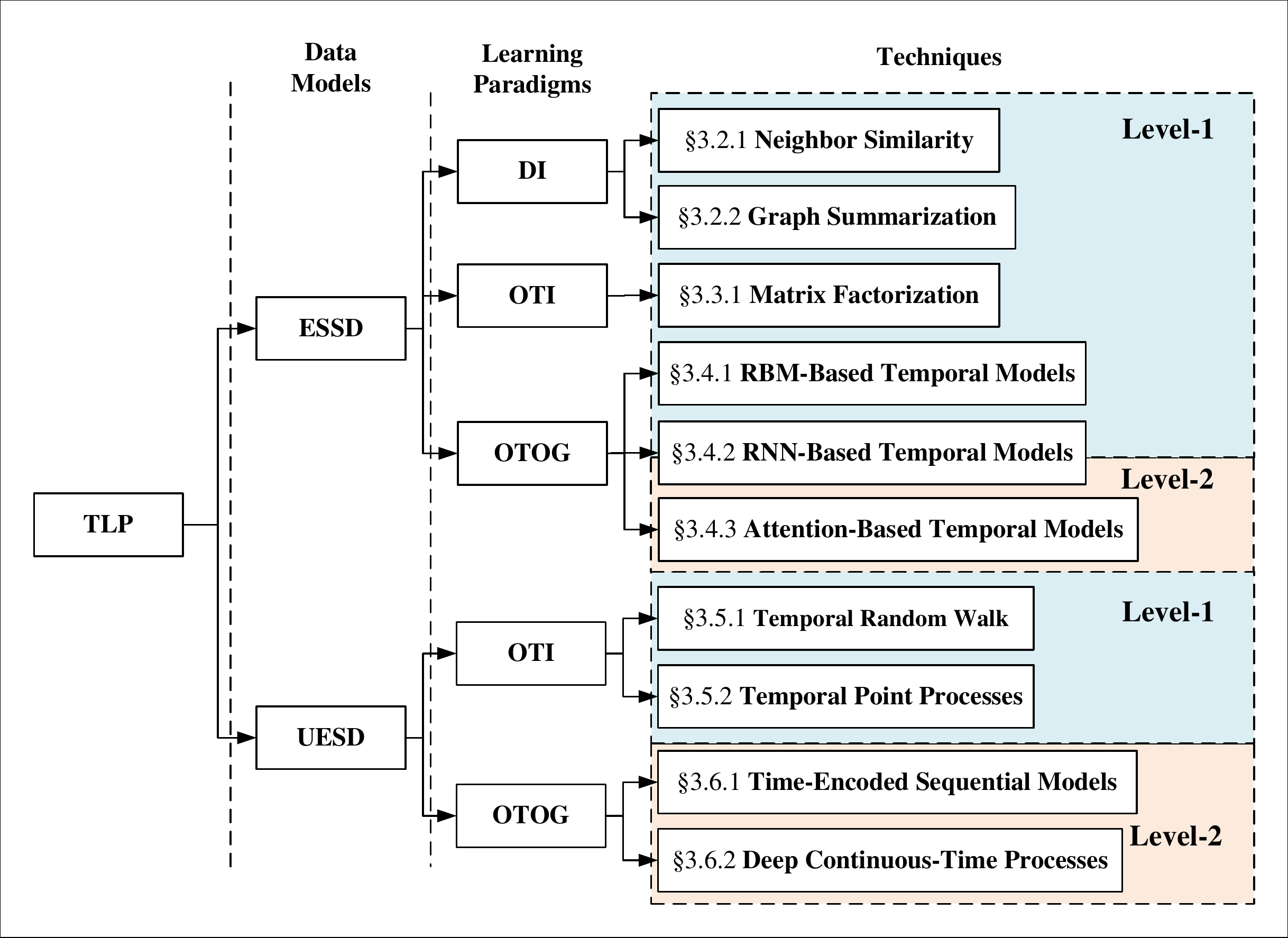}
\end{center}
\vspace{-0.25cm}
\caption{The proposed hierarchical fine-grained taxonomy of existing TLP methods.}\label{fig:taxon}
\vspace{-0.3cm}
\end{figure}

\begin{table}[t]\footnotesize
\centering
\caption{Summary of methods reviewed in this survey.}\label{tab:Meth}
\vspace{-0.25cm}
\begin{tabular}{l|l|l|l|l|l|l|l|l|l|l}
\hline
\textbf{Methods} & {\scriptsize \textbf{\begin{tabular}[c]{@{}l@{}}Data\\ Models\end{tabular}}} & {\scriptsize \textbf{\begin{tabular}[c]{@{}l@{}}Learning\\ Paradigms\end{tabular}}} & {\scriptsize \textbf{Level}} & {\scriptsize \textbf{Attributes}} & {\scriptsize \textbf{\begin{tabular}[c]{@{}l@{}}Link\\ Weights\end{tabular}}} & $L$ & $\Delta$ & $O_P$ & $O_M$ & $O_I$\\ \hline
\begin{tabular}[c]{@{}l@{}} \textit{Neighbor} \\ \textit{Similarity} \cite{liben2003link} \end{tabular} & ESSD & DI & 1 & N/A & Unable & 1 & 1 & $O_I$ & 0 & {\tiny $O(n^2)$ to $O(n^3)$}\\ \hline
\begin{tabular}[c]{@{}l@{}} \textit{Graph} \\ \textit{Summarization} \cite{sharan2008temporal} \end{tabular} & ESSD & DI & 1 & N/A & Able & $\tau$ & 1 & $O_I$ & 0 & $O(eL)$\\ \hline
\textit{CRJMF} \cite{gao2011temporal} & ESSD & OTI & 1 & Static & Able & $\tau$ & 1 & $O_M + O_I$ & {\tiny $O(eL + (e + nm)dI)$} & $O(n^2d)$\\ \hline
\textit{TLSI} \cite{zhu2016scalable} & ESSD & OTI & 1 & N/A & Able & $\tau$ & 1 & $O_M + O_I$ & {\tiny $O((e + nd)dIL)$} & $O(n^2d)$\\ \hline
\textit{MLjFE} \cite{ma2022joint} & ESSD & OTI & 1 & N/A & Able & $\tau$ & 1 & $O_M + O_I$ & {\tiny $O(eL + (e+nd)dI)$} & $O(n^2dL)$\\ \hline
\textit{GrNMF} \cite{ma2018graph} & ESSD & OTI & 1 & N/A & Able & $\tau$ & 1 & $O_M + O_I$ & {\tiny $O(eL + (e+nd)dI)$} & $O(n^2d)$\\ \hline
\textit{DeepEye} \cite{ahmed2018deepeye} & ESSD & OTI & 1 & N/A & Able & $\tau$ & 1 & $O_M + O_I$ & {\tiny $O((e + nd)dIL)$} & $O(n^2d)$\\ \hline
\textit{TMF} \cite{yu2017temporally} & ESSD & OTI & 1 & N/A & Able & $\tau$ & $\ge$1 & $O_M + O_I$ & {\tiny{$O((e + nd)dIL)$}} & $O(n^2d)$\\ \hline
\textit{LIST} \cite{yu2017link} & ESSD & OTI & 1 & N/A & Able & $\tau$ & $\ge$1 & $O_M + O_I$ & {\tiny{$O((e + nd)nIL)$}} & $O(n^2d)$\\ \hline
\textit{ctRBM} \cite{li2014deep} & ESSD & OTOG & 1 & N/A & Unable & Fixed & 1 & $O_I$ & $O_{\rm OPT}$ & $O_{\rm FFP}$\\ \hline
\textit{dyngraph2vec} \cite{goyal2020dyngraph2vec} & ESSD & OTOG & 1 & N/A & Able & Fixed & 1 & $O_I$ & $O_{\rm OPT}$ & $O_{\rm FFP}$\\ \hline
\textit{DDNE} \cite{li2018deep} & ESSD & OTOG & 1 & N/A & D/L-Dep & Fixed & 1 & $O_I$ & $O_{\rm OPT}$ & $O_{\rm FFP}$\\ \hline
\textit{EvolveGCN} \cite{pareja2020evolvegcn} & ESSD & OTOG & 2 & Dynamic & D/L-Dep & Fixed & 1 & $O_I$ & $O_{\rm OPT}$ & $O_{\rm FFP}$\\ \hline
\textit{GCN-GAN} \cite{lei2019gcn} & ESSD & OTOG & 1 & N/A & Able & Fixed & 1 & $O_I$ & $O_{\rm OPT}$ & $O_{\rm FFP}$\\ \hline
\textit{IDEA} \cite{qin2023high} & ESSD & OTOG & 2 & Static & Able & Fixed & 1 & $O_I$ & $O_{\rm OPT}$ & $O_{\rm FFP}$\\ \hline
\textit{STGSN} \cite{min2021stgsn} & ESSD & OTOG & 2 & Dynamic & D/L-Dep & Fixed & 1 & $O_I$ & $O_{\rm OPT}$ & $O_{\rm FFP}$\\ \hline
\textit{DySAT} \cite{sankar2020dysat} & ESSD & OTOG & 2 & Dynamic & Unable & Fixed & 1 & $O_I$ & $O_{\rm OPT}$ & $O_{\rm FFP}$\\ \hline
\textit{CTDNE} \cite{nguyen2018continuous} & UESD & OTI & 1 & N/A & Unable & $\tau$ & >0 & $O_M + O_I$ & $O_{\rm OPT}$ & $O(n^2d)$\\ \hline
\textit{HTNE} \cite{zuo2018embedding} & UESD & OTI & 1 & N/A & Unable & $\tau$ & >0 & $O_M + O_I$ & $O_{\rm OPT}$ & $O(n^2d)$\\ \hline
\textit{M2DNE} \cite{lu2019temporal} & UESD & OTI & 1 & N/A & Unable & $\tau$ & >0 & $O_M + O_I$ & $O_{\rm OPT}$ & $O(n^2d)$\\ \hline
\textit{TGAT} \cite{xu2020inductive} & UESD & OTOG & 2 & Dynamic & Unable & $\tau$ & >0 & $O_I$ & $O_{\rm OPT}$ & $O_{\rm FFP}$\\ \hline
\textit{CAW} \cite{wang2021inductive} & UESD & OTOG & 2 & Dynamic & Unable & $\tau$ & >0 & $O_I$ & $O_{\rm OPT}$ & $O_{\rm FFP}$\\ \hline
\textit{DyRep} \cite{trivedi2019dyrep} & UESD & OTOG & 2 & Dynamic & Unable & $\tau$ & >0 & $O_I$ & $O_{\rm OPT}$ & $O_{\rm FFP}$\\ \hline
\textit{TREND} \cite{wen2022trend} & UESD & OTOG & 2 & Static & Unable & $\tau$ & >0 & $O_I$ & $O_{\rm OPT}$ & $O_{\rm FFP}$\\ \hline
\textit{GSNOP} \cite{luo2023graph} & UESD & OTOG & 2 & Dynamic & Unable & $\tau$ & >0 & $O_I$ & $O_{\rm OPT}$ & $O_{\rm FFP}$ \\ \hline
\end{tabular}
\vspace{-0.3cm}
\end{table}

\begin{table}[t]\footnotesize
\centering
\caption{Summary of the advantages and disadvantages of different types of methods.}\label{tab:Pros-Cons}
\vspace{-0.25cm}
\begin{tabular}{c|c|c|l}
\hline
\multicolumn{1}{l|}{\textbf{\begin{tabular}[c]{@{}l@{}}Data\\ Models\end{tabular}}} & \multicolumn{1}{l|}{\textbf{\begin{tabular}[c]{@{}l@{}}Learning\\ Paradimgs\end{tabular}}} & \multicolumn{1}{l|}{} &  \\ \hline
\multirow{5}{*}{\textbf{ESSD}} & \multirow{5}{*}{\textbf{DI}} & \multirow{2}{*}{\textbf{+Pros.}} & (1) Easy to implement w/o loss functions for model optimization. \\
 &  &  & (2) Having the potential to satisfy \textbf{real-time constraints} of prediction. \\ \cline{3-4} 
 &  & \multirow{3}{*}{\textbf{-Cons.}} & (1) \textbf{Unable} to capture \textbf{complex non-linear characteristics} of topology. \\
 &  &  & (2) Only supporting \textbf{coarse-grained representations} of dynamic topology. \\
 &  &  & (3) \textbf{Unable} to support TLP in \textbf{level-2}. \\ \hline
\multirow{6}{*}{\textbf{ESSD}} & \multirow{6}{*}{\textbf{OTI}} & \multirow{2}{*}{\textbf{+Pros.}} & (1) Able to capture the \textbf{latest evolving patterns} of topology. \\
 &  &  & (2) Able to support TLP on \textbf{weighted graphs}. \\ \cline{3-4} 
 &  & \multirow{4}{*}{\textbf{-Cons.}} & (1) \textbf{Unable} to capture \textbf{non-linear characteristics} of topology. \\
 &  &  & (2) \textbf{Inefficient} for applications w/ \textbf{real-time constraints}. \\
 &  &  & (3) Only supporting \textbf{coarse-grained representations} of dynamic topology. \\
 &  &  & (4) \textbf{Unable} to support TLP in \textbf{level-2}. \\ \hline
\multirow{5}{*}{\textbf{ESSD}} & \multirow{5}{*}{\textbf{OTOG}} & \multirow{3}{*}{\textbf{+Pros.}} & (1) Having the potential to satisfy \textbf{real-time constraints} of prediction. \\
 &  &  & (2) Most of the methods or their modified versions can support TLP on \textbf{weighted graphs}. \\
 &  &  & (3) Some methods can even derive \textbf{high-quality weighted prediction results}. \\ \cline{3-4} 
 &  & \multirow{2}{*}{\textbf{-Cons.}} & (1) Only supporting \textbf{coarse-grained representations} of dynamic graphs. \\
 &  &  & (2) Having the risk of \textbf{failing} to capture the \textbf{latest evolving patterns}. \\ \hline
\multirow{5}{*}{\textbf{UESD}} & \multirow{5}{*}{\textbf{OTI}} & \multirow{2}{*}{\textbf{+Pros.}} & (1) Able to capture the \textbf{latest evolving patterns} of topology. \\
 &  &  & (2) Able to support \textbf{fine-grained representations} of dynamic topology. \\ \cline{3-4} 
 &  & \multirow{3}{*}{\textbf{-Cons.}} & (1) \textbf{Inefficient} for applications w/ \textbf{real-time constraints}. \\
 &  &  & (2) \textbf{Unable} to support TLP in \textbf{level-2}. \\
 &  &  & (3) \textbf{Unable} to support TLP on \textbf{weighted graphs}. \\ \hline
\multirow{5}{*}{\textbf{UESD}} & \multirow{5}{*}{\textbf{OTOG}} & \multirow{3}{*}{\textbf{+Pros.}} & (1) Able to support \textbf{fine-grained representations} of dynamic topology. \\
 &  &  & (2) Having the potential to satisfy \textbf{real-time constraints} of prediction. \\
 &  &  & (3) Able to support TLP in \textbf{level-2}. \\ \cline{3-4} 
 &  & \multirow{2}{*}{\textbf{-Cons.}} & (1) \textbf{Unable} to support TLP on \textbf{weighted graphs}. \\
 &  &  & (2) Having the risk of \textbf{failing} to capture the \textbf{latest evolving patterns}. \\ \hline
\end{tabular}
\vspace{-0.3cm}
\end{table}

As illustrated in Fig.~\ref{fig:taxon}, we propose a hierarchical taxonomy that can describe a TLP method in a finer-grained manner covering multiple aspects, compared with those in existing surveys. The proposed taxonomy first categorizes existing methods according to their data models (i.e., ESSD and UESD) as described in Section~\ref{Sec:Data-Model}.
Both the data models can be further categorized based on learning paradigms (i.e., DI, OTI, and OTOG) defined in Section~\ref{Sec:Learn-Parad}.
Finally, each method is characterized based on the techniques used to handle the dynamic topology.

Table~\ref{tab:Meth} summarizes details of all the methods reviewed in this survey. In addition to the data models and learning paradigms covered in our hierarchical taxonomy, we also highlight properties of the (\romannumeral1) ability to handle the variation of node sets (i.e., level-1 or -2), (\romannumeral2) availability of node attributes, (\romannumeral3) ability to capture and predict weighted links, (\romannumeral4) setting of window size $L$, and (\romannumeral5) number of future time steps or time interval $\Delta$ for prediction. Since the abilities of some \textit{task-dependent} DNE based methods to predict weighted links rely on the designs of their \textit{decoders} and \textit{loss functions}, we use `D/L-Dep' to denote that the original version of a method cannot support the weighted TLP but can be easily extended to tackle this setting by replacing the \textit{decoder} or \textit{loss function}.

The complexity of one prediction operation denoted as $O_P$ is also summarized in Table~\ref{tab:Meth}, where $O_M$ and $O_I$ denote the complexities of model optimization and inference; $n$ and $m$ are the numbers of nodes and attributes; $e$ is the maximum number of edges in training snapshots $G_{\tau-L}^{\tau}$ for ESSD; $d$ is the embedding dimensionality that is usually much smaller than $n$, $m$, and $e$; $I$ is the number of iterations in model optimization. As we consider the inference that estimates future links between all the possible $n^2$ node pairs, some methods are with $O_I \ge O(n^2)$. For DL-based methods, we could only use $O_{\rm OPT}$ and $O_{\rm FFP}$ to respectively represent their complexities of model optimization and inference (i.e., feedforward propagation of DL modules), because the complexities of some DL models are usually hard to analyze, which rely heavily on layer configurations, initialization, and optimization settings of optimizers, learning rates, and numbers of epochs. In general, $O_{\rm FFP}$ of a method is larger than $O(n^2dL)$. $O_{\rm OPT}$ is usually assumed to be time-consuming. We can further speed up the overall runtime of these DL-based methods using parallel implementations and GPUs. Consistent with our discussions in Section~\ref{Sec:Learn-Parad}, $O_P$ only includes $O_I$ (without time-consuming model optimization) for DI and OTOG approaches, thus having the potential to satisfy the real-time constraints of systems.

As the properties of a TLP method largely depend on its data model and learning paradigm, we also summarize the advantages and disadvantages of different types of approaches according to the two aspects in Table~\ref{tab:Pros-Cons}, which are detailed later at the end of each subsection (see Sections~\ref{Sec:Sum-ESSD-DI}, \ref{Sec:Sum-ESSD-OTI}, \ref{Sec:Sum-ESSD-OTOG}, \ref{Sec:Sum-UESD-OTI}, and \ref{Sec:Sum-UESD-OTOG}).

\subsection{ESSD-Based DI Methods}\label{Sec:SNAP-DI}

Some ESSD-based TLP methods adopt the following hypothesis about the evolving patterns of dynamic graphs.

\textbf{Hypothesis 3.1}. Given a dynamic graph $G=(G_1, G_2, \cdots, G_\tau, \cdots)$, \textit{snapshots close to the current time step $\tau$ should have more contributions than those far away from $\tau$} in the optimization and inference w.r.t. the prediction of ${\tilde G}_{\tau}^{\tau+\Delta}$.

Typical techniques used in existing DI methods include \textit{neighbor similarity} and \textit{graph summarization}.

\subsubsection{\textbf{Neighbor Similarity}}~

\begin{table}[t]\small
\centering
\caption{Some commonly-used similarity measures ${\rm{sim}}(v_i^{\tau}, v_j^{\tau})$ for \textit{neighbor similarity} based methods.}\label{tab:Neigh-Sim}
\vspace{-0.25cm}
\begin{tabular}{ll|ll}
\hline
\textbf{Similarity Measures} & \textbf{Definitions} & \textbf{Similarity Measures} & \textbf{Definitions} \\ \hline
\textit{Shortest Path} & $- |{\mathop{\rm SPth}\nolimits} (v_i^\tau ,v_j^\tau )|$ & \textit{Common Neighbor} & $|{\mathop{\rm Nei}\nolimits}(v_i^\tau ) \cap {\mathop{\rm Nei}\nolimits}(v_j^\tau )|$ \\ \hline
\textit{Jaccard Coefficient} & $\frac{{|{\mathop{\rm Nei}\nolimits} (v_i^\tau ) \cap {\mathop{\rm Nei}\nolimits} (v_j^\tau )|}}{{|{\mathop{\rm Nei}\nolimits} (v_i^\tau ) \cup {\mathop{\rm Nei}\nolimits} (v_j^\tau )|}}$ & \textit{Adamic-Adar} & $\sum\nolimits_{v_k^\tau  \in {\mathop{\rm Nei}\nolimits} (v_i^\tau ) \cap {\mathop{\rm Nei}\nolimits} (v_j^\tau )} {\frac{1}{{\ln |{\mathop{\rm Nei}\nolimits} (v_k^\tau )|}}}$ \\ \hline
\textit{Preferential Attachment} & $|{\mathop{\rm Nei}\nolimits}(v_i^\tau )| \cdot |{\mathop{\rm Nei}\nolimits}(v_j^\tau )|$ & \textit{Katz Index} & ${[{({\bf{I}} - \theta {{\bf{A}}_\tau })^{ - 1}} - {\bf{I}}]_{ij}}$ \\ \hline
\end{tabular}
\vspace{-0.3cm}
\end{table}

Some classic TLP methods are based on \textit{neighbor similarities} \cite{liben2003link} of dynamic graphs. These approaches use several similarity measures between nodes in current snapshot $G_\tau$ to predict next snapshot $G_{\tau+1}$, with the window size set to $L=1$. Consistent with \textbf{Hypothesis 3.1}, $G_\tau$ is assumed to have properties closest to the snapshot $G_{\tau+1}$ to be predicted. In our unified framework, the encoder and decoder of \textit{neighbor similarity} based methods can be described as
\begin{equation}
    {\mathop{\rm Enc}\nolimits} ({G_{\tau-L}^{\tau}}) \equiv G_{\tau} {~~\rm{and}~~} ({\bf{\tilde A}}_{\tau+1})_{ij} = {\mathop{\rm Dec}\nolimits} (v_i^{\tau}, v_j^{\tau}) \equiv {\mathop{\rm sim}\nolimits} (v_i^{\tau}, v_j^{\tau}),
\end{equation}
where ${\mathop{\rm sim}\nolimits} (v_i^{\tau}, v_j^{\tau})$ denotes a \textit{neighbor similarity} measure between nodes $(v_i^{\tau}, v_j^{\tau})$, which is also the output of decoder. The derived prediction result $({\bf{\tilde A}}_{\tau+1})_{ij} = {\mathop{\rm sim}\nolimits} (v_i^{\tau}, v_j^{\tau})$ is directly proportional to the probability that there is an edge between $(v_i^{\tau+1}, v_j^{\tau+1})$ in the next snapshot.

Let ${\mathop{\rm Nei}\nolimits}(v_i^{\tau})$ and ${\mathop{\rm SPth}\nolimits}(v_i^{\tau}, v_j^{\tau})$ be the set of neighbors of $v_i^{\tau}$ and the shortest path between $(v_i^{\tau}, v_j^{\tau})$. Some commonly-used \textit{neighbor similarity} measures are summarized in Table~\ref{tab:Neigh-Sim}.
These measures are based on several intuitions that nodes $v_i^{\tau}$ and $v_j^{\tau}$ are more likely to form a link if (\romannumeral1) their neighbors ${\mathop{\rm Nei}\nolimits}(v_i^{\tau})$ and ${\mathop{\rm Nei}\nolimits}(v_j^{\tau})$ have a large overlap (with several normalization strategies) or (\romannumeral2) the length of shortest path $|{\mathop{\rm SPth}\nolimits}(v_i^{\tau}, v_j^{\tau})|$ is small.
Let ${{\mathop{\rm Pth}\nolimits}_{k}}(v_i^\tau ,v_j^\tau )$ be the set of paths between $(v_i^{\tau}, v_j^{\tau})$ with length $k$. \textit{Katz index} \cite{katz1953new} in Table~\ref{tab:Neigh-Sim} is equivalent to $\sum\nolimits_{k = 1}^\infty  {{\theta ^k}|{{\mathop{\rm Pth}\nolimits}_{k}}(v_i^\tau ,v_j^\tau )|}$, which is the discounted sum of the number of paths between $(v_i^{\tau}, v_j^{\tau})$ with length $k \in [1, \infty)$. $\theta \in (0,1)$ is a pre-set decaying factor.

Since there is no model optimization procedure for \textit{neighbor similarity} based methods due to the DI paradigm, we do not need to define their loss functions.
However, they derive prediction results only based on current snapshot $G_{\tau}$ (with $L=1$), failing to explore the dynamic topology across snapshots. Furthermore, their similarity measures are usually designed only for unweighted topology, indicating that they cannot tackle the prediction of weighted links.

\subsubsection{\textbf{Graph Summarization}}~

\textit{Graph summarization} \cite{hill2006building,sharan2008temporal} is another classic technique for inference tasks on dynamic graphs. It collapses historical snapshots $G_{\tau-L}^{\tau}$ into an auxiliary weighted snapshot $G_C$ via linear combination, which is expected to preserve key properties of dynamic topology. In our encoder-decoder framework, \textit{graph summarization} can be described as
\begin{equation}
     {\mathop{\rm Enc}\nolimits} ({G_{\tau-L}^{\tau}}) \equiv {{\bf{W}}_\tau } {~~\rm{and}~~} {\bf{\tilde A}}_{\tau+1} = {\mathop{\rm Dec}\nolimits}({\bf{W}}_\tau) \equiv {\bf{W}}_\tau,
\end{equation}
where ${\bf{W}}_{\tau}$ is the adjacency matrix of the collapsed snapshot $G_C$ and is directly adopted as the prediction result ${\bf{\tilde A}}_{\tau+1}$.
For the prediction of unweighted graphs, each element $({\bf{\tilde A}}_{\tau+1})_{ij} = ({\bf{W}}_{\tau})_{ij}$ is directly proportional to the probability that there is an edge between $(v_i^{\tau+1}, v_j^{\tau+1})$. The normalized value of $({\bf{W}}_{\tau})_{ij}$ (e.g., averaging ${\bf{W}}_{\tau}$ over the window size $L$) can also be the predicted edge weight of $(v_i^{\tau+1}, v_j^{\tau+1})$ for weighted graphs.

Different variants of \textit{graph summarization} only differ in terms of their encoders. Typical variants include the \textit{exponential kernel} $K_E(\cdot)$ \cite{hill2006building}, \textit{inverse linear kernel} $K_{IL}(\cdot)$ \cite{sharan2008temporal}, and \textit{linear kernel} $K_L(\cdot)$ \cite{sharan2008temporal}, which are defined as
\begin{equation}\label{Eq:K-E}
    {\rm{Enc}}(G_{\tau  - L}^\tau ) = {K_E}({\bf{A}}_{\tau  - L}^\tau ) \equiv \sum\nolimits_{t = \tau  - L + 1}^\tau  {{{(1 - \theta )}^{\tau  - t}}\theta {{\bf{A}}_t}},
\end{equation}
\begin{equation}\label{Eq:K-IL}
    {\rm{Enc}}(G_{\tau  - L}^\tau ) = {K_{IL}}({\bf{A}}_{\tau  - L}^\tau ) \equiv \sum\nolimits_{t = \tau  - L + 1}^\tau  {\frac{1}{{(\tau  - t) + 1}}\theta {{\bf{A}}_t}},
\end{equation}
\begin{equation}\label{Eq:K-L}
    {\rm{Enc}}(G_{\tau  - L}^\tau ) = {K_L}({\bf{A}}_{\tau  - L}^\tau ) \equiv \sum\nolimits_{t = \tau  - L + 1}^\tau  {\frac{{L + 1}}{{(\tau  - t) + 1}}\theta {{\bf{A}}_t}},
\end{equation}
where $\theta \in [0, 1]$ is a tunable parameter. The three kernels adopt different decaying rates for adjacency matrices ${\bf{A}}_{\tau-L}^{\tau}$ w.r.t. historical topology to ensure \textbf{Hypothesis 3.1}.
Due to the DI paradigm, \textit{graph summarization} does not have the loss function for model optimization.
In some cases, one can combine \textit{neighbor similarity} with \textit{graph summarization} by applying \textit{neighbor similarities} (see Table~\ref{tab:Neigh-Sim}) to the predicted adjacency matrix ${\bf{\tilde A}}_{\tau+1}$ to refine the prediction result.

\subsubsection{\textbf{Summary of ESSD-Based DI Methods}}\label{Sec:Sum-ESSD-DI}~

For DI methods, there are no loss functions defined for model optimization in our encoder-decoder framework. Therefore, they are easy to implement and have the potential to satisfy the real-time constraints of applications.
However, these methods cannot fully capture the complex non-linear characteristics of dynamic graphs because they still rely on simple intuitions and linear models. Furthermore, they can only support coarse-grained representations of dynamic graphs due to the limitations of ESSD.
As the aforementioned approaches are based on heuristics designed for graphs with fixed node sets, they can only support the TLP in level-1, failing to tackle the variation of node sets.

\subsection{ESSD-Based OTI Methods}\label{Sec:SNAP-OTI}

\subsubsection{\textbf{Matrix Factorization}}\label{sec:OTI-SNAP-MF}~

Most ESSD-based OTI methods combine matrix factorization techniques \cite{huang2012non,de2021survey} with \textbf{Hypothesis 3.1}.
They decompose adjacency matrices ${\bf{A}}_{\tau-L}^{\tau}$ or their transformations into low-dimensional matrices (e.g., $\arg {\min} _{{{\bf{U}}_t},{{\bf{V}}_t}} || {{{\bf{A}}_t} - {{\bf{U}}_t}{\bf{V}}_t^T} ||_F^2$) with key properties of historical topology preserved.
The derived latent matrices can be used to `reconstruct' the adjacency matrix of a future snapshot via an inverse process of matrix factorization (e.g., ${{{\bf{\tilde A}}}_{\tau  + 1}} = {\bf{U}}_\tau{\bf{V}}_\tau ^{T}$).

Non-negative matrix factorization (NMF) \cite{lee1999learning,huang2012non} is a typical technique adopted by related methods, where there are non-negative constraints on the latent matrices to be learned (e.g., $\arg {\min} _{{{\bf{U}}_t} \ge 0,{{\bf{V}}_t} \ge 0} || {{{\bf{A}}_t} - {{\bf{U}}_t}{\bf{V}}_t^T} ||_F^2$). NMF-based approaches are usually optimized via specific multiplicative procedures \cite{seung2001algorithms} instead of classic additive optimization algorithms (e.g., gradient descent). We leave details regarding the model optimization of NMF in supplementary materials. In our encoder-decoder framework, most NMF-based methods have similar definitions of decoders but differ from their encoders and loss functions.

(1) \textbf{\textit{CRJMF}}.
Gao et al. \cite{gao2011temporal} proposed \textit{CRJMF}, which extends \textit{graph summarization} to incorporate additional node attributes and neighbor similarity using NMF. It assumes that node attributes, described by a matrix ${\bf{X}} \in \Re^{N \times M}$, are available and fixed for all snapshots. A similarity matrix ${\bf{S}} \in \Re^{N \times N}$ is also introduced to describe the neighbor similarity, where ${\bf{S}}_{ij}$ is the \textit{common neighbor} measure between $(v_i^{\tau}, v_j^{\tau})$ as defined in Table~\ref{tab:Neigh-Sim}. Given historical snapshots $G_{\tau-L}^{\tau}$ and fixed node attributes $\mathcal{A}$, the encoder and loss function of \textit{CRJMF} can be described as
\begin{equation}\label{Eq:Enc-CRJMF}
    {\mathop{\rm Enc}\nolimits} (G_{\tau  - L}^\tau , \mathcal{A}) \equiv \mathop {\arg \min }\limits_{{\bf{U}} \ge 0,{\bf{V}} \ge 0,{\bf{Y}} \ge 0} \mathcal{L}(G_{\tau  - L}^\tau , \mathcal{A}; {\bf{U}},{\bf{V}},{\bf{Y}}) \equiv \left\| {{{\bf{W}}_\tau } - {\bf{UY}}{{\bf{U}}^T}} \right\|_F^2 + \alpha \left\| {{\bf{X}} - {\bf{U}}{{\bf{V}}^T}} \right\|_F^2 + \beta {\mathop{\rm tr}\nolimits} ({{\bf{U}}^T}{{\bf{L}}_{\bf{S}}}{\bf{U}}),
\end{equation}
where $\{ {\bf{U}} \in \Re^{N \times d}, {\bf{V}} \in \Re^{N \times d}, {\bf{Y}} \in \Re^{d \times d}\}$ are latent matrices to be optimized and outputs of the encoder; ${\bf{W}}_{\tau}$ is derived from the \textit{exponential kernel} of \textit{graph summarization} defined in (\ref{Eq:K-E}); ${\bf{L}}_{{\bf{S}}} = {\bf{D}}_{{\bf{S}}} - {\bf{S}}$ is the Laplacian matrix of ${\bf{S}}$; ${\bf{D}}_{\bf{S}} = {\rm{diag}}(d_1^{\bf{S}}, d_2^{\bf{S}}, \cdots, d_N^{\bf{S}})$ is the degree diagonal matrix of ${\bf{S}}$ with $d_i^{\bf{S}} = \sum\nolimits_j {{{\bf{S}}_{ij}}}$; $\alpha$ and $\beta$ are parameters to adjust the second and third terms.
In particular, the third term is defined as \textit{graph regularization} \cite{cai2010graph}, which can be rewritten as ${\mathop{\rm tr}\nolimits} ({{\bf{U}}^T}{{\bf{L}}_{\bf{S}}}{\bf{U}}) = 0.5 \cdot \sum\nolimits_{i,j = 1}^N {{{\bf{S}}_{ij}}|{{\bf{U}}_{i,:}} - {{\bf{U}}_{j,:}}|_2^2}$. It can apply additional regularization encoded in ${\bf{S}}$ (i.e., the second-order neighbor similarity) to ${\bf{U}}$, where larger ${\bf{S}}_{ij}$ forces $( {\bf{U}}_{i, :}, {\bf{U}}_{j, :} )$ to be closer in the latent space.

Since ${\bf{U}}$ is shared by all the terms in (\ref{Eq:Enc-CRJMF}), it is expected to jointly encode the properties of historical topology, node attributes, and neighbor similarity after model optimization, while $\{{\bf{V}}, {\bf{Y}}\}$ are auxiliary variables.
Let $\{{{\bf{U}}^*},{{\bf{V}}^*},{{\bf{Y}}^*}\}$ be the solution to objective (\ref{Eq:Enc-CRJMF}). The decoder executes an inverse process of the first term in (\ref{Eq:Enc-CRJMF}), which is defined as
\begin{equation}
    {{{\bf{\tilde A}}}_{\tau  + 1}} = {\mathop{\rm Dec}\nolimits} ({{\bf{U}}^*},{{\bf{Y}}^*}) \equiv {{\bf{U}}^*}{{\bf{Y}}^*}{{\bf{U}}^*}^T.
\end{equation}
However, \textit{CRJMF} still relies on \textit{graph summarization} to explore dynamic topology, which simply collapses historical snapshots into an auxiliary weighted snapshot described by ${\bf{W}}_{\tau}$. In contrast, some other NMF-based approaches directly extract latent characteristics from the raw dynamic topology.

(2) \textbf{\textit{TLSI}}.
Zhu et al. \cite{zhu2016scalable} proposed \textit{TLSI}, which automatically learns latent representations regarding dynamic topology based on the \textit{temporal smoothness intuition} that \textit{nodes change their representations smoothly over time}.
Given historical snapshots $G_{\tau-L}^{\tau}$, the encoder and loss function of \textit{TLSI} are defined as
\begin{equation}\label{Eq:Enc-BCGD}
    {\mathop{\rm Enc}\nolimits} (G_{\tau  - L}^\tau ) \equiv \mathop {\arg \min }\limits_{{\bf{U}}_{\tau - L}^\tau  \ge 0} \mathcal{L}(G_{\tau  - L}^\tau ;{\bf{U}}_{\tau  - L}^\tau ) \equiv \sum\limits_{t = \tau  - L+1}^\tau  {\left\| {{{\bf{A}}_t} - {{\bf{U}}_t}{\bf{U}}_t^T} \right\|_F^2}  + \beta \sum\limits_{t = \tau  - L + 2}^\tau  {{\mathop{\rm tr}\nolimits} ({{\bf{U}}_t}{\bf{U}}_{t - 1}^T - {\bf{I}})} \rm{~s.t.~} {\mathop{\rm tr}\nolimits} ({{\bf{U}}_t}{\bf{U}}_t^T - {\bf{I}}) = 0,
\end{equation}
where $\{ {\bf{U}}_t \in \Re^{N \times d}\}$ are latent matrices to be learned and outputs of the encoder, with $({\bf{U}}_t)_{i,:}$ as the representation of node $v_i^t$.
The first term is the standard symmetric NMF, which enables ${\bf{U}}_t$ to encode the structural properties of snapshot $G_t$. Namely, nodes $(v_i^t, v_j^t)$ close to each other in $G_t$ should have close representations $(({\bf{U}}_t)_{i,:}, ({\bf{U}}_t)_{j,:})$. ${{\mathop{\rm tr}\nolimits} ({{\bf{U}}_t}{\bf{U}}_{t - 1}^T - {\bf{I}})}$ is the \textit{temporal smoothness term} of time step $t$ that penalizes each node for suddenly changing its representation, following the \textit{temporal smoothness intuition}. In this setting, $\{ {\bf{U}}_t\}$ can also capture temporal characteristics, where successive snapshots have close latent representations.
$\beta$ is a parameter to balance the objectives of NMF and \textit{temporal smoothness}. The constraint ${\mathop{\rm tr}\nolimits} ({\bf{U}}_t^T{{\bf{U}}_t} - {\bf{I}}) = 0$ ensures that ${\bf{U}}_{\tau-L}^{\tau}$ are normalized for each node.

Let $\{ {\bf{U}}_t^* \}$ be the solution to the aforementioned objective. The decoder derives the prediction result ${\bf{\tilde A}}_{\tau+1}$ by executing an inverse process of symmetric NMF, which can be described as
\begin{equation}\label{Eq:Dec-BCGD}
    {{{\bf{\tilde A}}}_{\tau  + 1}} = {\mathop{\rm Dec}\nolimits} ({\bf{U}}_\tau ^*) \equiv {\bf{U}}_\tau ^*{\bf{U}}{_\tau^*}^T.
\end{equation}

(3) \textbf{\textit{MLjFE}}. Also based on \textit{temporal smoothness}, Ma et al. \cite{ma2022joint} proposed \textit{MLjFE}. In addition to dynamic topology, a point-wise mutual information matrix ${{\bf{M}}_t} = \ln (\sum\nolimits_{r = 1}^h {{{({{\bf{A}}_t})}^h}} /h)$ is introduced to encode the high-order proximity of each snapshot $G_t$ (i.e., multi-step neighbor similarities beyond the observable topology described by ${\bf{A}}_t$), with $h$ as the order to be specified.
Given historical snapshots $G_{\tau-L}^{\tau}$, the encoder and loss function can be described as
\begin{equation}\label{Eq:Enc-MLjFE}
\begin{array}{l}
    {\mathop{\rm Enc}\nolimits} (G_{\tau  - L}^\tau ) \equiv {\arg \min }_{{\bf{U}}_{\tau  - L}^\tau  \ge 0,{\bf{V}}_{\tau  - L}^\tau  \ge 0, {\bf{Y}}_{\tau  - L}^\tau \ge 0} ~ \mathcal{L}(G_{\tau  - L}^\tau ;{\bf{U}}_{\tau  - L}^\tau ,{\bf{V}}_{\tau  - L}^\tau ,{\bf{Y}}_{\tau  - L}^\tau )\\
    \equiv \sum\limits_{t = \tau - L+1}^\tau  {[\left\| {{{\bf{A}}_t} - {{\bf{U}}_t}{\bf{V}}_t^T} \right\|_F^2 + \alpha \left\| {{{\bf{M}}_t} - {{\bf{V}}_t}{\bf{Y}}_t^T} \right\|_F^2]}  + \beta \sum\limits_{t = \tau  - L + 2}^\tau  {\left\| {{{\bf{U}}_t} - {{\bf{U}}_{t - 1}}} \right\|_F^2} {\rm{~s.t.~}} {\mathop{\rm tr}\nolimits} ({\bf{V}}_t^T{{\bf{V}}_t} - {\bf{I}}) = 0
\end{array},
\end{equation}
where $\{ {\bf{U}}_t \in \Re^{N \times d}, {\bf{V}}_t \in \Re^{N \times d}, {\bf{Y}}_t \Re^{N \times d}\}$ are latent matrices to be learned; ${|| {{{\bf{U}}_t} - {{\bf{U}}_{t - 1}}} ||_F^2}$ is the \textit{temporal smoothness term} with the same physical meaning as that in (\ref{Eq:Enc-BCGD}); $\{\alpha, \beta \}$ are parameters to adjust the contributions of high-order proximity and temporal smoothness. Similar to (\ref{Eq:Enc-BCGD}), the constraint ${\mathop{\rm tr}\nolimits} ({\bf{V}}_t^T{{\bf{V}}_t} - {\bf{I}}) = 0$ normalizes $\{ {\bf{V}}_t \}$ for each node.

After model optimization, $ {\bf{V}}_{\tau-L}^{\tau} $ shared by the first and second terms can comprehensively encode the temporal characteristics and high-order proximity of successive snapshots, while ${\bf{U}}_{\tau-L}^{\tau}$ and ${\bf{Y}}_{\tau-L}^{\tau}$ are auxiliary variables. Let $\{ {\bf{U}}_t^*, {\bf{V}}_t^*, {\bf{Y}}_t^* \}$ be the solution to objective (\ref{Eq:Enc-MLjFE}). The decoder of \textit{MLjFE} derives the prediction result ${\bf{\tilde A}}_{\tau+1}$ based on $\{ {\bf{U}}_t^*, {\bf{V}}_t^* \}$:
\begin{equation}
    {{{\bf{\tilde A}}}_{\tau  + 1}} = {\mathop{\rm Dec}\nolimits} ({\bf{U}}_t^*,{\bf{V}}_t^*) \equiv \sum\nolimits_{t = \tau  - L+1}^\tau  {{\theta ^{\tau  - t}}{\bf{U}}_t^*{\bf{V}}{{_t^*}^T}},
\end{equation}
where $\theta \in [0, 1]$ is a pre-set time decaying factor to ensure \textbf{Hypothesis 3.1}.

(4) \textbf{\textit{GrNMF}}.
In addition to \textit{temporal smoothness}, \textit{GrNMF} \cite{ma2018graph} uses the NMF-based \textit{graph regularization} \cite{cai2010graph} to explore evolving patterns of dynamic graphs. Given historical snapshots $G_{\tau-L}^{\tau}$, the encoder and loss function of \textit{GrNMF} are
\begin{equation}\label{Eq:Enc-GrNMF}
    {\mathop{\rm Enc}\nolimits} (G_{\tau  - L}^\tau ) \equiv {\arg \min }_{{\bf{U}} \ge 0,{\bf{V}} \ge 0}~ \mathcal{L}(G_{\tau-L}^{\tau}; {\bf{U}}, {\bf{V}}) \equiv \left\| {{{\bf{A}}_\tau } - {\bf{U}}{{\bf{V}}^T}} \right\|_F^2 + \alpha \sum\nolimits_{t = \tau  - L + 1}^{\tau  - 1} {{\theta ^{\tau  - t}}{\rm{tr}}({{\bf{V}}^T}{{\bf{L}}_t}{\bf{V}})},
\end{equation}
where $\{ {\bf{U}} \in \Re^{N \times d}, {\bf{V}} \in \Re^{N \times d}\}$ are latent matrices to be optimized; ${\bf{L}}_t$ is the Laplacian matrix of ${\bf{A}}_t$ with the same definition as ${\bf{L}}_{\bf{S}}$ in (\ref{Eq:Enc-CRJMF}); $\alpha>0$ and $\theta \in [0, 1]$ are parameters to be specified.
The first term is the standard NMF objective to learn $\{ {\bf{U}}, {\bf{V}}\}$ that preserve structural properties of current snapshot $G_\tau$. The second \textit{graph regularization} term \cite{cai2010graph}, with a physical meaning similar to that in (\ref{Eq:Enc-CRJMF}), incorporates regularization regarding previous snapshots $G_{\tau-L}^{\tau-1}$ to ${\bf{V}}$, enabling ${\bf{V}}$ to capture evolving patterns of successive snapshots. In this setting, $\theta$ and $\alpha$ are used to ensure \textbf{Hypothesis 3.1} and adjust the \textit{graph regularization} term, respectively.

Let $\{ {\bf{U}}^*, {\bf{V}}^* \}$ be the learned latent matrices. The decoder executes an inverse process of NMF such that
\begin{equation}\label{Eq:Dec-GrNMF}
    {{{\bf{\tilde A}}}_{\tau  + 1}} = {\mathop{\rm Dec}\nolimits} ({{\bf{U}}^*},{{\bf{V}}^*}) \equiv {{\bf{U}}^*}{{\bf{V}}^{*T}}.
\end{equation}

(5) \textbf{\textit{DeepEye}}.
Ahmed et al. \cite{ahmed2018deepeye} developed \textit{DeepEye} that explores dynamic topology via the linear combination of multiple NMF components w.r.t. historical snapshots. Given $G_{\tau-L}^{\tau}$, the encoder and loss function are defined as
\begin{equation}\label{Eq:loss-DeepEye}
    \begin{array}{l}
    {\mathop{\rm Enc}\nolimits} (G_{\tau  - L}^\tau ) \equiv \mathop {\arg \min }_{{\bf{U}}_{\tau  - L}^\tau \ge 0,{\bf{V}}_{\tau  - L}^\tau \ge 0,{\bf{U} \ge 0} ,{\bf{V}} \ge 0}~\mathcal{L} (G_{\tau  - L}^\tau ;{\bf{U}}_{\tau  - L}^\tau ,{\bf{V}}_{\tau  - L}^\tau ,{\bf{U}},{\bf{V}})\\
    \equiv \sum\nolimits_{t = \tau  - L + 1}^\tau  {{\theta ^{\tau  - t}}[\left\| {{{\bf{A}}_t} - {{\bf{U}}_t}{\bf{V}}_t^T} \right\|_F^2 + \left\| {{{\bf{U}}_t} - {\bf{U}}} \right\|_F^2 + \left\| {{{\bf{V}}_t} - {\bf{V}}} \right\|_F^2]}
    \end{array},
\end{equation}
where $\{ {\bf{U}} \in \Re^{N \times d}, {\bf{V}} \in \Re^{N \times d}, {\bf{U}}_t \in \Re^{N \times d}, {\bf{V}}_t \in \Re^{N \times d} \}$ are latent matrices to be learned; $\theta \in [0, 1]$ is the decaying factor to ensure \textbf{Hypothesis 3.1}.
In (\ref{Eq:loss-DeepEye}), each NMF component ${|| {{{\bf{A}}_t} - {{\bf{U}}_t}{\bf{V}}_t^T} ||_F^2}$ corresponds to a unique snapshot $G_t$, enabling $\{ {\bf{U}}_t, {\bf{V}}_t\}$ to capture structural properties of $G_t$. ${|| {{{\bf{U}}_t} - {\bf{U}}} ||_F^2}$ and ${|| {{{\bf{V}}_t} - {\bf{V}}} ||_F^2}$ further force $\{{\bf{U}}, {\bf{V}} \}$ to comprehensively capture the dynamic structural properties w.r.t. successive snapshots $G_{\tau-L}^{\tau}$.
Let $\{{\bf{U}}^*, {\bf{V}}^*, {\bf{U}}_t^*, {\bf{V}}_t^*\}$ be the solution to the aforementioned objective, \textit{DeepEye} has the same decoder as \textit{GrNMF} defined in (\ref{Eq:Dec-GrNMF}).

(6) \textbf{\textit{TMF}}.
In addition to NMF, some other related methods are based on the general matrix factorization without non-negative constraints.
Yu et al. \cite{yu2017temporally} formulated the dynamic topology as a function of time with learnable parameters and introduced \textit{TMF}.
Given historical snapshots $G_{\tau-L}^{\tau}$, the encoder and loss function of \textit{TMF} are defined as
\begin{equation}\label{Eq:Enc-TMF}
    \begin{array}{l}
    {\mathop{\rm Enc}\nolimits} (G_{\tau  - L}^\tau ) \equiv {\arg \min }_{{\bf{U}}, \{{{\bf{V}}^{(i)}}\} } ~\mathcal{L}(G_{\tau  - L}^\tau ;{\bf{U}},\{ {{\bf{V}}^{(i)}} \})\\
    \equiv \sum\limits_{t = \tau - L+1}^\tau  {{D_t}\left\| {{{\bf{E}}_t} \odot [{{\bf{A}}_t} - {\bf{U}}{{[\sum\nolimits_{i = 0}^h {{{\bf{V}}^{(i)}}{{(t - (\tau - L) )}^i}} ]}^T}]} \right\|_F^2}  + \alpha \left\| {\bf{U}} \right\|_F^2 + \sum\limits_{i = 0}^h {{\beta _i}\left\| {{{\bf{V}}^{(i)}}} \right\|_F^2} 
\end{array},
\end{equation}
where $\{ {\bf{U}} \in \Re^{N \times d}, {\bf{V}}^{(i)} \in \Re^{N \times d} (0 \le i \le h) \}$ are model parameters to be optimized, with $h$ as a tunable feature order; $D_t = {e^{ - \theta (\tau  - t)}}$ ($\theta>0$) is the time decaying factor to ensure \textbf{Hypothesis 3.1};
$\alpha$ and $\beta_i$ are parameters for the $l_2$-regularization regarding $\{ {\bf{V}}, {\bf{V}}^{(i)}\}$ to avoid overfitting.
${\bf{E}}_t \in \Re^{N \times N}$ is an auxiliary matrix such that ${({{\bf{E}}_t})_{ij}} = 1$ if $({\bf{A}}_t)_{ij}>0$ and ${({{\bf{E}}_t})_{ij}} = 0$ otherwise. It ensures that only the pair of nodes $(v_i^t, v_j^t)$ with an edge at time step $t$ can contribute to the model optimization.
Different from NMF-based methods (e.g., \textit{TLSI}, \textit{GrNMF}, and \textit{DeepEye}) that encode structural properties of each snapshot in one or more matrices (e.g., ${\bf{Y}}_t$ in ${\min} _{{{\bf{Y}}_t} \ge 0} || {{{\bf{A}}_t} - {{\bf{Y}}_t}{\bf{Y}}_t^T} ||_F^2$), the model parameters $\{ {\bf{U}}, {\bf{V}}^{(i)} \}$ of \textit{TMF} are shared by all time steps.
In the first term, ${{\bf{V}}_t} \equiv \sum\nolimits_i {{{\bf{V}}^{(i)}}{{[(t - (\tau - L)]}^i}}$ is a time-induced representation w.r.t. snapshot $G_t$, which is also a function regarding time index $t$.
\textit{TMF} is optimized via the classic gradient descent algorithm.
Let $\{ {\bf{U}}^*, {\bf{V}}^{(i)*}\}$ be the learned model parameters. The decoder of \textit{TMF} is defined as
\begin{equation}\label{Eq:Dec-TMF}
    {{{\bf{\tilde A}}}_{\tau  + r }} = {\rm{Dec}}({{\bf{U}}^*}, \{{{\bf{V}}^{(i)*}}\} ) \equiv {{\bf{U}}^*}{[\sum\nolimits_{i = 0}^h {{{\bf{V}}^{(i)}}^*{{[(\tau  + r ) - (\tau  - L)]}^i}} ]^T} = {{\bf{U}}^*}{[\sum\nolimits_{i = 0}^h {{{\bf{V}}^{(i)*}}{{(L + r )}^i}} ]^T},
\end{equation}
which uses the time-induced representation ${\bf{V}}_{\tau+r}^{*}$ w.r.t. a future time step $(\tau+r)$ to derive prediction result ${\bf{\tilde A}}_{\tau+r}$. In this setting, \textit{TMF} can support the multi-step prediction with $1 \le r \le \Delta$.

(7) \textbf{\textit{LIST}}.
Also based on the motivation of modeling dynamic topology as a function of time, Cheng et al. \cite{yu2017link} proposed \textit{LIST} that extends \textit{TMF} to incorporate multi-step label propagation on each snapshot.
Given historical snapshots $G_{\tau-L}^{\tau}$, the encoder and loss function of \textit{LIST} can be described as
\begin{equation}
    {\mathop{\rm Enc}\nolimits} (G_{\tau  - L}^\tau ) \equiv \mathop {\arg \min }\limits_{ \{ {{\bf{V}}^{(i)}} \} } \mathcal{L}(G_{\tau  - L}^\tau ;\{ {{\bf{V}}^{(i)}} \} ) \equiv \sum\limits_{t = \tau - L+1}^\tau  {{D_t}\left\| {{{\bf{E}}_t} \odot ({{\bf{A}}_t} - {{\bf{U}}_t}{{\bf{V}}_t}{\bf{V}}_t^T{\bf{U}}_t^T)} \right\|_F^2}  + \sum\limits_{i = 0}^h {{\beta _i}\left\| {{{\bf{V}}^{(i)}}} \right\|_F^2},
\end{equation}
where ${{\bf{V}}_t} \equiv \sum\nolimits_i {{{\bf{V}}^{(i)}}{{[t - (\tau  - L)]}^i}}$ is the time-induced representation w.r.t. time step $t$; $\{ {\bf{V}}^{(i)} \in \Re^{N \times d} (0 \le i \le h)\}$ are model parameters to be optimized with a tunable order $h$; $D_t$ and ${\bf{E}}_t$ are with the same definitions and physical meanings as those of \textit{TMF} described in (\ref{Eq:Enc-TMF}). Furthermore, ${{\bf{U}}_t} \equiv (1 - \hat \theta ){({\bf{I}} - \hat \theta {{\bf{\hat A}}_t})^{ - 1}}$ (with ${{{\bf{\hat A}}}_t} = {\bf{D}}_t^{ - 1/2}{{\bf{A}}_t}{\bf{D}}_t^{ - 1/2}$ and $\hat \theta \in (0, 1)$ as a pre-set parameter) is an auxiliary variable regarding the analytical solution of multi-step label propagation \cite{cheng2016ranking} on each snapshot $G_t$ (see \cite{yu2017link} for its details), which captures the high-order proximity of $G_t$. Similar to \textit{TMF}, \textit{LIST} is also optimized via gradient descent.
Let $\{ {{{\bf{V}}^{(i)*}}} \}$ be the learned model parameters. The decoder of \textit{LIST} is defined as
\begin{equation}
   {{{\bf{\tilde A}}}_{\tau  + r }} = {\rm{Dec}}(\{ {{\bf{V}}^{(i)}}^*\} ) \equiv {\bf{V}}_{\tau  + r}^*{\bf{V}}_{\tau  + r}^{*T} = [\sum\nolimits_{i = 0}^h {{{\bf{V}}^{(i)*}}{{(L + r )}^i}} ]{[\sum\nolimits_{i = 0}^h {{{\bf{V}}^{(i)*}}{{(L + r )}^i}} ]^T},
\end{equation}
which uses a strategy similar to that of \textit{TMF} in (\ref{Eq:Dec-TMF}) to derive the prediction result ${\bf{\tilde A}}_{\tau+r}$ with $1 \le r \le \Delta$.

\subsubsection{\textbf{Summary of ESSD-Based OTI Methods}}\label{Sec:Sum-ESSD-OTI}~

Compared with conventional \textit{neighbor similarity} and \textit{graph summarization} techniques based on manually designed heuristics, the aforementioned matrix factorization methods can automatically extract latent characteristics from dynamic topology. Some of them can also incorporate additional information (e.g., node attributes and high-order proximity) beyond the observable topology described by adjacency matrices.
Moreover, all the ESSD-based OTI methods reviewed in this subsection can support TLP on weighted graphs by using adjacency matrices to describe weighted topology.
However, they are still based on linear models (e.g., NMF) that cannot capture the non-linear characteristics of dynamic graphs. Due to the limitations of ESSD, they can only support coarse-grained representations of dynamic topology but may fail to handle the rapid evolution of systems. Following the OTI paradigm, they are designed and optimized for one prediction operation. Although this paradigm can capture the latest evolving patterns, when it comes to a new time step, we still need to optimize the model from scratch, which is inefficient for applications with real-time constraints. Since the dimensionality of model parameters is related to the number of nodes, the aforementioned approaches can only support the TLP in level-1 but fail to handle the variation of node sets in level-2.

\subsection{ESSD-Based OTOG Methods}

Most ESSD-based OTOG methods use deep learning (DL) techniques to handle dynamic topology.
We categorize this type of method based on the DL module used to explore the evolving patterns across snapshots, including \textit{restricted Boltzmann machines} (RBM) \cite{zhang2018overview,ghojogh2021restricted}, \textit{recurrent neural networks} (RNNs) \cite{gers2000learning,chung2014empirical}, and \textit{attention mechanisms} \cite{vaswani2017attention,niu2021review}.

\begin{figure}[t]
\begin{center}
 \includegraphics[width=0.65\linewidth, trim=18 18 18 18,clip]{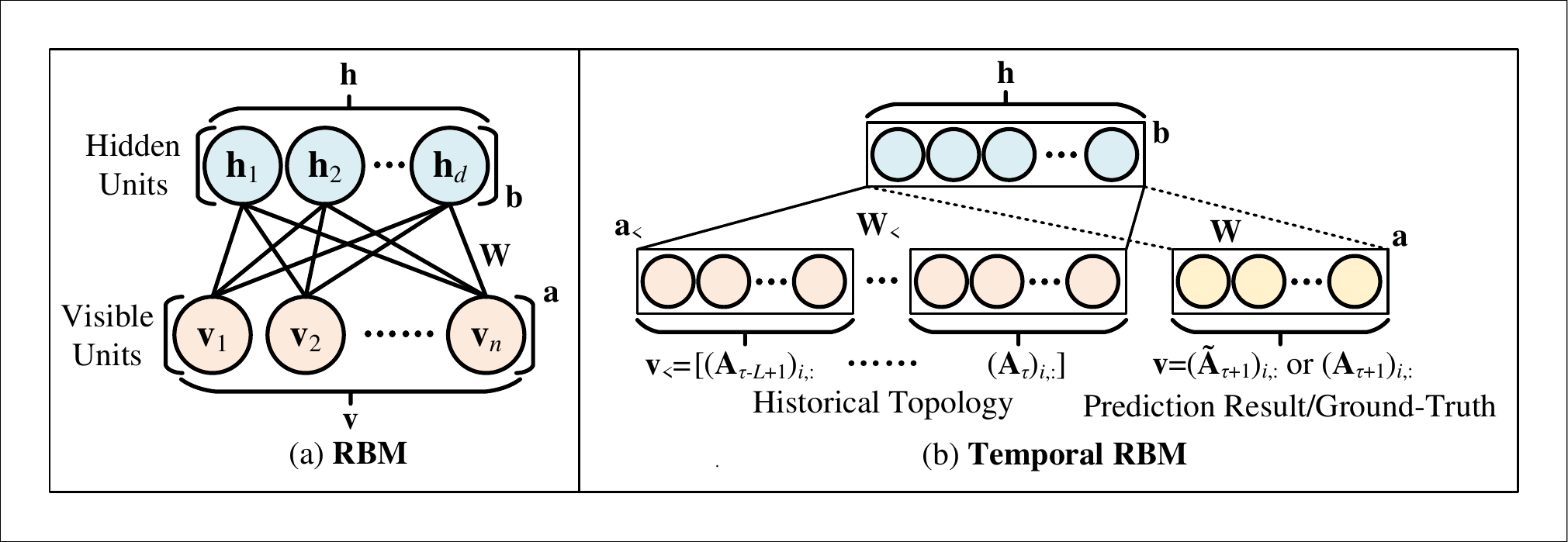}
\end{center}
\vspace{-0.25cm}
\caption{Overviews of the (a) standard RBM and (b) temporal RBM.}\label{fig:rbm}
\vspace{-0.35cm}
\end{figure}

\subsubsection{\textbf{RBM-Based Temporal Models}}~

RBM \cite{zhang2018overview,ghojogh2021restricted}, with an overview depicted in Fig.~\ref{fig:rbm} (a), is a DL structure that contains a layer of $N$ visible units ${\bf{v}} \in \Re^{N \times 1}$ and a layer of $d$ hidden units ${\bf{h}} \in \Re^{d \times 1}$, forming a fully-connected network between the two layers. ${\bf{v}}$ and ${\bf{h}}$ are stochastic binary units (i.e., ${\bf{v}}_i \in \{ 0, 1\}$ and ${\bf{h}}_j \in \{ 0, 1 \}$) that encode the observable data and latent representations. In addition, there is a weight matrix ${\bf{W}} \in \Re^{N \times d}$ and two bias vectors $\{ {\bf{a}} \in \Re^{N \times 1}, {\bf{b}} \in \Re^{d \times 1} \}$ for $\{ {\bf{v}}, {\bf{h}} \}$. Such a structure defines a joint distribution over $\bf{h}$ and ${\bf{v}}$ that
\begin{equation}
    P({\bf{v}},{\bf{h}}) \equiv \frac{1}{Z}\exp \{  - E({\bf{v}},{\bf{h}})\}  = \frac{1}{Z}\exp \{ {{\bf{v}}^T}{\bf{Wh}} + {{\bf{a}}^T}{\bf{v}} + {{\bf{b}}^T}{\bf{h}}\},
\end{equation}
where $E({\bf{v}},{\bf{h}}) \equiv  - {{\bf{v}}^T}{\bf{Wh}} - {{\bf{a}}^T}{\bf{v}} - {{\bf{b}}^T}{\bf{h}}$ is the energy function; $Z$ is a normalizing factor to ensure the normalization constraint of probability (i.e., $\sum {P({\bf{v}},{\bf{h}})} = 1$); $\{ {\bf{W}}, {\bf{a}}, {\bf{b}} \}$ are model parameters to be optimized. One can further derive the following conditional probabilities from the joint distribution:
\begin{equation}
    P({{\bf{h}}_j} = 1|{\bf{v}}) = \sigma ({\bf{W}}_{:,j}^T{\bf{v}} + {{\bf{b}}_j})
    {\rm{~~and~~}}
    P({{\bf{v}}_i} = 1|{\bf{h}}) = \sigma ({{\bf{W}}_{i,:}}{\bf{h}} + {{\bf{a}}_i}),
\end{equation}
where $\sigma (x) = {(1 + \exp \{  - x\} )^{ - 1}}$ is the sigmoid function. The optimization of RBM aims to maximize the likelihood (i.e., minimizing the negative log-likelihood) of training data via
\begin{equation}
    \min -\ln P({\bf{v}}) = -\ln (\sum\nolimits_{\bf{h}} {P({\bf{v}},{\bf{h}})} )
\end{equation}

(1) \textbf{\textit{ctRBM}}.
Li et al. \cite{li2014deep} developed \textit{ctRBM} by extending the standard RBM to handle dynamic topology. An overview of \textit{ctRBM} is shown in Fig.~\ref{fig:rbm} (b).
It has two independent layers of visible units (denoted as ${\bf{v}}_{<}$ and ${\bf{v}}$) fully connected to hidden units ${\bf{h}}$. ${\bf{v}}_{<}$ and ${\bf{v}}$ are used to encode historical topology $G_{\tau-L}^{\tau}$ and prediction result $\tilde G_{\tau+1}$ (or training ground-truth $G_{\tau+1}$), respectively. For each node $v_i$, we set ${{\bf{v}}_ < } = [{({{\bf{A}}_{\tau - L+1}})_{i,:}} || \cdots || {({{\bf{A}}_\tau })_{i,:}}]^{T} \in \Re^{NL \times 1}$ by concatenating the $i$-th rows of ${\bf{A}}_{\tau-L}^{\tau}$ and ${\bf{v}} = ({\bf{\tilde A}}_{\tau+1})^{T}_{i,:} \in \Re^{N \times 1}$ (or ${\bf{v}} = ({\bf{A}}_{\tau+1})^{T}_{i,:}$). Moreover, ${\bf{W}}_{<} \in \Re^{NL \times d}$ and ${\bf{a}}_{<} \in \Re^{NL \times 1}$ are the weight matrix and bias vector of the connection between ${\bf{v}}_{<}$ and ${\bf{h}}$, while ${\bf{W}} \in \Re^{N \times d}$ and ${\bf{a}} \in \Re^{N \times 1}$ are the weight and bias for $\bf{v}$.
The encoder of \textit{ctRBM} is defined as the following conditional probability:
\begin{equation}
    {\bf{h}} = {\mathop{\rm Enc}\nolimits} (G_{\tau  - L}^\tau ) \equiv P({\bf{h}}|{\bf{v}},{{\bf{v}}_ < };\delta ) = \alpha  \cdot \sigma ({\bf{W}}_ < ^T{{\bf{v}}_ < } + {\bf{b}}) + (1 - \alpha )\sigma ({{\bf{W}}^T}{\bf{v}} + {\bf{b}}),
\end{equation}
where $\delta = \{ {\bf{W}}, {\bf{W}}_{<}, {\bf{a}}_{<}, {\bf{a}}, {\bf{b}} \}$ denotes the set of model parameters to be optimized; $\alpha$ is a parameter to balance components w.r.t. the historical topology and prediction result; ${\bf{h}}$ is the learned latent representation encoding properties of dynamic topology.
In the offline training with a given training ground-truth ${\bf{A}}_{\tau+1}$, we set ${\bf{v}} = ({\bf{A}}_{\tau+1})_{i,:}^T$ for each node $v_i$, while we let ${\bf{v}} = [0.5, \cdots, 0.5]^T$ (i.e., a constant vector with all elements set to $0.5$) for the online generalization without available ground-truth.
Given ${\bf{h}}$, the decoder of \textit{ctRBM} is then defined as
\begin{equation}\label{Eq:Dec-ctRBM}
    {({{{\bf{\tilde A}}}_{\tau  + 1}})_{i,:}^T} = {\mathop{\rm Dec}\nolimits} ({\bf{h}}) \equiv P({\bf{v}}|{\bf{h}};\delta ) = \sigma ({\bf{Wh}} + {\bf{a}}),
\end{equation}
which derives the $i$-th row of the prediction result ${\bf{\tilde A}}_{\tau+1}$ for node $v_i$.

Note that the encoder and decoder are defined for each node $v_i \in V$, where $V$ is assumed to be fixed for all the snapshots. Li et al. \cite{li2014deep} suggested training a \textit{ctRBM} model for each node.
Due to the OTOG paradigm, the ground-truth $G_{\tau+1}$ (in terms of ${\bf{A}}_{\tau+1}$) w.r.t. the snapshot to be predicted is given in the offline training.
The loss function w.r.t. the prediction of a node $v_i$ is defined as
\begin{equation}
    {\min}_\delta ~\mathcal{L}(G_{\tau  - L}^\tau ,{G_{\tau  + 1}}, v_i;\delta ) \equiv -\ln P({\bf{v}} = {({{\bf{A}}_{\tau  + 1}})_{i,:}};\delta ) = -\ln \sum\nolimits_{{\bf{h}},{{\bf{v}}_ < }} {P({\bf{v}} = {({{\bf{A}}_{\tau  + 1}})_{i,:}}, {{\bf{v}}_ < },{\bf{h}};\delta )}.
\end{equation}
In most cases, directly optimizing this objective is intractable. Contrastive divergence \cite{hinton2002training}, an alternative approximated algorithm, is adopted to optimize \textit{ctRBM}.
Compared with conventional linear models (e.g., matrix factorization introduced in Section~\ref{sec:OTI-SNAP-MF}), RBM-based methods can capture non-linear characteristics of dynamic topology.
However, to enable RBM to handle dynamic topology with successive snapshots, we concatenate rows of historical adjacency matrices to a long vector.
In this setting, the dimensionality of model parameters is related to the number of nodes. Hence, RBM-based approaches can only tackle the TLP in level-1, where all the snapshots share a common node set. Such a setting of RBM also fails to utilize the sparsity of graph topology and usually has high complexity of model parameters.

\begin{figure}[t]
\begin{center}
 \includegraphics[width=0.65\linewidth, trim=18 18 18 18,clip]{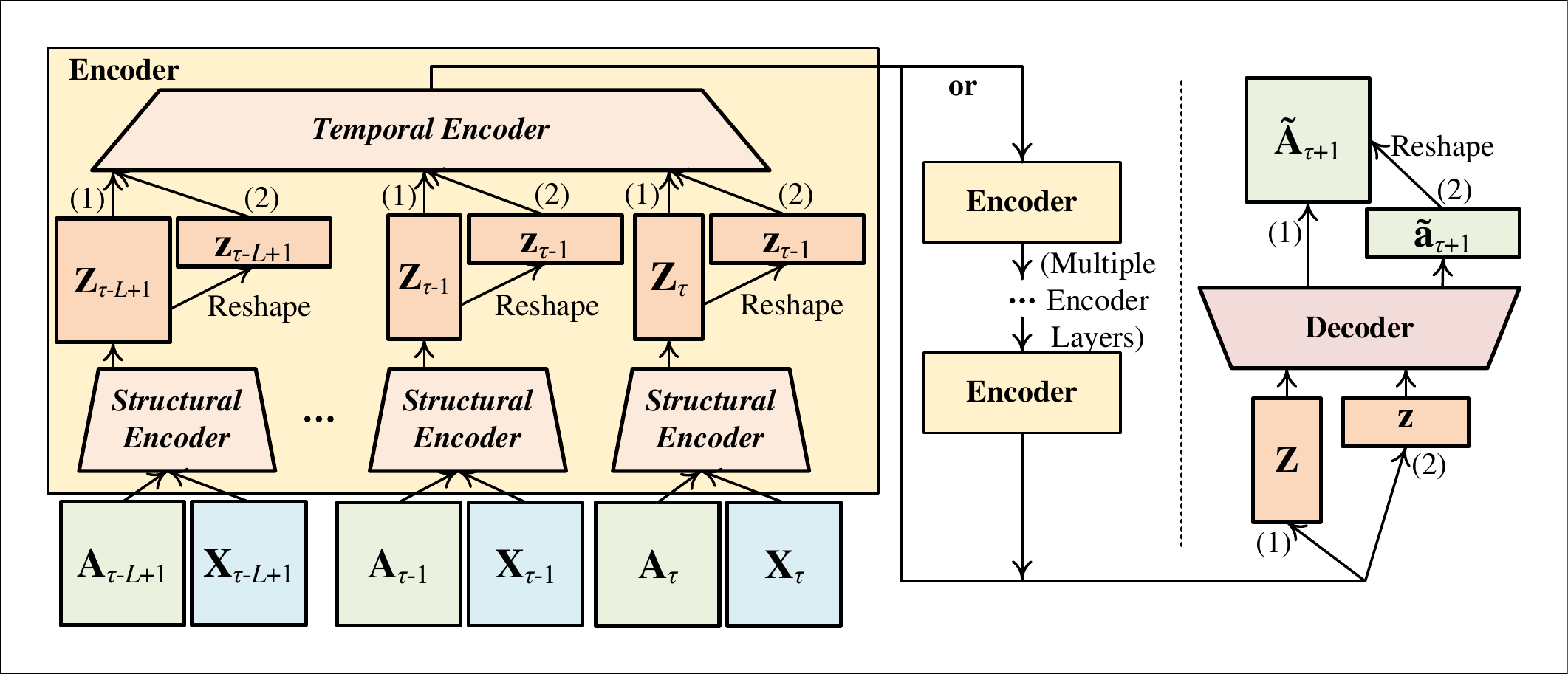}
\end{center}
\vspace{-0.25cm}
\caption{The extended encoder-decoder framework of some ESSD-based OTOG methods.}\label{fig:DeepDNE}
\vspace{-0.35cm}
\end{figure}

\subsubsection{\textbf{RNN-Based Temporal Models}}~

Most ESSD-based OTOG approaches that use RNNs to capture temporal characteristics, can be described by the framework depicted in Fig.~\ref{fig:DeepDNE}. It is an extension of our encoder-decoder framework introduced in Section~\ref{Sec:Frame}, where the encoder is further divided into (\romannumeral1) a \textit{structural encoder} and (\romannumeral2) a \textit{temporal encoder}. 

For each snapshot $G_t$, the \textit{structural encoder}
maps adjacency matrix ${\bf{A}}_t$ and attribute matrix ${\bf{X}}_t$ (if available) to latent embedding ${\bf{Z}}_t \in \Re^{N \times d}$, capturing the structural properties of each single snapshot $G_t$. The \textit{temporal encoder} then maps ${\bf{Z}}_{\tau-L}^{\tau}$ w.r.t. $G_{\tau-L}^{\tau}$ to another representation ${\bf{Z}} \in \Re^{N \times d}$, capturing the evolving patterns across successive snapshots.
In this setting, the $i$-th rows of ${\bf{Z}}_t$ and ${\bf{Z}}$ can be the embedding of node $v_i$. Finally, the decoder takes ${\bf{Z}}$ as input and derives prediction result ${\bf{\tilde A}}_{\tau+1}$ (i.e., mapping embedding ${\bf{Z}}_{i,:}$ of $v_i$ to the $i$-th row of ${\bf{\tilde A}}_{\tau+1}$).
In addition, some methods first reshape ${\bf{Z}}_t$ to a row-wise long vector ${\bf{z}}_t \in \Re^{Nd \times 1}$ before feeding it to the \textit{temporal encoder} and then obtain another long vector ${\bf{z}} \in \Re^{Nd \times 1}$ from the \textit{temporal encoder}, where ${\bf{z}}_t$ and ${\bf{z}}$ can be considered as snapshot-level embedding. The decoder takes ${\bf{z}}$ as input and outputs another long vector ${\bf{\tilde a}}_{\tau+1} \in \Re^{N^2 \times 1}$, which is further reshaped to a matrix form ${\bf{\tilde A}}_{\tau+1} \in \Re^{N \times N}$ as the prediction result. Some approaches also adopt a stacked multi-layer structure for their encoders, with each layer containing a \textit{structural encoder} and a \textit{temporal encoder}.

Some RNN structures (e.g., long short-term memory (LSTM) \cite{gers2000learning} and gated recurrent unit (GRU) \cite{chung2014empirical}) can be used to build the \textit{temporal encoders} of an ESSD-based OTOG approach.

(1) \textbf{\textit{Dyngraph2vec}}.
Based on the framework in Fig.~\ref{fig:DeepDNE}, Goyal et al. \cite{goyal2020dyngraph2vec} proposed \textit{dyngraph2vec} with three variants. One variant uses multi-layer perceptron (MLP) and LSTM to build its \textit{structural} and \textit{temporal encoders}. Due to space limit, we leave details of MLP and LSTM in supplementary materials.
The encoder of \textit{dyngraph2vec} can be described as
\begin{equation}\label{Eq:Enc-dyngraph2vec}
    {\bf{Z}} = {\rm{Enc}}(G_{\tau  - L}^\tau ) \equiv {\rm{LSTM}}{({\bf{Z}}_{\tau  - L}^\tau )_L}{\rm{~with~}}{{\bf{Z}}_t} \equiv {\mathop{\rm MLP}\nolimits} ({{\bf{A}}_t}) {\rm ~} (\tau  - L < t \le \tau ).
\end{equation}
In (\ref{Eq:Enc-dyngraph2vec}), the \textit{temporal encoder} (i.e., LSTM) in sequence takes ${\bf{Z}}_{\tau-L}^{\tau}$ as input and then outputs a series of hidden states $[{\bf{H}}_1, \cdots, {\bf{H}}_L] = {\rm{LSTM}}({\bf{Z}}_{\tau-L}^{\tau})$. Only the last state ${\bf{H}}_{L}$ is adopted as the final output of the \textit{temporal encoder} denoted as ${\bf{Z}} = {\bf{H}}_L = {\rm{LSTM}}({\bf{Z}}_{\tau-L}^{\tau})_L$.
The decoder further uses another MLP to map ${\bf{Z}}$ to the predicted adjacency matrix via
\begin{equation}\label{Eq:Dec-dyngrahp2vec}
    {{{\bf{\tilde A}}}_{\tau  + 1}} = {\mathop{\rm Dec}\nolimits} ({\bf{Z}}) \equiv {\mathop{\rm MLP}\nolimits} ({\bf{Z}}).
\end{equation}
For one prediction operation with $G_{\tau+1}$ as the ground-truth, the loss function of \textit{dyngraph2vec} is defined as
\begin{equation}\label{Eq:loss-dyngraph2vec}
    {\min}_\delta ~ {\mathcal{L}}(G_{\tau  - L}^\tau ,{G_{\tau  + 1}};\delta ) \equiv \left\| {({{{\bf{\tilde A}}}_{\tau  + 1}} - {{\bf{A}}_{\tau  + 1}}) \odot {{\bf{E}}_{\tau  + 1}}} \right\|_F^2,
\end{equation}
where ${\bf{E}}_{\tau+1}$ is an auxiliary variable to give different penalties to the observed and non-existent edges in $G_{\tau+1}$. We set $({\bf{E}}_{\tau+1})_{ij} =\beta$ if $({\bf{A}}_{\tau+1})_{ij} >0$ and ${\bf{E}}_{\tau+1}=1$ otherwise, with $\beta > 0$ as a tunable parameter. Chen et al. \cite{chen2019lstm} proposed \textit{E-LSTM-D} based on the same encoder, decoder, and loss function as the aforementioned variant of \textit{dyngraph2vec}.

(2) \textbf{\textit{DDNE}}. Li et al. \cite{li2018deep} introduced \textit{DDNE}, which does not specify its \textit{structural encoder} but directly uses adjacency matrices as snapshot-induced features (i.e., ${\bf{Z}}_t = {\bf{A}}_t$). The \textit{temporal encoder} contains two GRUs with different directions regarding the sequence ${\bf{Z}}_{\tau-L}^{\tau} \equiv {\bf{A}}_{\tau-L}^{\tau}$. Due to space limit, we leave details of GRU in supplementary materials.
For simplicity, the encoder of \textit{DDNE} can be described as
\begin{equation}\label{Eq:Enc-DDNE}
    {\bf{Z}} = {\mathop{\rm Enc}\nolimits} (G_{\tau  - L}^\tau ) \equiv [{\mathop{\rm GRU}\nolimits}^\rightarrow ({\bf{A}}_{\tau  - L}^\tau ) || {\mathop{\rm GRU}\nolimits}^\leftarrow ({\bf{A}}_{\tau - L} ^{\tau})].
\end{equation}
The first GRU in sequence takes $[{\bf{A}}_{\tau-L+1}, \cdots, {\bf{A}}_{\tau}]$ as input and derives states $[{\bf{H}}_1^\rightarrow, \cdots, {\bf{H}}_L^\rightarrow]$. In contrast, the input and output of the second GRU are denoted as $[{\bf{A}}_\tau, \cdots, {\bf{A}}_{\tau-L+1}]$ and $[{\bf{H}}_1^\leftarrow, \cdots, {\bf{H}}_L^\leftarrow]$. The output of the \textit{temporal encoder} is rearranged as ${\bf{Z}} = [{\bf{H}}_{1}^ \to ||{\bf{H}}_{1}^ \leftarrow  || \cdots || {\bf{H}}_L ^ \to ||{\bf{H}}_L ^ \leftarrow]$.
Given ${\bf{Z}}$, the decoder of \textit{DDNE} has the same definition as \textit{dyngraph2vec} in (\ref{Eq:Dec-dyngrahp2vec}). The original version of \textit{DDNE} considers TLP on unweighted graphs, treating it as a binary edge classification task. The loss function w.r.t. one prediction operation combines the classic cross-entropy loss with graph regularization regarding historical connection frequency, which is defined as
\begin{equation}\label{Eq:Loss-DDNE}
    {\min}_\delta  ~ {\mathcal L}(G_{\tau  - L}^\tau ,{G_{\tau  + 1}};\delta ) \equiv \sum\nolimits_{i,j = 1}^N {{{({{\bf{E}}_{\tau  + 1}})}_{ij}}{{({{\bf{A}}_{\tau  + 1}})}_{ij}}\ln {{({{{\bf{\tilde A}}}_{\tau  + 1}})}_{ij}}}  + \alpha {\rm{tr}}({{\bf{Z}}^T}{{\bf{L}}_{\bf{N}}}{\bf{Z}}),
\end{equation}
where ${\bf{E}}_{\tau+1}$ is an auxiliary variable for the first term (i.e., cross-entropy loss) with the same definition and physical meaning as that in (\ref{Eq:loss-dyngraph2vec}); ${{\bf{L}}_{\bf{N}}} = {{\bf{D}}_{\bf{N}}} - {\bf{N}}$ is the Laplacian matrix of an auxiliary matrix ${\bf{N}} = \sum\nolimits_{t = \tau - L+1}^\tau  {{{\bf{A}}_t}}$; $\alpha$ is a tunable parameter.
Since \textit{DDNE} assumes $({\bf{A}}_t)_{ij} \in \{ 0, 1\}$, which is used to describe unweighted topology, ${\bf{N}}_{ij}$ is the historical connection frequency between $(v_i, v_j)$. The second term of graph regularization can be rewritten as ${\mathop{\rm tr}\nolimits} ({{\bf{Z}}^T}{{\bf{L}}_{\bf{N}}}{\bf{Z}}) = 0.5{{\bf{N}}_{ij}}\left\| {{{\bf{Z}}_{i,:}} - {{\bf{Z}}_{j,:}}} \right\|_F^2$, which regularizes the embedding ${\bf{Z}}$ using connection frequency ${\bf{N}}$, with a physical meaning similar to that in (\ref{Eq:Enc-CRJMF}).

As \textit{DDNE} is a typical \textit{task-dependent} DNE-based method, it can also be extended to support the prediction of weighted topology by replacing the loss function (\ref{Eq:Loss-DDNE}) with that of \textit{dyngraph2vec} defined in (\ref{Eq:loss-dyngraph2vec}).

(3) \textbf{\textit{EvolveGCN}}.
Instead of MLP, \textit{EvolveGCN} \cite{pareja2020evolvegcn} uses GCN \cite{kipf2016semi}, a type of GNN, to build its \textit{structural encoder}.
RNN is adopted as the \textit{temporal encoder} to evolve parameters of GCN rather than handling snapshot-induced embedding ${\bf{Z}}_{\tau-L}^{\tau}$ like \textit{dyngraph2vec}. We leave details of GCN and RNN (e.g., LSTM and GRU) in supplementary materials. The encoder of \textit{EvolveGCN} is a multi-layer structure, with each layer containing a \textit{structural encoder} and a \textit{temporal encoder}.

Two variants of \textit{EvolveGCN} with two manners to evolve model parameters of GCN were proposed in \cite{pareja2020evolvegcn}.
Let ${\bf{Z}}_t^{(k-1)}$ and ${\bf{Z}}_t^{(k)}$ be the input and output of the $k$-th encoder layer w.r.t. each snapshot $G_t$, where ${\bf{Z}}_t^{(0)} \equiv {\bf{X}}_t$ (i.e., available node attributes in $G_t$). Let ${\bf{W}}_t^{(k-1)}$ be the weight (i.e., learnable model parameters) of GCN in the $k$-th layer at time step $t$. The $k$-th encoder layer of the first variant can be described as
\begin{equation}
    {\bf{Z}}_t^{(k)} = {\rm{En}}{{\rm{c}}^{(k)}}({G_t}) \equiv {\rm{GCN}}({{\bf{A}}_t},{\bf{Z}}_t^{(k - 1)};{\bf{W}}_t^{(k - 1)}), {\rm{~with~}}{\bf{W}}_t^{(k - 1)} = {\mathop{\rm GRU}\nolimits} ({\bf{Z}}_t^{(k - 1)},{\bf{W}}_{t - 1}^{(k - 1)}).
\end{equation}
In each time step $t$, the $k$-th encoder layer derives a new GCN weight ${\bf{W}}_t^{(k-1)}$ by letting ${\bf{Z}}_t^{(k-1)}$ and ${\bf{W}}_{t-1}^{(k-1)}$ be the feature input and the previous hidden state of GRU. The $k$-th encoder layer of the second variant is defined as
\begin{equation}
    {\bf{Z}}_t^{(k)} = {\rm{En}}{{\rm{c}}^{(k)}}({G_t}) \equiv {\rm{GCN}}({{\bf{A}}_t},{\bf{Z}}_t^{(k - 1)};{\bf{W}}_t^{(k - 1)}),{\rm{~with~}}{\bf{W}}_t^{(k - 1)} = {\rm{LSTM}}({\bf{W}}_{t - 1}^{(k - 1)},{\bf{W}}_{t - 1}^{(k - 1)}),
\end{equation}
which updates GCN weight ${\bf{W}}_t^{(k-1)}$ by letting ${\bf{W}}_{t-1}^{(k-1)}$ be both the feature input and previous hidden state of LSTM.
The encoder adopts the GCN output of the last encoder layer w.r.t. current time step $\tau$ as the temporal embedding ${\bf{Z}}$.

\textit{EvolveGCN} is a \textit{task-dependent DNE} method. One should specify the decoder and training loss related to the downstream task (i.e., TLP). Pareja et al. \cite{pareja2020evolvegcn} recommended using the following decoder
\begin{equation}\label{Eq:Dec-EvolveGCN}
    {({{\bf{\tilde A}}_{\tau  + 1}})_{ij}} = {\mathop{\rm Dec}\nolimits} ({\bf{Z}}) \equiv {\mathop{\rm MLP}\nolimits} ([{{\bf{Z}}_{i,:}}||{{\bf{Z}}_{j,:}}]).
\end{equation}
It concatenates embedding $({\bf{Z}}_{i,:}, {\bf{Z}}_{j,:})$ w.r.t. each node pair $(v_i, v_j)$ and applies an MLP to map $[{\bf{Z}}_{i,:} || {\bf{Z}}_{j,:}]$ to $({\bf{\tilde A}}_{\tau+1})_{ij}$.
The original version of \textit{EvolveGCN} only considers the TLP on unweighted graphs. One can use the following binary cross-entropy loss w.r.t. one prediction operation to train the model in an end-to-end manner:
\begin{equation}\label{Eq:loss-EvolveGCN}
    {\min}_\delta  ~ \mathcal{L}(G_{\tau  - L}^\tau ,{G_{\tau  + 1}};\delta ) \equiv \sum\nolimits_{i,j = 1}^{{N_t}} { - [{{({{\bf{A}}_{\tau  + 1}})}_{ij}}\ln {{({{{\bf{\tilde A}}}_{\tau  + 1}})}_{ij}} + (1 - {{({{\bf{A}}_{\tau  + 1}})}_{ij}})\ln (1 - {{({{{\bf{\tilde A}}}_{\tau  + 1}})}_{ij}})]} /{N_t}.
\end{equation}
To extend \textit{EvolveGCN} to support the prediction of weighted topology, we can replace (\ref{Eq:loss-EvolveGCN}) with the loss function of \textit{dyngraph2vec} defined in (\ref{Eq:loss-dyngraph2vec}).

(4) \textbf{\textit{GCN-GAN}}. Most existing methods merely consider inference tasks on unweighted graphs, while the TLP on weighted graphs is seldom studied. Some approaches (e.g., \textit{neighbor similarity} and \textit{ctRBM}) may even fail to capture and predict weighted topology. Although several methods (e.g., \textit{GrNMF}, \textit{DeepEye}, \textit{DDNE}, and \textit{dyngraph2vec}) can still tackle weighted TLP, they can only derive low-quality prediction results.
We elaborate on this advanced topic regarding the TLP on weighted graphs later in Section~\ref{Sec:Adv-Topic}.

Inspired by the high-resolution video prediction \cite{mathieu2015deep} using generative adversarial network (GAN) \cite{goodfellow2014generative,arjovsky2017wasserstein}, Lei et al. \cite{lei2019gcn} focused on the weighted TLP and proposed \textit{GCN-GAN}. It combines the extended framework in Fig.~\ref{fig:DeepDNE} with GAN and can derive high-quality prediction results for weighted graphs.
Following GAN, the model contains a generator $\mathcal{G}$ and a discriminator $\mathcal{D}$. $\mathcal{G}$ adopts the encoder-decoder framework to generate prediction results while $\mathcal{D}$ is an auxiliary structure to refine the generated results.
In addition to the historical topology described by ${\bf{A}}_{\tau-L}^{\tau}$, \textit{GCN-GAN} also generates random noise via $U[0, 1]$ to support GAN (i.e., generating plausible samples from noise), which are treated as attribute inputs described by ${\bf{X}}_{\tau-L}^{\tau}$. Given ${\bf{A}}_{\tau-L}^{\tau}$ and ${\bf{X}}_{\tau-L}^{\tau}$ w.r.t. $G_{\tau-L}^{\tau}$, the encoder of \textit{GCN-GAN} is defined as
\begin{equation}\label{Eq:Enc-GCN-GAN}
    {\bf{z}} = {\mathop{\rm Enc}\nolimits} (G_{\tau  - L}^\tau ) \equiv {\mathop{\rm LSTM}\nolimits} {({\bf{z}}_{\tau  - L}^\tau )_L}{\rm{~with~}}{{\bf{z}}_t} \equiv {r_ > }({{\bf{Z}}_t}) {\rm{~and~}} {{\bf{Z}}_t} \equiv {\mathop{\rm GCN}\nolimits} ({{\bf{A}}_t},{{\bf{X}}_t})~(\tau  - L < t \le \tau ),
\end{equation}
where GCN and LSTM (see supplementary materials for their details) are used to build the \textit{structural} and \textit{temporal encoders}. For each snapshot $G_t$, GCN takes ${\bf{A}}_t$ and ${\bf{X}}_t$ as inputs and derives embedding ${\bf{Z}}_t \in \Re^{N \times d}$. A function $r_>(\cdot)$ is then applied to reshape ${\bf{Z}}_t$ to a row-wise long vector ${\bf{z}}_t \in \Re^{Nd \times 1}$ (i.e., second strategy in Fig.~\ref{fig:DeepDNE}) before feeding it to the LSTM. In this setting, the final output of the encoder is also a vector ${\bf{z}} \in \Re^{Nd \times 1}$ (i.e., the last hidden state of LSTM), preserving the evolving patterns of successive snapshots.
Given ${\bf{z}}$, the decoder of \textit{GCN-GAN} is defined as
\begin{equation}\label{Eq:Dec-GCN-GAN}
    {{{\bf{\tilde A}}}_{\tau  + 1}} = {\rm{Dec}}({\bf{z}}) \equiv {r_ < }({{{\bf{\tilde a}}}_{\tau  + 1}}){\rm{~with~}}{{{\bf{\tilde a}}}_{\tau  + 1}} \equiv {\mathop{\rm MLP}\nolimits} ({\bf{z}}).
\end{equation}
It uses an MLP to map ${\bf{z}}$ to a row-wise long vector ${\bf{\tilde a}}_{\tau+1} \in \Re^{N^2 \times 1}$. Another function $r_<(\cdot)$ is applied to reshape ${\bf{\tilde a}}_{\tau+1}$ to the matrix form ${\bf{\tilde A}}_{\tau+1} \in \Re^{N \times N}$ as the prediction result.
In addition to $\mathcal{G}$ with an encoder and a decoder, $\mathcal{D}$ is an auxiliary structure defined as
\begin{equation}\label{Eq:D-GCN-GAN}
    p = {\mathcal{D}}({\bf{M}}) \equiv {\mathop{\rm MLP}\nolimits} ({\bf{m}}){\rm{~with~}}{\bf{m}} \equiv {r_ > }({\bf{M}}),
\end{equation}
where ${\bf{M}} \in \{ {\bf{A}}_{\tau+1}, {\bf{\tilde A}}_{\tau+1}\}$; $r_>(\cdot)$ is a function to reshape ${\bf{M}} \in \Re^{N \times N}$ to a row-wise long vector ${\bf{m}} \in \Re^{N^2 \times 1}$; $p$ denotes the probability that ${\bf{M}} = {\bf{A}}_{\tau+1}$ rather than ${\bf{M}} = {\bf{\tilde A}}_{\tau+1}$.

Let $\delta_{\mathcal{D}}$ and $\delta_{\mathcal{G}}$ be sets of model parameters in $\mathcal{D}$ and $\mathcal{G}$. The optimization of \textit{GCN-GAN} includes the (\romannumeral1) pre-training of $\mathcal{G}$ and (\romannumeral2) joint optimization of $\mathcal{D}$ and $\mathcal{G}$. The pre-training loss of $\mathcal{G}$ is with the same definition as (\ref{Eq:loss-dyngraph2vec}). After pre-training, the model can preliminarily generate predicted snapshot ${\bf{\tilde A}}_{\tau+1}$ consistent with ground-truth ${\bf{A}}_{\tau+1}$.
In formal optimization, \textit{GCN-GAN} adopts the losses of GAN. On the one hand, $\mathcal{D}$ tries to distinguish ${\bf{A}}_{\tau+1}$ from ${\bf{\tilde A}}_{\tau+1}$ via
\begin{equation}\label{Eq:loss-D-GCN-GAN}
    {\min}_{\delta_\mathcal{D}} ~ {\mathcal{L}_\mathcal{D}}(G_{\tau  - L}^\tau ,{G_{\tau  + 1}};{\delta _\mathcal{D}}) \equiv  - [\ln (1 - \mathcal{D}({{{\bf{\tilde A}}}_{\tau  + 1}})) + \ln \mathcal{D}({{\bf{A}}_{\tau  + 1}})].
\end{equation}
On the other hand, $\mathcal{G}$ tries to generate a plausible snapshot ${\bf{\tilde A}}_{\tau+1}$ to fool $\mathcal{D}$ via
\begin{equation}\label{Eq:loss-G-GCN-GAN}
    {\min }_{\delta _\mathcal{G}} ~ {\mathcal{L}_\mathcal{G}}(G_{\tau  - L}^\tau ,{G_{\tau  + 1}};{\delta _\mathcal{G}}) \equiv  - \ln \mathcal{D}({{{\bf{\tilde A}}}_{\tau  + 1}}).
\end{equation}
During the joint optimization between $\mathcal{D}$ and $\mathcal{G}$, ${\bf{\tilde A}}_{\tau+1}$ can be further refined and is expected to be a high-quality prediction result for weighted graphs. As the model parameters of RNN in the encoder and MLP in the decoder are related to the number of nodes, \textit{GCN-GAN} can only support TLP in level-1, failing to tackle the variation of nodes.

(5) \textbf{\textit{IDEA}}.
Qin et al. \cite{qin2023high} introduced \textit{IDEA} that extends \textit{GCN-GAN} to the weighted TLP in level-2. Similar to \textit{GCN-GAN}, \textit{IDEA} also contains (\romannumeral1) a generator $\mathcal{G}$ following the encoder-decoder framework and (\romannumeral2) a discriminator $\mathcal{D}$.

The encoder in ${\mathcal{G}}$ is a multi-layer structure, with the GCN and a modified GRU (see supplementary materials for details of GCN and GRU) used to build the \textit{structural} and \textit{temporal encoders} of each layer. In addition to adjacency matrices ${\bf{A}}_{\tau-L}^{\tau}$ and attribute matrices ${\bf{X}}_{\tau-L}^{\tau}$ that describe topology and attributes of $G_{\tau-L}^{\tau}$, \textit{IDEA} also maintains an aligning matrix ${\bf{B}}_t \in \Re^{N_{t} \times N_{t+1}}$ for successive snapshots $\{ G_t, G_{t+1} \}$ to encode the variation of node sets. $({\bf{B}}_t)_{ij} = 1$ if $v_i^t$ corresponds to $v_j^{t+1}$ and  $({\bf{B}}_t)_{ij} = 0$ otherwise. Let ${\bf{Z}}_t^{(k-1)}$ and ${\bf{Z}}_t^{(k)}$ be the input and output of the $k$-th encoder layer at time step $t$, where ${\bf{Z}}_t^{(0)} \equiv {\mathop{\rm MLP}\nolimits} ({\bf{X}}_t)$. For each time step $t$, the $k$-th encoder layer takes adjacency matrix ${\bf{A}}_t$, attribute matrices $\{ {\bf{X}}_{t-1}, {\bf{X}}_t \}$, aligning matrix ${\bf{B}}_t$, and previous encoder output ${\bf{Z}}_{t-1}^{(k)} \in \Re^{N_{t-1} \times d}$ (in the same layer) as the joint inputs, and then derives embedding ${\bf{Z}}_t^{(k)} \in \Re^{N_t \times d}$, which can be described as
\begin{equation}\label{Eq:Enc-IDEA}
\begin{array}{l}
{\bf{Z}}_t^{(k)} = {{\mathop{\rm Enc}\nolimits} ^{(k)}}({{\bf{A}}_t},{\bf{X}}_{t-1}, {\bf{X}}_t, {{\bf{B}}_t},{\bf{Z}}_{t - 1}^{(k)}) \equiv {\mathop{\rm GRU}\nolimits} ({\bf{G}}_t^{(k)},{\bf{\hat Z}}_{t - 1}^{(k)}), {\rm{~with~}}{\bf{G}}_t^{(k)} \equiv {\mathop{\rm GCN}\nolimits} ({{\bf{A}}_t},{\bf{Z}}_t^{(k - 1)})\\
{\bf{\hat Z}}_{t - 1}^{(k)} \equiv {[{{\bf{B}}_{t-1}} + a ({{\bf{X}}_{t - 1}},{{\bf{X}}_t})]^T}{\bf{Z}}_{t - 1}^{(k)}, {\rm{~and~}} a ({{\bf{X}}_{t - 1}},{{\bf{X}}_t}) \equiv {\mathop{\rm MLP}\nolimits} ({{\bf{X}}_{t - 1}}){\mathop{\rm MLP}\nolimits} {({{\bf{X}}_t})^T}.
\end{array}
\end{equation}
In (\ref{Eq:Enc-IDEA}), we first obtain auxiliary embedding ${\bf{G}}_t^{(k)} \equiv {\mathop{\rm GCN}\nolimits} ({{\bf{A}}_t},{\bf{Z}}_t^{(k - 1)}) \in \Re^{N_t \times d}$ that preserves structural properties of snapshot $G_t$. Before feeding ${\bf{G}}_t^{(k)}$ and the hidden state ${\bf{Z}}_{t-1}^{(k)} \in \Re^{N_{t-1} \times d}$ that matches with the node set $V_{t-1}$ of $G_{t-1}$ to GRU, we derive the aligned state ${\bf{\hat Z}}_{t - 1}^{(k)} \equiv {[{{\bf{B}}_{t-1}} + a ({{\bf{X}}_{t - 1}},{{\bf{X}}_t})]^T}{\bf{Z}}_{t - 1}^{(k)} \in \Re^{N_t \times d}$ that matches with $V_t$ by mapping the rows of ${\bf{Z}}_{t-1}^{(k)}$ to those of ${\bf{\hat Z}}_{t - 1}^{(k)}$. In addition to the node correspondence encoded in ${\bf{B}}_t$, an attentive aligning unit $a({\bf{X}}_{t-1}, {\bf{X}}_t)$ is introduced to extract additional aligning relations from attributes $\{ {\bf{X}}_{t-1}, {\bf{X}}_t\}$. In this setting, the output ${\bf{Z}}_t^{(k)} \in \Re^{N_t \times d}$ can match with $V_t$ to handle the variation of node sets (e.g., $V_{t-1} \ne V_t$) while preserving the evolving patterns across snapshots.

For current time step $\tau$, let the aligned embedding ${\bf{\hat Z}}_{\tau} \in \Re^{N_{\tau+1} \times d}$ be the final output of the encoder, which matches with $V_{\tau+1}$. Different from \textit{dyngraph2vec} and \textit{DDNE} that directly map the derived embedding to the prediction result ${\bf{\tilde A}}_{\tau+1}$ via an MLP in (\ref{Eq:Dec-dyngrahp2vec}), the decoder of \textit{IDEA} use the aggregation of ${\bf{\hat Z}}_{\tau}$ to generate each element $({\bf{\tilde A}}_{\tau+1})_{ij}$ via
\begin{equation}
\begin{array}{l}
{({{{\bf{\tilde A}}}_{\tau  + 1}})_{ij}} = {\mathop{\rm Dec}\nolimits} ({{\bf{\hat Z}}_\tau }) \equiv 1 + \tanh ( - {\varsigma _{ij}}|{({{\bf{U}}_\tau })_{i,:}} - {({{\bf{U}}_\tau })_{j,:}}|_2^2),\\
{\rm{with~}}{\varsigma _{ij}} \equiv {({{\bf{V}}_\tau }{\bf{V}}_\tau ^T)_{ij}},~{{\bf{V}}_\tau } \equiv {\mathop{\rm MLP}\nolimits} ({{\bf{\hat Z}}_\tau }),{\rm{~and~}}{{\bf{U}}_\tau } \equiv {\mathop{\rm MLP}\nolimits} ({{\bf{\hat Z}}_\tau }).
\end{array}
\end{equation}
$\mathcal{D}$ is an auxiliary structure defined as ${\bf{p}} = \mathcal{D}({\bf{M}}, {\bf{X}}_{\tau+1}) \equiv {\mathop{\rm MLP}\nolimits} ({\mathop{\rm GCN}\nolimits} ({\bf{M}, {\bf{X}}_{\tau+1}}))$, where ${\bf{M}} \in \{ {\bf{A}}_{\tau+1}, {\bf{\tilde A}}_{\tau+1}\}$; ${\bf{p}}$ is an $N_{\tau+1}$-dimensional vector with ${\bf{p}}_i$ as the probability that ${\bf{M}}_{i,:} = ({\bf{A}}_{\tau+1})_{i,:}$ rather than $({\bf{\tilde A}}_{\tau+1})_{i,:}$. Given historical snapshots $G_{\tau-L}^{\tau}$, IDEA can in sequence generate prediction results ${\bf{\tilde A}}_{\tau-L+1}^{\tau+1}$ w.r.t. $G_{\tau-L+1}^{\tau+1}$. All the results in ${\bf{\tilde A}}_{\tau-L+1}^{\tau+1}$ are used for model optimization. In particular, \textit{IDEA} combines the adversarial learning loss $\mathcal{L}_{\rm AL}$ of GAN with the conventional error minimization loss $\mathcal{L}_{\rm EM}$ and a novel scale difference minimization loss $\mathcal{L}_{\rm SDM}$ to optimize $\mathcal{G}$, which are defined as ${\mathcal{L}_{{\rm{AL}}}}({G_t}) \equiv  - \sum\nolimits_i {\ln \mathcal{D}({{\bf{A}}_t},{{\bf{X}}_t})/{N_t}}$, ${\mathcal{L}_{{\rm{EM}}}}({G_t}) \equiv || {{{\bf{A}}_t} - {{{\bf{\tilde A}}}_t}} ||_F^2 + \sum\nolimits_{ij} {|{{({{\bf{A}}_t})}_{ij}} - {{({{{\bf{\tilde A}}}_t})}_{ij}}|} $, and ${\mathcal{L}_{{\rm{SDM}}}}({G_t}) \equiv \sum\nolimits_{ij} {|{{\log }_{10}}[{{({{\bf{\hat U}}_t})}_{ij}}/{{({{\bf{\hat V}}_t})}_{ij}}]|}$ for each snapshot $G_t$. ${\mathcal{L}_{{\rm{EM}}}}$ minimizes the reconstruction errors measured by the F-norm and $l_1$-norm to help $\mathcal{G}$ derive prediction result ${\bf{\tilde A}}_t$ consistent with ground-truth ${\bf{A}}_t$. In addition to ${\mathcal{L}_{{\rm{AL}}}}$, ${\mathcal{L}_{{\rm{SDM}}}}$ can also refine the generated prediction results by using $\log_{10}(\cdot)$ to minimize the scale difference between $\{ {\bf{A}}_t, {\bf{\tilde A}}\}$, where $\{ {\bf{\hat U}}_t, {\bf{\hat V}}_t \}$ are applied to clip $\{ {\bf{A}}_t, {\bf{\tilde A}}_t \}$ with the same motivations and definitions as the \textit{MLSD} metric (see Section~\ref{Sec:Q-Eva} and supplementary materials for its details). The loss functions to optimize $\mathcal{G}$ and $\mathcal{D}$ are then defined as
\begin{equation}\label{Eq:Loss-G-IDEA}
{\min}_{{\delta _G}}~{\mathcal{L}_\mathcal{G}}(G_{\tau  - L + 1}^{\tau  + 1}; \delta_\mathcal{G}) \equiv \sum\nolimits_{t = \tau  - L + 1}^{\tau  + 1} {{D_t}[{\mathcal{L}_{{\rm{AL}}}}({G_t}) + \alpha {\mathcal{L}_{{\rm{EM}}}}({G_t}) + \beta {\mathcal{L}_{{\rm{SDM}}}}({G_t})]}, 
\end{equation}
\begin{equation}\label{Eq:Loss-D-IDEA}
    {\min}_{{\delta _D}}~{\mathcal{L}_\mathcal{D}}(G_{\tau  - L + 1}^{\tau  + 1}; \delta_\mathcal{D}) \equiv  - \sum\nolimits_{t = \tau  - L + 1}^{\tau  + 1} {{D_t}\sum\nolimits_i {[\ln (1 - \mathcal{D}({{{\bf{\tilde A}}}_t},{{\bf{X}}_t})) + \ln \mathcal{D}({{\bf{A}}_t},{{\bf{X}}_t})]}}/N_t,
\end{equation}
where $D_t \equiv (1 - \theta)^{\tau+1-t}$ (with $\theta \in [0, 1]$ as a tunable parameter) is the decaying factor integrating \textbf{Hypothesis 3.1}; $\alpha$ and $\beta$ are pre-set parameters to adjust the contributions of $\mathcal{L}_{\rm EM}$ and $\mathcal{L}_{\rm SDM}$.

Compared with RBM-based approaches (e.g., \textit{ctRBM}) that concatenate historical adjacency matrices, the aforementioned RNN-based methods, with model parameters shared by successive time steps, are more space-efficient to handle dynamic topology.
As the dimensionality of the \textit{temporal encoders} (i.e., RNN) and decoders (i.e., MLP) of \textit{dyngraph2vec}, \textit{DDNE}, and \textit{GCN-GAN} is related to the number of nodes $N$, these approaches can only deal with the TLP in level-1, assuming that all the snapshots share a common node set.
In contrast, RNN in \textit{EvolveGCN} is used to evolve parameters of GNN that are not related to $N$. \textit{IDEA} adopts a modified RNN that aligns the non-fixed node sets between successive snapshots. The decoders of \textit{EvolveGCN} and \textit{IDEA} are also not related to $N$. Therefore, \textit{EvolveGCN} and \textit{IDEA} can support level-2 and handle the variation of node sets.

\subsubsection{\textbf{Attention-Based Temporal Models}}~

Most existing ESSD-based OTOG methods, which use attention mechanisms \cite{vaswani2017attention,niu2021review} to capture evolving patterns of successive snapshots, also follow the extended encoder-decoder framework in Fig.~\ref{fig:DeepDNE}, with attention as building blocks of the \textit{temporal encoder}. Due to space limit, we elaborate on the general form of attention in supplementary materials.

(1) \textbf{\textit{STGSN}}.
Min et al. \cite{min2021stgsn} proposed \textit{STGSN} using GCN and attention to build the \textit{structural} and \textit{temporal encoders}. In addition to the historical topology described by ${\bf{A}}_{\tau-L}^{\tau}$, an auxiliary adjacency matrix ${{\bf{A}}_{\rm{gbl}}} \equiv \sum\nolimits_{t = \tau  - L+1}^\tau  {{{\bf{A}}_t}}$ is introduced to encode the `global' topology of successive snapshots. Node attributes described by ${\bf{X}}_{\tau-L}^{\tau}$ are also assumed to be available.
Given the (\romannumeral1) historical topology ${\bf{A}}_{\tau-L}^{\tau}$ and (\romannumeral2) historical attributes ${\bf{X}}_{\tau-L}^{\tau}$, (\romannumeral3) `global' topology ${\bf{A}}_{\rm{gbl}}$, and (\romannumeral4) attributes ${\bf{X}}_{\tau+1}$ of the next snapshot, the encoder that derives the embedding ${\bf{Z}}_{i,:}$ for each node $v_i$ is defined as
\begin{equation}\label{Eq:Enc-STGSN}
\begin{array}{l}
    {{\bf{Z}}_{i,:}} = {\mathop{\rm Enc}\nolimits} (G_{\tau  - L}^\tau , {\bf{X}}_{\tau+1}) \equiv [{\mathop{\rm Att}\nolimits} ({\bf{q}},{\bf{K}},{\bf{V}})||{({{\bf{Z}}_{{\rm{gbl}}}})_{i,:}}]{\rm{~with~}}{\bf{q}} = {({{\bf{Z}}_{{\rm{gbl}}}})_{i,:}},~{{\bf{Z}}_{\rm gbl}} \equiv {\mathop{\rm GCN}\nolimits} ({{\bf{A}}_{\rm gbl}},{{\bf{X}}_{\tau  + 1}}),\\
    {{\bf{K}}_{t,:}} = {{\bf{V}}_{t,:}} = {({{\bf{Z}}_t})_{i,:}}, {\rm{~and~}} {{\bf{Z}}_t} \equiv {\mathop{\rm GCN}\nolimits} ({{\bf{A}}_t},{{\bf{X}}_t})~ (\tau  - L < t \le \tau ).
\end{array}
\end{equation}
We first derive the snapshot-induced embedding ${{\bf{Z}}_t} = {\mathop{\rm GCN}\nolimits} ({{\bf{A}}_t},{{\bf{X}}_t})$ ($\tau-L < t \le \tau$) and auxiliary `global' embedding ${{\bf{Z}}_{\rm{gbl}}} = {\mathop{\rm GCN}\nolimits} ({{\bf{A}}_{\rm{gbl}}},{{\bf{X}}_{\tau  + 1}})$ via GCN. An attention unit ${\mathop{\rm Att}\nolimits} ({\bf{q}},{\bf{K}},{\bf{V}})$ is then applied to capture evolving patterns across snapshots, where ${\bf{q}} \in \Re^{1 \times d}$, ${\bf{K}} \in \Re^{L \times d}$, and ${\bf{V}} \in \Re^{L \times d}$ are inputs of query, key, and value. We leave details of GCN and attention in supplementary materials. For each node $v_i$, we let ${\bf{q}}$ be the global embedding of $v_i$ (i.e., ${\bf{q}} = ({\bf{Z}}_{\rm{gbl}})_{i,:}$). ${\bf{K}}$ and ${\bf{V}}$ are set to be the concatenation of snapshot-induced embedding $\{ {\bf{Z}}_t \}$ of $v_i$ over historical snapshots $G_{\tau-L}^{\tau}$, with the $t$-th row as the embedding w.r.t. time step $t$ (i.e., ${\bf{K}}_{t,:} = {\bf{V}}_{t,:} = ({\bf{Z}}_t)_{i,:}$). Accordingly, the output of ${\mathop{\rm Att}\nolimits} ({\bf{q}},{\bf{K}},{\bf{V}})$ is a vector associated with $v_i$. The encoder treats the concatenated vector ${\bf{Z}}_{i,:} = [{\mathop{\rm Att}\nolimits} ({\bf{q}},{\bf{K}},{\bf{V}}) || ({\bf{Z}}_{\rm{gbl}})_{i,:}]$ as the final embedding output of $v_i$.
As \textit{STGSN} is a \textit{task-dependent DNE} method, one can use the same decoder and training loss as \textit{EvolveGCN} defined in (\ref{Eq:Dec-EvolveGCN}), (\ref{Eq:loss-EvolveGCN}), and (\ref{Eq:loss-dyngraph2vec}).

(2) \textbf{\textit{DySAT}}. Sankar et al. \cite{sankar2020dysat} developed \textit{DySAT} that adopts GAT \cite{velivckovic2017graph}, a type of GNN, and self-attention as building blocks of the \textit{structural} and \textit{temporal encoders}. Given historical topology ${\bf{A}}_{\tau-L}^{\tau}$ and attributes ${\bf{X}}_{\tau-L}^{\tau}$, the encoder of \textit{DySAT} derives corresponding embedding ${{\bf{Z}}_{i,:}}$ for each node $v_i$ via the following procedure:
\begin{equation}
\begin{array}{l}
{{\bf{Z}}_{i,:}} = {\mathop{\rm Enc}\nolimits} (G_{\tau  - L}^\tau) \equiv {r_ > }({\mathop{\rm Att}\nolimits} ({\bf{Q}},{\bf{K}},{\bf{V}})){\rm{~with~}}{\bf{Q}} = {\bf{K}} = {\bf{V}} = {[{({\bf{Z}}_{\tau  - L + 1})_{i,:}^T}|| \cdots ||({{\bf{Z}}_\tau })_{i,:}^T]^T}\\
{\rm{and~}} {{\bf{Z}}_t} \equiv {\mathop{\rm GAT}\nolimits} ({{\bf{A}}_t},{{\bf{X}}_t})~(\tau  - L < t \le \tau ).
\end{array}
\end{equation}
The \textit{structural encoder} first generates the snapshot-induced embedding ${{\bf{Z}}_t} \equiv {\mathop{\rm GAT}\nolimits} ({{\bf{A}}_t},{{\bf{X}}_t})$ for each historical snapshot $G_t$ using GAT (see supplementary materials for its details). For each node $v_i$, the \textit{temporal encoder} applies attention to derive embedding ${\bf{Z}}_{i,:}$ that preserves the evolving patterns across snapshots, where the query, key, and value are set to be the concatenation of snapshot-induced embedding of $v_i$ over $G_{\tau-L}^{\tau}$ (i.e., ${\bf{Q}} = {\bf{K}} = {\bf{V}} = {[{({\bf{Z}}_{\tau  - L + 1}^T)_{i,:}}|| \cdots ||({{\bf{Z}}_\tau })_{i,:}^T]^T} \in \Re^{L \times d}$). Accordingly, the output of ${\mathop{\rm Att}\nolimits} ({\bf{Q}},{\bf{K}},{\bf{V}})$ is also an $L \times d$ matrix. Another function $r_>(\cdot)$ is introduced to reshape ${\mathop{\rm Att}\nolimits} ({\bf{Q}},{\bf{K}},{\bf{V}})$ to a row-wise long vector ${\bf{Z}}_{i,:} \in \Re^{Ld \times 1}$ as the dynamic embedding of $v_i$.

\textit{DySAT} is a \textit{task-independent} DNE approach. In \cite{sankar2020dysat}, random walks on each snapshot were used to optimize the model.
Let ${\mathcal{W}}_i^t$ and ${\mathcal{P}}_i^t$ be the sets of (\romannumeral1) nodes that co-occur with $v_i$ in fixed-length random walks and (\romannumeral2) negative samples of $v_i$ on $G_t$. The loss of \textit{DySAT} maximizes the likelihood (i.e., minimizing the negative log-likelihood) formulated by embedding ${\bf{Z}}$ w.r.t. the sampled random walks. The approximated loss with negative sampling is defined as
\begin{equation}\label{Eq:loss-DySAT}
    \mathop {\min }\limits_\delta ~ \mathcal{L}(G_{\tau  - L}^\tau;\delta ) \equiv \sum\limits_{t = \tau  - L + 1}^\tau  {\sum\limits_{{v_i} \in V_t } {[\sum\limits_{{v_j} \in {\mathcal{W}}_i^t} { - \ln \sigma ({{\bf{Z}}_{i,:}}{\bf{Z}}_{j,:}^T)}  - n_s \sum\limits_{{v_k} \in {\mathcal{P}}_i^t} {\ln (1 - \sigma ({{\bf{Z}}_{i,:}}{\bf{Z}}_{k,:}^T))} ]} },
\end{equation}
where $n_s$ is the number of negative samples w.r.t. each node on a snapshot; $\sigma (\cdot)$ is the sigmoid function. To derive the prediction result, one can adopt one of the strategies illustrated in Table~\ref{tab:DNE-TLP} to build the decoder of \textit{DySAT}.

Different from some RNN-based methods with the dimensionality of model parameters related to the number of nodes (e.g., \textit{dyngraph2vec}, \textit{DDNE}, and \textit{GCN-GAN}), the aforementioned attention-based approaches (i.e., \textit{STGSN} and \textit{DySAT}) can be generalized to new unseen nodes and handle the variation of node sets (i.e., TLP in level-2), based on the inductive nature of GNNs \cite{hamilton2017inductive,qin2022trading} and attentive combination of snapshot-induced embedding ${\bf{Z}}_{\tau-L}^{\tau}$.

\subsubsection{\textbf{Summary of ESSD-Based OTOG Methods}}\label{Sec:Sum-ESSD-OTOG}~

Compared with conventional linear models (e.g., \textit{neighbor similarity}, \textit{graph summarization}, and \textit{matrix factorization} introduced in Section~\ref{Sec:SNAP-OTI}), the aforementioned ESSD-based OTOG methods can explore the non-linear characteristics of dynamic graphs via DL structures (e.g., RBM, MLP, GNN, RNN, and attention).
Following the OTOG paradigm, these methods have the potential to satisfy the real-time constraints of systems, because there is no additional optimization in online generalization.
Most of the methods (except \textit{ctRBM}) can directly (or be easily adapted to) support the TLP on weighted graphs by using adjacency matrices to describe weighted topology. Some approaches (e.g., \textit{GCN-GAN} and \textit{IDEA}) can even derive high-quality weighted prediction results.
However, they may suffer from the limitations of ESSD which can only support the coarse-grained representations of dynamic graphs. The adopted OTOG paradigm may also have the risk of failing to capture the latest evolution of dynamic graphs in online generalization.

\subsection{UESD-Based OTI Methods}\label{Sec:Edge-OTI}

Existing UESD-based OTI approaches usually follow the embedding lookup scheme of classic network embedding techniques \cite{perozzi2014deepwalk,grover2016node2vec}. In this scheme, there is an embedding lookup table ${\bf{Z}} \in \Re^{N \times d}$ shared by all the time steps, with ${\bf{Z}}_{i,:}$ mapping node $v_i$ to its embedding.
${\bf{Z}}$ is also the model parameter to be optimized, whose dimensionality is related to the number of nodes $N$, and thus can only be used to support the TLP in level-1. In general, the encoder of this type of method can be described as
\begin{equation}\label{Eq:Enc-CTDNE}
    {\mathop{\rm Enc}\nolimits} (G_\Gamma) \equiv {\bf{Z}} \in {\Re ^{N \times d}}.
\end{equation}
We divide related methods into two categories according to their techniques used to capture the evolving patterns of UESD-based topology, which are \textit{temporal random walk} (TRW) and \textit{temporal point process} (TPP).

\begin{figure}[t]
\begin{center}
 \begin{minipage}{0.35\linewidth}
 \subfigure[TRW of \textit{CTDNE}]{
  \frame{
\includegraphics[width=\textwidth,trim=15 15 15 15,clip]{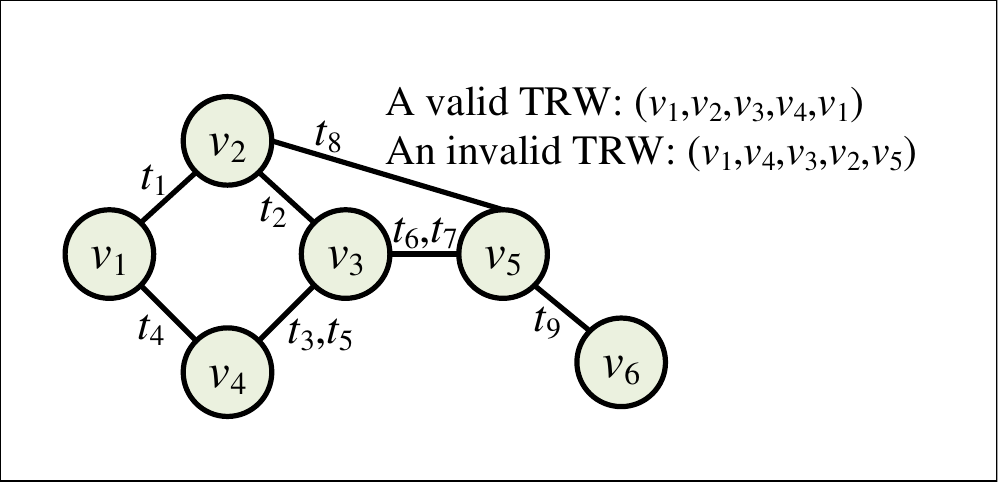}}
  }
 \end{minipage}
 \begin{minipage}{0.50\linewidth}
 \subfigure[Inverse TRW and set-based anonymization features of \textit{CAW}]{
  \frame{
\includegraphics[width=\textwidth,trim=15 15 15 15,clip]{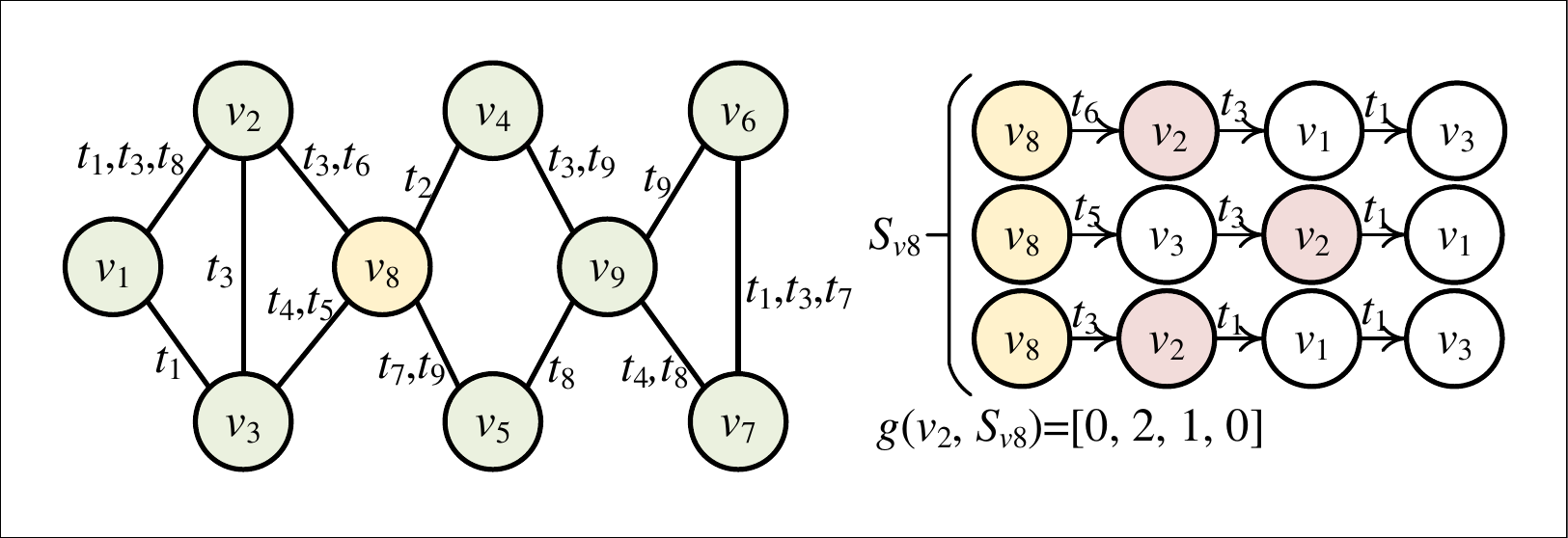}}
  }
 \end{minipage}
\end{center}
\vspace{-0.25cm}
\caption{Examples regarding (a) temporal random walk (TRW) of \textit{CTDNE} as well as (b) inverse TRW of \textit{CAW}.}\label{fig:TRW}
\vspace{-0.35cm}
\end{figure}

\subsubsection{\textbf{Temporal Random Walk} (TRW)}\label{Sec:Edge-OTI-TRW}~

Inspired by the random walk on static graphs \cite{perozzi2014deepwalk,grover2016node2vec}, TRW is an extension to dynamic graphs with UESD.
A TRW with length $K$ can be defined as $\omega  = ({v^{(0)}},{v^{(1)}}, \cdots ,{v^{(K)}})$ such that ${v^{(r)}} \in {V_\Gamma }$, $(({v^{(r-1)}},{v^{(r)}}),{t^{(r)}}) \in {E_\Gamma }$, and $t^{(r)} \in \Gamma$. To enable TRWs to capture the evolution of topology, we also assume that $t^{(r-1)} \le t^{(r)}$. Namely, each TRW is sampled in ascending order of time steps. For the example in Fig.~\ref{fig:TRW} (a), $(v_1, v_2, v_3, v_4, v_1)$ is a valid TRW but $(v_1, v_4, v_3, v_2, v_5)$ is not valid because $t^{(1)}=t_4 > t^{(2)}=t_3$.

(1) \textbf{\textit{CTDNE}}.
Based on TRW, Nguyen et al. \cite{nguyen2018continuous} introduced \textit{CTDNE} following the embedding lookup scheme. The encoder of \textit{CTDNE} is already defined in (\ref{Eq:Enc-CTDNE}).
Let $\mathcal{W}$ be the set of sampled TRWs. The loss function can be described as
\begin{equation}
    \mathop {\min }\limits_{\bf{Z}} ~ \mathcal{L}(\mathcal{W};{\bf{Z}}) \equiv  - \sum\limits_{\omega  \in \mathcal{W}} {\sum\limits_{{v_i} \in \omega } {\sum\limits_{{v_j} \in \omega \backslash \{ {v_i}\} } \ln {P({v_j}|{v_i};{\bf{Z}})} } }, {\rm{~with~}}P({v_j}|{v_i};{\bf{Z}}) = \frac{{\exp \{ {{\bf{Z}}_{j,:}}{\bf{Z}}_{i,:}^T\} }}{{\sum\nolimits_{{v_k} \in {V_\Gamma}} {\exp \{ {{\bf{Z}}_{k,:}}{\bf{Z}}_{i,:}^T\} } }},
\end{equation}
which maximizes the likelihood (i.e., minimizing the negative log-likelihood) of each TRW $\omega \in \mathcal{W}$. Given a node $v_i$ selected in a TRW $\omega$, it maximizes the co-occurrence probability $P(v_j|v_i;{\bf{Z}})$ for each rest node $v_j \in \omega \backslash \{v_i\}$, which is derived by the softmax of embedding w.r.t. associated nodes. Directly computing $P(v_j|v_i;{\bf{Z}})$ via softmax is usually time-consuming due to the summation of all nodes in the denominator. Instead, negative sampling \cite{mikolov2013distributed} can be used to derive an approximated loss similar to that of \textit{DySAT} in (\ref{Eq:loss-DySAT}). As \textit{CTDNE} is a \textit{task-independent DNE} method, we can use one of the strategies in Table~\ref{tab:DNE-TLP} to define its decoder.

\subsubsection{\textbf{Temporal Point Processes} (TPP)}\label{Sec:Edge-OTI-TPP}~

TPP is a continuous-time stochastic process that can also be used to formulate the UESD-based dynamic topology. Assuming that an event happens in a tiny period $[t, t+dt)$, TPP represents the conditional probability of this event given historical events as $\lambda(t)dt$. Hawkes process \cite{yuan2019multivariate} is a typical TPP with $\lambda(t)$ defined as
\begin{equation}\label{Eq:Hawkes-Process}
    \lambda (t) = \mu (t) + \int_{ - \infty }^t {\kappa (t - s)dn(s)},
\end{equation}
where $\lambda (t)$ is the conditional intensity; $\mu (t)$ is the base intensity describing the arrival rate of a spontaneous event at time $t$; $\kappa (t - s)$ is the kernel modeling the time decay of past events; $n(t)$ denotes the number of events until $t$. Methods based on the Hawkes process usually use dynamic embedding ${\bf{Z}}$ to formulate $\{ \lambda (t), \mu (t), \kappa (t - s), n(t)\}$.

(1) \textbf{\textit{HTNE}}.
Zuo et al. \cite{zuo2018embedding} proposed \textit{HTNE} based on the Hawkes process and embedding lookup scheme, with the encoder defined in (\ref{Eq:Enc-CTDNE}).
Let $H_i = ((v^{(1)}, t^{(1)}), (v^{(2)}, t^{(2)}), \ldots)$ be the sequence of historical neighbors of node $v_i$ describing the formation of local topology centered at $v_i$, where $((v_i, v^{(s)}), t^{(s)}) \in E_\Gamma$ and $t^{(s)} \le t^{(s+1)}$. For simplicity, we denote the sequence of historical neighbors of $v_i$ before time step $t$ as $H_i(t)$.
For each edge $((v_i, v_j), t) \in E_\Gamma$, \textit{HTNE} defines the conditional intensity ${\lambda _{ij}}(t)$ as
\begin{equation}\label{Eq:Int-HTNE}
    {\lambda _{ij}}(t) \equiv {\mu _{ij}} + \sum\nolimits_{({v_p},{t_p}) \in {H_i}(t)} {{\alpha _{pj}}\kappa (t - {t_p})},
\end{equation}
where ${\mu _{ij}} \equiv  - |{{\bf{Z}}_{i,:}} - {{\bf{Z}}_{j,:}}|_2^2$ is the base intensity; $\kappa (t - {t_p}) = \exp \{  - {\delta _j}(t - {t_p})\}$ is the kernel function with $\delta_j$ as a learnable parameter w.r.t. node $v_j$; $\alpha_{pj}$ is a weight adjusting the contribution of each historical neighbor $(v_p, t_p)$, which is determined by an attention unit applied to ${\bf{Z}}$ (see \cite{zuo2018embedding} for its details).
\textit{HTNE} is then optimized by maximizing the likelihood (i.e., minimizing the negative log-likelihood) w.r.t. historical topology $\{ H_i | v_i \in V\}$ via the following loss:
\begin{equation}\label{Eq:loss-HTNE}
    \mathop {\min }\limits_{\bf{Z}} ~ {\mathcal{L}}({G_{\Gamma (\tau  - L,\tau )}};{\bf{Z}}) \equiv  - \sum\limits_{{v_i} \in V_\Gamma} {\sum\limits_{({v_j},t) \in {H_i}} {\ln P({v_j}|{v_i},{H_i}(t);{\bf{Z}})} }, {\rm{~with~}}P({v_j}|{v_i},{H_i}(t);{\bf{Z}}) \equiv \frac{{\exp \{ {\lambda _{ij}}(t)\} }}{{\sum\nolimits_{{v_p} \in V_\Gamma} {\exp \{ {\lambda _{ip}}(t)\} } }},
\end{equation}
where the likelihood $P({v_j}|{v_i},{H_i}(t);{\bf{Z}})$ w.r.t. each edge $((v_i, v_j), t)$ is formulated by the softmax of ${\lambda _{ij}}(t)$. Negative sampling \cite{mikolov2013distributed} is also used to derive an approximated version of the aforementioned loss similar to that of \textit{DySAT} in (\ref{Eq:loss-DySAT}).
\textit{HTNE} is a \textit{task-independent} DNE approach that needs a user-defined decoder for TLP. As recommended in \cite{zuo2018embedding}, one can use the weighted-L1 norm strategy in Table~\ref{tab:DNE-TLP} to derive edge embedding $\{ {\bf{e}}_{ij} \}$ and train a downstream logistic regression classifier on $\{ {\bf{e}}_{ij} \}$ to build the decoder.

(2) \textbf{\textit{M2DNE}}.
Lu et al. \cite{lu2019temporal} proposed \textit{M2DNE} by extending \textit{HTNE} to explore the micro- and macro-dynamics that describe the (\romannumeral1) formation of graph topology and (\romannumeral2) evolution of graph scale in terms of the number of edges. The encoder of \textit{M2DNE} is already defined in (\ref{Eq:Enc-CTDNE}) following the embedding lookup scheme.

To optimize the embedding lookup table ${\bf{Z}}$, \textit{M2DNE} first formulates the micro-dynamics of topology using a strategy similar to that of \textit{HTNE}.
For each edge $((v_i, v_j), t) \in E_\Gamma$, it defines the conditional intensity $\lambda_{ij}(t)$ as
\begin{equation}
    {\lambda _{ij}}(t) \equiv {\mu _{ij}} + {\beta _{ij}}[\sum\limits_{({v_p},{t_p}) \in {H_i}(t)} {{\alpha _{pi}}(t){\mu _{pj}}\kappa (t - {t_p})} ] + (1 - {\beta _{ij}})[\sum\limits_{({v_q},{t_q}) \in {H_j}(t)} {{\alpha _{qj}}(t){\mu _{qi}}\kappa (t - {t_q})} ],
\end{equation}
where $\mu_{ij}$, $H_i(t)$, and $\kappa (t - {t_p})$ are with the same definitions as those of \textit{HTNE} in (\ref{Eq:Int-HTNE}); $\alpha_{pi}(t)$ and $\beta_{ij}$ are weights determined by two attention modules applied to ${\bf{Z}}$ (see \cite{lu2019temporal} for their details).
Let $e(t)$ be the number of edges at time step $t$ and $\Delta e(t_s) \equiv e(t_{s+1}) - e(t_s)$ with $t_s$ and $t_{s+1}$ as two successive time steps. For simplicity, let $\Omega$ be the set of time steps w.r.t. the training set. The loss function of \textit{M2DNE} can be described as
\begin{equation}\label{Eq:loss-M2DNE}
\begin{array}{l}
\mathop {\min }\limits_{\bf{Z}}~\mathcal{L}({G_\Omega };{\bf{Z}}) \equiv  - \sum\limits_{t \in \Omega } {\sum\limits_{(({v_i},{v_j}),t) \in E_\Omega } {\ln P({v_i},{v_j}|{H_i}(t), {H_j}(t); {\bf{Z}})} }  + \theta \sum\limits_{t \in \Omega } {{{(\Delta e(t) - \Delta \tilde e(t))}^2}} ,\\
{\rm{with~}}P({v_i},{v_j}|{H_i}(t), {H_j}(t); {\bf{Z}}) \equiv {\lambda _{ij}}(t)/[\sum\limits_{({v_p},{t_p}) \in {H_j}(t)} {{\lambda _{pj}}(t)}  + \sum\limits_{({v_q},{t_q}) \in {H_i}(t)} {{\lambda _{iq}}(t)} ],
\end{array}
\end{equation}
where the first and second terms are losses of micro- and macro-dynamics; $\theta$ is a tunable parameter to balance the two losses. The first term maximizes the likelihood $P({v_i},{v_j}|{H_i}(t), {H_j}(t); {\bf{Z}})$ w.r.t. each edge $((v_i, v_j), t) \in E_\Omega$ based on $\{\lambda_{ij}(t)\}$. The second term minimizes the error between the real number of new edges $\Delta e(t)$ at time step $t$ and the predicted value $\Delta {\tilde{e}}(t)$ derived from embedding ${\bf{Z}}$ (see \cite{lu2019temporal} for its details). Negative sampling \cite{mikolov2013distributed} is also applied to approximate the first term with a form similar to that of \textit{DySAT} in (\ref{Eq:loss-DySAT}). Based on the aforementioned settings, \textit{M2DNE} is a \textit{task-independent} DNE approach and has the same decoder as \textit{HTNE}.

\subsubsection{\textbf{Summary of UESD-Based OTI Methods}}\label{Sec:Sum-UESD-OTI}~

Although \textit{CDTNE}, \textit{HTNE}, and \textit{M2DNE} can support the fine-grained representations of dynamic topology via UESD, they still follow the embedding lookup scheme of \textit{task-independent} DNE and OTI paradigm, where the embedding lookup table ${\bf{Z}}$ is the model parameter optimized for a fixed period and one prediction operation.
When it comes to a new time step, we need to optimize ${\bf{Z}}$ from scratch to support TLP on the latest topology, which is inefficient for applications with real-time constraints.
The embedding lookup scheme, where the dimensionality of ${\bf{Z}}$ is related to the number of nodes $N$, indicates that these methods can only support the TLP in level-1 but cannot tackle the deletion and addition of nodes in level-2. As the decoders of most \textit{task-independent} DNE techniques (see Table~\ref{tab:DNE-TLP}) merely consider the TLP on unweighted graphs, the aforementioned methods also fail to support the prediction of weighted topology.

\subsection{UESD-Based OTOG Methods}
Most UESD-based OTOG approaches utilize the inductive nature of some DL structures to qualify the embedding of each node as a function of time. In this subsection, we denote the embedding of node $v_i$ at time step $t$ as ${\bf{z}}_i(t) \in \Re^{1 \times d}$. This type of method can be further categorized based on the DL technique used to capture the evolving patterns of UESD-based topology, which include the \textit{time-encoded sequential models} and \textit{deep continuous-time processes}.

\subsubsection{\textbf{Time-Encoded Sequential Models}}~

Some related methods use DL structures designed for discrete sequential data (e.g., RNN and attention) to handle UESD-based topology, where the continuous time difference between edges is preserved via specific temporal encoding.

(1) \textbf{\textit{TGAT}}.
Xu et al. \cite{xu2020inductive} proposed \textit{TGAT} that combines attention with temporal encoding based on Bochner’s theorem \cite{loomis2013introduction} in harmonic analysis. It defines a translation-invariant kernel $\left\langle {\Phi ({t_1}),\Phi ({t_2})} \right\rangle$ using a continuous functional mapping $\Phi (t) \equiv [\cos ({\omega _1}t),\sin ({\omega _1}t), \cdots ,\cos ({\omega _d}t),\sin ({\omega _d}t)]/\sqrt d$ with ${\omega _1}, \ldots ,{\omega _d}{ \sim _{\rm{i.i.d~}}}p(\omega )$ (i.e., independently sampled via a common distribution). $\Phi (\Delta t)$ is then used to encode the information of continuous time difference $\Delta t$. \textit{TGAT} also assumes that node attributes are available and fixed for all the time steps, which can be described by an attribute matrix ${\bf{X}}$ with the $i$-th row denoting the attributes of node $v_i$.

\textit{TGAT} follows a multi-layer structure. For each node $v_i$ at time step $t$, let ${\bf{z}}_i^{(k-1)}(t)$ and ${\bf{z}}_i^{(k)}(t)$ be the input and output of the $k$-th layer. $H_i(t)$ denotes the historical neighbors of $v_i$ before $t$, with the same definition as that in (\ref{Eq:Int-HTNE}).
For $v_i$ and its neighbor $(v_j, t_j) \in H_i(t)$, we let ${\bf{\hat z}}_i^{(k - 1)}(t) = [{\bf{z}}_i^{(k - 1)}(t)|| {\Phi (0)} ]$ and ${\bf{\hat z}}_j^{(k - 1)}(t) = [{\bf{z}}_j^{(k - 1)}({t_j})|| {\Phi (t - {t_j})}]$ by concatenating the node embedding and corresponding time difference encoding $\Phi (\Delta t)$. Given a dynamic graph $G_\Gamma$ with fixed attributes ${\bf{X}}$, the encoder w.r.t. the $k$-th \textit{TGAT} layer to obtain embedding ${\bf{z}}_t^{(k)}$ can be formulated as
\begin{equation}
\begin{array}{l}
{\bf{z}}_i^{(k)}(t) = {\mathop{\rm Enc}\nolimits} ^{(k)} ({G_\Gamma }) \equiv {\rm{MLP}}([{{\bf{h}}^{(k - 1)}}(t)||{{\bf{X}}_{i,:}}]),{\rm{~with~}}{{\bf{h}}^{(k - 1)}}(t) \equiv \sum\limits_{({v_j}, {t_j}) \in {H_i}(t)} {{a_{ij}}({{{\bf{\hat z}}}_j^{(k-1)}}(t){{\bf{W}}_v})}\\
{\rm{and~}} {a_{ij}} \equiv \exp \{ ({{\bf{\hat z}}_i^{(k-1)}}(t){{\bf{W}}_q}){({{\bf{\hat z}}_j^{(k-1)}}(t){{\bf{W}}_k})^T}\} /\sum\limits_{{v_s} \in H_i(t)} {\exp \{ ({{\bf{\hat z}}_i^{(k-1)}}(t){{\bf{W}}_q}){{({{\bf{\hat z}}_s^{(k-1)}}(t){{\bf{W}}_k})}^T}\} },
\end{array}
\end{equation}
where ${\bf{W}}_q$, ${\bf{W}}_k$, and ${\bf{W}}_v$ are learnable parameters for the linear mapping of query, key, and value in the attention unit. Concretely, the $k$-th layer first derives an intermediate representation ${\bf{h}}^{(k-1)}(t)$ via the attentive combination of embedding ${\bf{\hat z}}_j^{(k-1)}$ w.r.t. each historical neighbor $(v_j, t_j) \in H_i(t)$, from which the temporal local topology of $v_i$ can be preserved. ${\bf{h}}^{(k-1)}(t)$ is then concatenated with the attributes ${\bf{X}}_{i,:}$ of target node $v_i$ and further fed into an MLP.

We denote the embedding given by the last layer as $\{ {\bf{z}}_i(t) \}$. Let $\Omega$ denote the set of time steps w.r.t. the training set. \textit{TGAT} is a  \textit{task-independent} DNE method that can be trained by maximizing the likelihood of historical topology $G_\Omega$. As recommended in \cite{xu2020inductive}, we can apply the following approximated loss function with negative sampling for the model optimization of \textit{TGAT}:
\begin{equation}\label{Eq:Loss-TGAT}
    {\min}_\delta ~ {\mathcal{L}}({G_\Omega };\delta ) \equiv \sum\nolimits_{(({v_i},{v_j}),t) \in {E_\Omega }} { - \ln (\sigma ( - {{\bf{z}}_i}(t){\bf{z}}_j^T(t))) - n_s {\mathbb{E}_{{v_p}\sim{\mathcal{P}_n}}}[\ln (\sigma ({{\bf{z}}_i}(t){\bf{z}}_p^T(t)))} ],
\end{equation}
where $n_s$ is the number of negative samples; $\mathcal{P}_n$ is the distribution of negative sampling; $\sigma (\cdot)$ is the sigmoid function. One of the strategies described in Table~\ref{tab:DNE-TLP} can be used to build the decoder of \textit{TGAT}.
By utilizing the availability of node attributes and the inductive nature of attention, \textit{TAGT} is able to tackle the TLP in level-2 with non-fixed node sets.

(2) \textbf{\textit{CAW}}.
Wang et al. \cite{wang2021inductive} extended the TRW (used by \textit{CTDNE} as described in Section~\ref{Sec:Edge-OTI-TRW}) to a set-based anonymized version, which removes node identities to support inductive learning while preserving the local topology structures. \textit{CAW} was proposed to handle the anonymized features of TRWs using RNNs.
Different from that of \textit{CTDNE} as illustrated in Fig.~\ref{fig:TRW} (a), \textit{CAW} adopts the reverse TRW such that $\omega  = ({v^{(0)}},{v^{(1)}}, \cdots ,{v^{(K)}})$ with $(({v^{(r-1)}},{v^{(r)}}),{t^{(r)}}) \in {E_\Gamma }$ and ${t^{(r-1)}} \ge {t^{(r)}}$. Namely, each TRW should be sampled in descending order of the time steps. For instance, in Fig.~\ref{fig:TRW} (b), $(v_8, v_3, v_2, v_1)$ and $(v_8, v_2, v_1, v_3)$ are valid TRWs for \textit{CAW}.

Let $S_{v_i}$ be the set of TRWs starting from node $v_i$. For each edge $(({v_i},{v_j}),t) \in {E_\Gamma }$, consider a node ${v_p} \in {S_{v_i}} \cup {S_{v_j}}$. \textit{CAW} defines a set-based anonymized feature $ I({v_p};{S_{{v_i}}},{S_{{v_j}}}) \equiv [g({v_p},{S_{{v_i}}}),g({v_p},{S_{{v_j}}})]$, where $g({v_p},{S_{v_i}}) \equiv [c_0^{p,{S_{v_i}}}, \cdots ,c_K^{p,{S_{v_i}}}]$; $c_r^{p,{S_{v_i}}}$ denotes the frequency that $v_p$ appears at the $r$-th position among all the TRWs in ${S_{v_i}}$. For the example in Fig.~\ref{fig:TRW} (b), there are $3$ TRWs in $S_{v_8}$. For node $v_2 \in S_{v_8}$, one can obtain $g(v_2, S_{v_8}) = [0, 2, 1, 0]$ because the frequencies that $v_2$ appears at positions $[0, 1, 2, 3]$ in $S_{v_8}$ are $[0, 2, 1, 0]$.
Accordingly, the set-based anonymization of a TRW $\omega = (v^{(0)}, v^{(1)}, \cdots)$ is defined as $\hat \omega  \equiv (I({v^{(0)}};{S_{{v^{(0)}}}},{S_{{v^{(1)}}}}),I({v^{(1)}};{S_{{v^{(1)}}}},{S_{{v^{(2)}}}}), \cdots )$.
For a node pair $(v_i, v_j)$, let ${{\mathcal{\hat W}}_{ij}}$ be the set of anonymized TRWs $\hat \omega$ w.r.t. ${S_{v_i}} \cup {S_{v_j}}$. Given ${{\mathcal{\hat W}}_{ij}}$, the encoder of \textit{CAW} is defined as
\begin{equation}
    \begin{array}{l}
    {{\bf{e}}_{ij} (t^{(1)})} = {\mathop{\rm Enc}\nolimits} ({{\mathcal{\hat W}}_{ij}}) \equiv \frac{1}{{|{{\mathcal{\hat W}}_{ij}}|}}\sum\nolimits_{\hat \omega  \in {{\mathcal{\hat W}}_{ij}} } {{\mathop{\rm VanRNN}\nolimits} ({{ [ {f}({v^{(r)}})||{\Phi}({t^{(r+1)}} - {t^{(r+2)}}) ] }_{r = 0, \cdots ,K-2}})}, \\
    {f}({v^{(r)}}) \equiv {\mathop{\rm MLP}\nolimits} (g({v^{(r)}},{S_{{v^{(r)}}}})) + {\mathop{\rm MLP}\nolimits} (g({v^{(r)}},{S_{{v^{(r + 1)}}}})) {\rm{~w.r.t.~}} I(v^{(r)}; S_{v^{(r)}}, S_{v^{(r+1)}}),\\
    {\Phi }(\Delta t) \equiv [\cos ({\omega _1} \Delta t),\sin ({\omega _1} \Delta t), \cdots ,\cos ({\omega _d} \Delta t),\sin ({\omega _d} \Delta t)],
    \end{array}
\end{equation}
where ${\bf{e}}_{ij} (t)$ is the auxiliary edge embedding of node pair $(v_i, v_j)$ at time step $t$; $f(v^{(r)})$ represents a function to process $I(v^{(r)}; S_{v^{(r)}}, S_{v^{(r+1)}})$ (i.e., the anonymized features of the $r$-th node in a TRW), which uses two MLPs with shared parameters to encode the effects of both source and destination nodes of an associated edge $(v^{(r)}, v^{(r+1)})$; ${\Phi }(\Delta t)$ is the temporal encoding that preserves the information of continuous time difference $\Delta t$ with the same definition as that of \textit{TGAT}; ${\mathop{\rm VanRNN}\nolimits} (\cdot)$ denotes the vanilla RNN (see supplementary materials for its details) used to handle features $[ {f}({v^{(r)}})||{\Phi}({t^{(r)}} - {t^{(r+1)}}) ] _{r = 0, \cdots ,K}$ w.r.t. discrete TRWs.

One can apply an MLP to ${\bf{e}}_{ij} (t)$ to derive the estimated probability $\mathcal{\tilde E}_{ij}^{\tau+r} \equiv P((({v_i},{v_j}), t)|{G_\Gamma })$ that an edge $(v_i, v_j)$ appears at a future time step $(\tau + r)$ with $0< r \le \Delta$. Namely, the decoder of \textit{CAW} is defined as
\begin{equation}
    {\mathcal{{\tilde E}}_{ij}^{\tau + r}} = {\mathop{\rm Dec}\nolimits} ({\bf{e}}_{ij} (\tau + r)) \equiv {\mathop{\rm MLP}\nolimits} ({{\bf{e}}_{ij}} (\tau + r)).
\end{equation}
Let $\bar \Omega$ be a sampled training set with both positive and negative samples in terms of edges $\{ ((v_i, v_j), t) \}$. \textit{CAW} is a \textit{task-dependent} method optimized by the following cross-entropy loss regarding TLP:
\begin{equation}\label{Eq:loss-CAW}
    \mathop {\min }\limits_\delta ~ \mathcal{L}({\bar \Omega} ;\delta ) \equiv \frac{1}{{|{\bar \Omega} |}}\sum\nolimits_{(({v_i},{v_j}), t) \in {\bar \Omega} } { - {\mathcal{E}_{ij}^{t}}\ln {\mathcal{{\tilde E}}_{ij}^{t}} + (1 - {\mathcal{E}_{ij}^{tr}})\ln (1 - {\mathcal{{\tilde E}}_{ij}^{t}})},
\end{equation}
where $\mathcal{E}_{ij}^{t} \in \{ 0, 1\}$ is the corresponding ground-truth.
Since the set-based anonymized features $\{ I(v_p; S_{v_i}, S_{v_j}) \}$ are shared by all the possible dynamic topology, \textit{CAW} can support the TLP in level-2, handling the variation of node sets.

\subsubsection{\textbf{Deep Continuous-Time Processes}}~

In addition to using deep sequential models to handle the UESD-based topology, other methods directly formulate continuous-time processes (e.g., Hawkes process described in (\ref{Eq:Hawkes-Process}) and ordinary differential equation) using DL structures.

(1) \textbf{\textit{DyRep}}.
Trivedi et al. \cite{trivedi2019dyrep} adopted a DL model to formulate TPP and introduced \textit{DyRep}.
For edge $e=((v_i, v_j), t)$, let $\bar t$ denote the time step of another edge observed just before $t$. Similarly, let ${\bar t}_i$ be the time step that node $v_i$ is observed just before $t$. When $v_i$ is first added (i.e., not previously observed), we let ${{\bar t}_i} = 0$ and set the initial embedding ${{\bf{z}}_i}({{\bar t}_i})$ to be its node attributes (if available) or a random vector. $H_i(t)$ represents the set of historical neighbors of $v_i$ before $t$, with the same definition as that of \textit{HTNE} in (\ref{Eq:Int-HTNE}). For each edge $((v_i, v_j), t) \in E_\Gamma$, the encoder of \textit{DyRep} derives the embedding $({\bf{z}}_i(t), {\bf{z}}_j(t))$ w.r.t. the two induced nodes $(v_i, v_j)$ via
\begin{equation}
\begin{array}{l}
{{\bf{z}}_i}(t) = {\mathop{\rm Enc}\nolimits} ({G_\Gamma }) \equiv \sigma ({{\bf{h}}_i}(\bar t){{\bf{W}}_l} + {\bf{z}}({{\bar t}_i}){{\bf{W}}_s} + (t - {{\bar t}_i}){{\bf{W}}_t})\\
{\rm{with~}}{{\bf{h}}_i}(t) \equiv \max \{ \sigma (a_{ij}^{t_j} \cdot {\mathop{\rm MLP}\nolimits} ({{\bf{z}}_j}(t_j)) | \forall (v_j, t_j) \in {H_i}(t)\}.
\end{array}
\end{equation}
The first term ${{\bf{h}}_i}(\bar t){{\bf{W}}_l}$ explores the local second-order proximity of $v_i$, where ${\bf{h}}_i(\bar t)$ is the attentive aggregation of embedding $\{ {\bf{z}}_j(t_j) \}$ w.r.t. historical neighbors $H_i(t)$; $a_{ij}^{t_j}$ is the weight determined by an attention unit (see \cite{trivedi2019dyrep} for its details) to adjust the contribution of each historical neighbor $(v_i, t_j) \in H_i(t)$. The second term ${{\bf{z}}_i}({{\bar t}_i}){{\bf{W}}_s}$ encodes the self-propagation that ${\bf{z}}_i (t)$ evolves w.r.t. its previous position ${\bf{z}}_i ({\bar t}_i)$. The third term $(t - {{\bar t}_i}){{\bf{W}}_t}$ ensures that ${\bf{z}}_i (t)$ is updated smoothly during interval $(t - {\bar t}_i)$. $\{ {\bf{W}}_l, {\bf{W}}_s, {\bf{W}}_t\}$ are learnable parameters. $\sigma (\cdot)$ is the sigmoid function.


In \textit{DyRep}, $k=0$ and $k=1$ are used to denote the cases that (\romannumeral1) a new edge $(v_i, v_j)$ is first added and (\romannumeral2) an old edge $(v_i, v_j)$ is observed again. To formulate the TPP using embedding $\{ {\bf{z}}_i(t) \}$, \textit{DyRep} defines the conditional intensity for each edge $((v_i, v_j), t)$ with type $k$ as $\lambda _{ij}^k(t) \equiv {f_k}([{{\bf{z}}_i}(\bar t)||{{\bf{z}}_j}(\bar t)]{\bf{w}}_k^T)$, where ${f_k}(x) \equiv {\psi _k}\ln (1 + \exp \{ x/{\psi _k}\} )$ is a non-linear function ensuring that the intensity $\lambda _{ij}^k(t)$ is positive; $\{ \psi_k, {\bf{w}}_k \}$ are learnable parameters associated with $k \in \{ 0, 1\}$.
Let $\Omega$ be the collection of time steps associated with the training set. \textit{DyRep} maximizes the likelihood (i.e., minimizing the negative log-likelihood) w.r.t. edges in $E_\Omega$ via the following loss:
\begin{equation}
   \mathop {\min }\limits_\delta ~ \mathcal{L}({G_\Omega };\delta ) \equiv  - \sum\limits_{e=(({v_i},{v_j}),t) \in {E_\Omega }} {\ln (\lambda _{ij}^{\pi (e)}(t))}  + \int_0^{t}  {(\sum\limits_{{v_i} \in {V_\Omega }} {\sum\limits_{{v_j} \in {V_\Omega }} {\sum\limits_{k \in \{ 0,1\} } {\lambda _{ij}^k(s)} } } )ds},
\end{equation}
where $\pi(e) \in \{ 0, 1 \}$ maps $e$ to its type; the second term is the survival probability for events that do not happen.
The decoder estimates the probability $\mathcal{\tilde E}_{ij}^{\tau+r} \equiv P((({v_i},{v_j}), \tau + r)|{G_\Gamma })$ that an edge $(v_i, v_j)$ appears at a time step $(\tau+r)$ via
\begin{equation}
    {\mathcal{\tilde E}}_{ij}^{\tau + r} \propto {\mathop{\rm Dec}\nolimits} (\{ {{\bf{z}}_i}(\tau+r)\} ) \equiv \lambda _{ij}^k(\tau+r) \cdot \exp \{ \int_{\tau}^{\tau+r} {\lambda _{ij}^k(s)ds} \}.
\end{equation}
As \textit{DyRep} initializes the embedding of each newly added node by setting ${\bf{z}}_i(\bar t_i = 0)$, it can handle the TLP in level-2, deriving prediction results for new unseen nodes.

(2) \textbf{\textit{TREND}}.
In \cite{wen2022trend}, \textit{TREND} was proposed to formulate Hawkes process using a multi-layer GNN. Let ${\bf{z}}_i^{(k-1)}(t)$ and ${\bf{z}}_i^{(k)}(t)$ be the input and output of the $k$-th layer for node $v_i$ at time step $t$. As node attributes are assumed to be available and fixed for all time steps, we set ${\bf{z}}_i^{(0)}(t) = {\bf{X}}_{i,:}$. $H_i(t)$ and $\kappa (t - {t_j})$ denote the (\romannumeral1) sequence of historical neighbors and (\romannumeral2) decaying kernel with the same definitions as those in (\ref{Eq:Int-HTNE}). The $k$-th encoder layer can be described as
\begin{equation}
    {\bf{z}}_i^{(k)}(t) = {\rm{Enc}} ^{(k)} ({G_\Gamma },{\bf{X}}) \equiv \sigma ({\bf{z}}_i^{(k - 1)}(t){\bf{W}}_s^{(k - 1)} + \sum\nolimits_{({v_j},{t_j}) \in {H_i}(t)} {{\bf{z}}_j^{(k - 1)}(t_j){\bf{W}}_h^{(k - 1)}\tilde \kappa (t - {t_j})} ),
\end{equation}
where $\tilde \kappa (t - {t_j}) = \kappa (t - {t_j})/\sum\nolimits_{({v_k},{t_k}) \in {H_i}(t)} {\kappa (t - {t_k})}$ is the time decaying factor normalized over historical neighbors $H_i (t)$; $\{ {\bf{W}}_s, {\bf{W}}_h\}$ are learnable parameters; $\sigma (\cdot)$ is the sigmoid function. The first and second terms are used for receiving the self-information and aggregating the historical neighbors of $v_i$ before $t$.
Based on the embedding $\{ {\bf{z}}_i(t)\}$ given by the last encoder layer, the conditional intensity of Hawkes process w.r.t. each edge $((v_i, v_j), t) \in E_\Gamma$ is formulated as
\begin{equation}
\begin{array}{l}
    {\lambda _{ij}}(t) \equiv {\mathop{\rm MLP}\nolimits} ({({{\bf{z}}_i}(t) - {{\bf{z}}_j}(t))^{ \circ 2}};\theta ({v_i},{v_j},t)), {\rm{~with~}}\theta ({v_i},{v_j},t) = [\alpha ({v_i},{v_j},t) + 1] \odot {\theta _e} + \beta ({v_i},{v_j},t),\\
    \alpha ({v_i},{v_j},t) \equiv {\mathop{\rm MLP}\nolimits} ([{{\bf{z}}_i}(t)||{{\bf{z}}_j}(t)]), {\rm{~and~}} \beta ({v_i},{v_j},t) \equiv {\mathop{\rm MLP}\nolimits} ([{{\bf{z}}_i}(t)||{{\bf{z}}_j}(t)]),
\end{array}
\end{equation}
where $\circ 2$ denotes the element-wise square; $\theta ({v_i},{v_j},t)$ is the set of model parameters of ${\lambda _{ij}}(t)$; $\theta_e$ is a learnable event prior; $\alpha ({v_i},{v_j},t)$ and $\beta ({v_i},{v_j},t)$ are defined as the scaling and shifting operators, which are MLPs with $[{\bf{z}}_i(t)||{\bf{z}}_j(t)]$ as inputs, following the feature-wise linear modulation \cite{perez2018film} in meta-learning. In this setting, the model parameters $\theta ({v_i},{v_j},t)$ of ${\lambda _{ij}}(t)$ are set to be automatically adjusted according to the information encoded in $({\bf{z}}_i(t), {\bf{z}}_j(t))$.

Let $\Omega$ be the set of time steps associated with the training set. \textit{TREND} is optimized via the following loss integrating both (\romannumeral1) \textit{event dynamics} (i.e., formation of edges) and (\romannumeral2) \textit{node dynamics} (i.e., growth of edges w.r.t. each node):
\begin{equation}
    \begin{array}{l}
\mathop {\min }\limits_\delta  \mathcal{L}({G_\Omega };\delta ) \equiv \sum\limits_{(({v_i},{v_j}),t) \in {E_\Omega }} {[{\mathcal{L}_e}({v_i},{v_j},t) + {\theta _1}({\mathcal{L}_n}({v_i},t) + {\mathcal{L}_n}({v_j},t)) + {\theta _2}(\left\| {\alpha ({v_i},{v_j},t)} \right\|_2^2 + \left\| {\beta ({v_i},{v_j},t)} \right\|_2^2)]}, \\
{\rm{with~}} {\mathcal{L}_e}({v_i},{v_j},t) \equiv  - \ln ({\lambda _{ij}}(t)) - n_s{\mathbb{E}_{{v_k} \sim {\mathcal{P}_n}}}[\ln (1 - {\lambda _{ik}}(t))], {\rm{~and~}}
{\mathcal{L}_n}({v_i},t) \equiv {[{e_i}(t) - {\mathop{\rm MLP}\nolimits} ({{\bf{z}}_i}(t))]^2}.
\end{array}
\end{equation}
${\mathcal{L}_e}({v_i},{v_j},t)$ is the loss to capture \textit{event dynamics} via a strategy similar to (\ref{Eq:Loss-TGAT}), where $n_s$ and $\mathcal{P}_n$ also have the same definitions as those in (\ref{Eq:Loss-TGAT}).
${\mathcal{L}_n}({v_i},t)$ is the loss incorporating \textit{node dynamics} that minimizes the error between (\romannumeral1) the number of links $e_i(t)$ formed by $v_i$ at time step $t$ and (\romannumeral2) the predicted value $\tilde e_i(t) \equiv {\mathop{\rm MLP}\nolimits} ({{\bf{z}}_i}(t))$ given by an MLP, with motivations similar to the optimization of \textit{M2DNE} in (\ref{Eq:loss-M2DNE}). Moreover, the $l_2$-regularization is applied to $\alpha(v_i, v_j, t)$ and $\beta(v_i, v_j, t)$ to avoid over-fitting. $\{ \theta_1, \theta_2 \}$ are tunable hyper-parameters.

The aforementioned definitions of encoder and loss function indicate that \textit{TREND} is a \textit{task-independent} DNE approach. As recommended in \cite{wen2022trend}, one can use the same decoder as \textit{HTNE} to derive prediction results.
Furthermore, \textit{TREND} can also tackle the TLP in level-2 because it utilizes the inductive nature of GNNs and the availability of node attributes.

(3) \textbf{\textit{GSNOP}}. Neural ordinary differential equation (NODE) \cite{chen2018neural} is another continuous-time model that can be used to handle the UESD-based topology. Let ${\bf{u}} (t)$ a variable at time step $t$. NODE formulates the derivative of ${\bf{u}} (t)$ w.r.t. $t$ using a DL module denoted as ${f_{{\rm{ODE}}}}({\bf{u}}(t),t) \equiv d{\bf{u}}(t)/dt$. Given a start time step $t_0$ and initial value ${\bf{u}} (t_0)$, NODE can derive ${\bf{u}} (t)$ at any time step $t > t_0$ by solving the following equation
\begin{equation}
    {\bf{u}}(t) = {\mathop{\rm NODE}\nolimits} ({f_{{\mathop{\rm ODE}\nolimits} }},{\bf{u}}({t_0}),{t_0},t) \equiv {\bf{u}}({t_0}) + \int_{{t_0}}^t {{f_{{\rm{ODE}}}}({\bf{u}}(s), s)} ds.
\end{equation}
One can obtain the approximated value of ${\bf{u}}(t)$ by applying a numeric solver (e.g., Euler or Runge-Kutta solver \cite{garrappa2018numerical}) denoted as ${\bf{u}}(t) \approx {\mathop{\rm ODESolver}\nolimits} ({f_{{\mathop{\rm ODE}\nolimits} }},{\bf{u}}({t_0}),{t_0},t)$. Luo et al. \cite{luo2023graph} combined NODE with existing DNE models and proposed \textit{GSNOP}, which is a \textit{task-dependent} DNE approach for TLP.

Let $g_{\rm DNE} (\cdot)$ be the encoder of an existing DNE-based method (e.g., \textit{DySAT} and \textit{TGAT} reviewed in this survey) that can derive the embedding ${\bf{z}}_i (t)$ for each node $v_i$ at time step $t$. The encoder of \textit{GSNOP} generates auxiliary edge embedding ${\bf{e}}_{ij} (t)$ for each node pair $(v_i, v_j)$ at a previous time step $t \le \tau$ based on $({\bf{z}}_i (t), {\bf{z}}_j (t))$, which can be described as
\begin{equation}\label{Eq:Enc-GSNOP-1}
\begin{array}{l}
{{\bf{e}}_{ij}}(t) = {\mathop{\rm Enc}\nolimits} ({G_\Gamma },{g_{\rm DNE}},{v_i},{v_j},t) \equiv {\mathop{\rm MLP}\nolimits} ([{{\bf{z}}_i}(t)||{{\bf{z}}_j}(t)||y_{ij}^t]) + \Phi (t),\\{\rm{with~}}{{\bf{z}}_i}(t) \equiv {g_{\rm DNE}}({G_\Gamma },{v_i},t),
\end{array}
\end{equation}
where $y_{ij}^{t} = 1$ if $((v_i, v_j), t) \in E_\Gamma$ and $y_{ij}^{t} = 0$ otherwise (i.e., an auxiliary variable denoting the existence of an edge); $\Phi (t) \in \Re^d$ is the temporal encoding with the same definition as that of \textit{TGAT}. For a future time step $(\tau + r)$, the encoder further generates ${\bf{e}}_{ij} (\tau+r)$ based on NODE with ${\bf{e}}_{ij} (\tau)$ as the initial state, which can be formulated as
\begin{equation}\label{Eq:Enc-GSNOP-2}
\begin{array}{l}
{{\bf{e}}_{ij}}(\tau  + r) = {\mathop{\rm Enc}\nolimits} ({f_{{\rm{ODE}}}},{{\bf{e}}_{ij}}(\tau ),\tau  + r) \equiv {{\bf{e}}_{ij}}(\tau ) + \int_\tau ^{\tau  + r} {{f_{{\rm{ODE}}}}({{\bf{e}}_{ij}}(t),\tau )dt},\\
{\rm{~with~}}{f_{{\rm{ODE}}}}({{\bf{e}}_{ij}}(t),t) \equiv {\mathop{\rm MLP}\nolimits} ({{\bf{e}}_{ij}}(t) + \Phi (t)).
\end{array}
\end{equation}
The decoder of \textit{GSNOP} takes $({\bf{z}}_i (\tau), {\bf{z}}_j (\tau))$ and ${\bf{e}}_{ij} (\tau+r)$ as inputs and estimates the probability $\mathcal{E}_{ij}^{\tau+r} \equiv P(((v_i, v_j), \tau+r)|G_\Gamma)$ that an edge $(v_i, v_j)$ appears at a future time step $(\tau + r)$ via
\begin{equation}
\begin{array}{l}
 \mathcal{\tilde E}_{ij}^{\tau+r} = {\mathop{\rm Dec}\nolimits} ({{\bf{z}}_i}(\tau), {{\bf{z}}_j}(\tau), {\bf{e}}_{ij} (\tau+r)) \equiv {\mathop{\rm MLP}\nolimits} ([{{{\bf{\hat z}}}_i}(\tau+r)||{{{\bf{\hat z}}}_j}(\tau+r)]),\\
{\rm{with~}}{{{\bf{\hat z}}}_i}(\tau+r) \equiv {\mathop{\rm MLP}\nolimits} ([{{\bf{z}}_i}(\tau)||{{\bf{\hat e}} }_{ij}(\tau+r)]), ~{{\bf{\hat e}}}_{ij}(\tau+r) \sim \mathcal{N}(\mu_{ij}^{\tau+r}, \sigma_{ij}^{\tau+r}),\\
\mu_{ij}^{\tau+r} \equiv {\mathop{\rm MLP}\nolimits} ({{\bf{e}}_{ij}}(\tau  + r)) {\rm{~and~}} {\sigma_{ij}^{\tau+r}} \equiv 0.1 + 0.9{\mathop{\rm MLP}\nolimits} ({{\bf{e}}_{ij}}(\tau  + r)).
\end{array}
\end{equation}
Following a structure similar to the variational autoencoder (VAE) \cite{kingma2013auto}, the decoder first constructs a Gaussian distribution $\mathcal{N}(\mu_{ij}^{\tau+r}, \sigma_{ij}^{\tau+r})$, with the mean $\mu_{ij}^{\tau+r}$ and variance $\sigma_{ij}^{\tau+r}$ formulated by ${\bf{e}}_{ij} (\tau+r)$. Another embedding ${\bf{\hat e}}_{ij} (\tau+r)$ is then sampled from $\mathcal{N}(\mu_{ij}^{\tau+r}, \sigma_{ij}^{\tau+r})$ and concatenated with $\{ {\bf{z}}_i (\tau), {\bf{z}}_j (\tau) \}$ to derive $\mathcal{\tilde E}_{ij}^{\tau+r}$ using MLPs. Let $\Omega$ be the set of time steps w.r.t. the training set. Similar to VAE, the loss function of \textit{GSNOP} maximizes the evidence lower bound (ELBO) \cite{kingma2013auto} (i.e., minimizing the negative ELBO) to optimize model parameters $\delta$, which can be described as
\begin{equation}
\begin{array}{l}
\mathop {\min }\limits_\delta ~ \mathcal{L}({G_\Omega };\delta ) \equiv  - {\mathbb{E}_{Q({\bf{e}}_{ij}^ > | {\bf{e}}_{ij}^ <)}}[\ln \mathcal{\tilde E}_{ij}^{\tau+r}] + \sum\limits_{ij} {{\mathop{\rm KL}\nolimits} [Q({\bf{e}}_{ij}^ >| {\bf{e}}_{ij}^ < )||P({\bf{e}}_{ij}^ >)]} ,\\
{\rm{with~}} Q({\bf{e}}_{ij}^ > | {\bf{e}}_{ij}^ <) \approx \mathcal{N}(\mu_{ij}^{\tau+r},{\sigma}_{ij}^{\tau+r}) {\rm{~and~}} P({\bf{e}}_{ij}^ >) \approx {{\bf{e}}_{ij}}(\tau  + r).
\end{array}
\end{equation}
The first term maximizes the expectation of $\ln \mathcal{\tilde E}_{ij}^{\tau+r}$, which is equivalent to minimizing the cross-entropy between $\{ \mathcal{\tilde E}_{ij}^{\tau+r} \}$ and ground-truth, with a form similar to (\ref{Eq:loss-CAW}). The second term minimizes the KL-divergence between $Q({\bf{e}}_{ij}^ > | {\bf{e}}_{ij}^ <)$ and $P({\bf{e}}_{ij}^ >)$, where $Q({\bf{e}}_{ij}^ > | {\bf{e}}_{ij}^ <)$ is the posterior distribution of ${\bf{e}}_{ij} (\tau+r)$ given previous embedding ${\bf{e}}_{ij} (t)$ with $t \le \tau$, which is estimated by $\mathcal{N}(\mu_{ij}^{\tau+r},{\sigma}_{ij}^{\tau+r})$; $P({\bf{e}}_{ij}^ >)$ is the prior distribution of ${{\bf{e}}_{ij}}(\tau  + r)$ estimated via the NODE in (\ref{Eq:Enc-GSNOP-2}).

\subsubsection{\textbf{Summary of UESD-Based OTOG Methods}}\label{Sec:Sum-UESD-OTOG}~

In summary, the aforementioned approaches (e.g., \textit{TGAT}, \textit{CAW}, \textit{DyRep}, \textit{TREND}, and \textit{GSNOP}) can support the fine-grained representation of dynamic topology to handle the rapid topology variation via UESD.
As they adopt the OTOG paradigm based on the inductive nature of DL structures and attributes/features shared by all the possible nodes, they can tackle the TLP in level-2 with non-fixed node sets while having the potential to satisfy the real-time constraints of systems.
However, since they still rely on stochastic processes (e.g., TPP and TRW) on unweighted graphs, they cannot explore the weighted topology and support the advanced TLP on weighted graphs. Due to limitations of the OTOG paradigm, this kind of method may also have the risk of failing to capture the latest evolving patterns of graphs.

\section{Advanced Topics \& Future Directions}\label{Sec:Topic-Future}

In this section, we first summarize some advanced topics in recent research based on our review of existing TLP methods in Section~\ref{Sec:Meth}. Furthermore, several possible future directions are highlighted at the end of this section.

\subsection{Advanced Research Topics}\label{Sec:Adv-Topic}

\subsubsection{\textbf{Prediction of Weighted Topology}}~

Most existing TLP methods merely focus on the prediction of unweighted topology. Some of them are inapplicable to the TLP on weighted graphs.
On the one hand, the encoders and loss functions of some methods cannot capture the variation of weighted topology. For instance, most UESD-based approaches rely on stochastic processes defined on unweighted graphs (e.g., TRW and TPP introduced in Sections~\ref{Sec:Edge-OTI-TRW} and \ref{Sec:Edge-OTI-TPP}), which do not have the hypotheses regarding the evolution of edge weights.
On the other hand, the decoders of some approaches (e.g., \textit{ctRBM}, \textit{DyRep}, and \textit{TGAT}) are only designed for unweighted graphs, treating TLP as the binary edge classification, which can only derive the probability that an edge will appear in a time step but cannot predict the corresponding edge weight.

Although some ESSD-based methods (e.g., \textit{GrNMF}, \textit{TMF}, \textit{dyngraph2vec}, and \textit{DDNE}) can still support the TLP on weighted graphs by using adjacency matrices $\{{\bf{A}}_t\}$ to describe the weighted topology, they can only derive low-quality prediction results. Most of them are optimized via error minimization objectives that minimize the reconstruction error between the training ground-truth ${\bf{A}}_{\tau+1}$ and prediction result ${\bf{\tilde A}}_{\tau+1}$ as illustrated in (\ref{Eq:Enc-TMF}) and (\ref{Eq:loss-dyngraph2vec}). Qin et al. \cite{qin2023high} argued that these objectives cannot tackle the following \textit{wide-value-range} and \textit{sparsity} issues.

\textbf{\textit{Wide-Value-Range Issue}}. In weighted graphs, edge weights may have a wide value range (e.g., $[0, 2000]$). There may also be a non-ignorable portion of edges with small weights. However, error minimization objectives are only sensitive to large edge weights but fail to distinguish the scale difference between small weights. For instance, the scale difference between (1, 2) is larger than that between (1990, 2000), although the latter case has a larger error. From the view of pre-allocating system resources, failing to distinguish the scale difference of edge weights may have the risks of (\romannumeral1) allocating much more resources than the real demand of a link or (\romannumeral2) not allocating enough resources for a link.

\textbf{\textit{Sparsity Issue}}. In an adjacency matrix ${\bf{A}}_t$, small and zero elements have different physical meanings. $({\bf{A}}_t)_{ij} = 0$ indicates that there is no edge between $(v_i^t, v_j^t)$. A small element $({\bf{A}}_t)_{ij} > 0$ implies that there is still an edge between $(v_i^t , v_j^t)$ but the edge weight is small. The topology of some real-world systems may be sparse with a non-ignorable portion of zeros in ${\bf{A}}_t$. Since error minimization objectives are only sensitive to large weights, methods that are only optimized via these objectives may also fail to distinguish the difference between zero and small weights in ${\bf{A}}_t$. For resource pre-allocation, failing to distinguish between zero and small weights may also have the risks of (\romannumeral1) allocating resources for non-existent links or (\romannumeral2) not allocating resources for existing links.

Due to the aforementioned issues, most existing methods (e.g., \textit{GrNMF}, \textit{TMF}, \textit{dyngraph2vec}, and \textit{DDNE}) can only generate low-quality prediction results $\{ {\bf{\tilde A}}_{\tau+1} \}$ (in terms of adjacency matrices) that fail to distinguish between small and zero weights for weighted graphs, and thus can only support the coarse-grained resource allocation. 

\begin{figure}[t]
\begin{center}
 \includegraphics[width=0.75\linewidth, trim=18 32 18 18,clip]{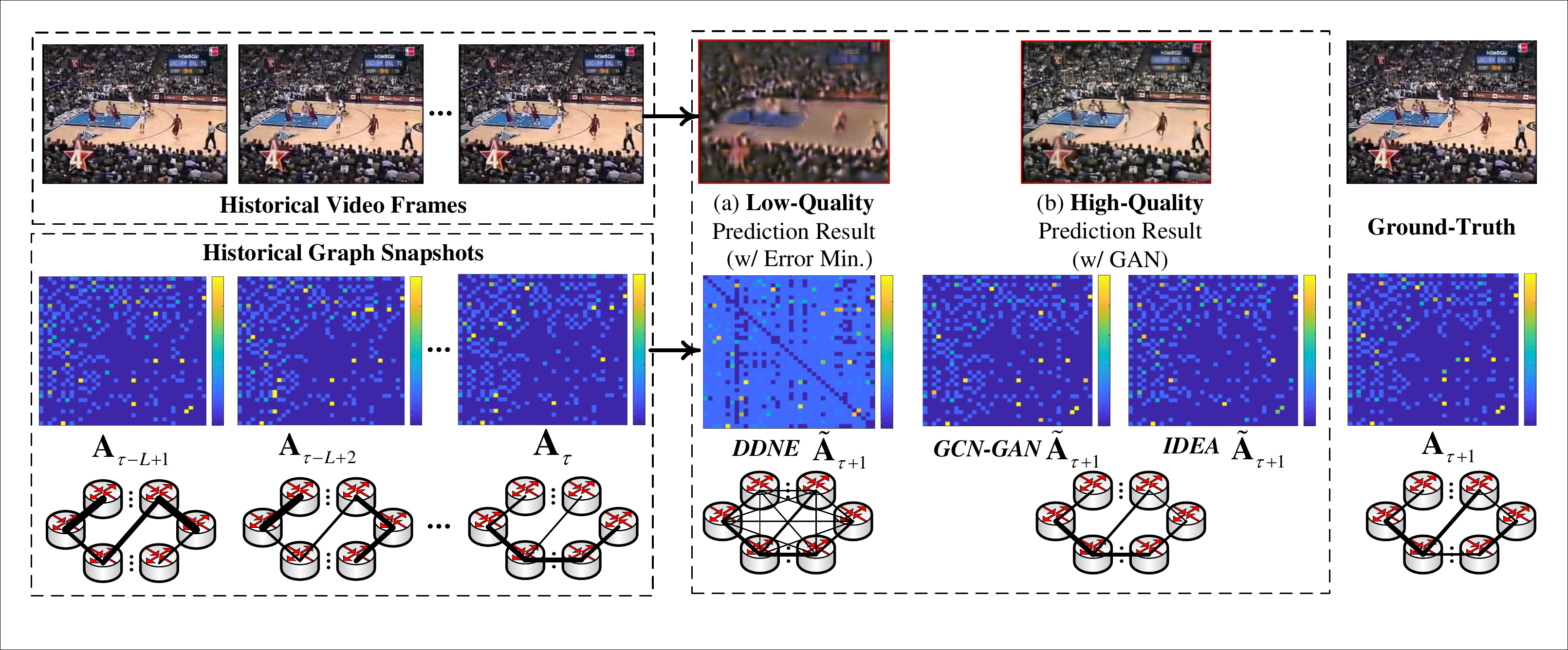}
\end{center}
\vspace{-0.25cm}
\caption{Examples of high-resolution video prediction and high-quality weighted TLP with adversarial learning.}\label{fig:HQ-TLP}
\vspace{-0.35cm}
\end{figure}

To derive high-quality prediction results for weighted snapshots, some advanced methods (e.g., \textit{GCN-GAN} and \textit{IDEA}) combine error minimization objectives with adversarial learning. Recent progress in high-resolution video prediction \cite{mathieu2015deep} has demonstrated that GAN can help minimize the scale difference between pixel values in images via its adversarial process and thus can derive high-resolution predicted video frames. Intuitively, it is also expected that \textit{adversarial learning can help distinguish between the scale difference of weights in adjacency matrices}.

Fig.~\ref{fig:HQ-TLP} illustrates the weighted TLP on the \textit{UCSB-MeshNet} dataset from a campus wireless mesh network (see supplementary materials for its details), where we visualize adjacency matrices of (\romannumeral1) historical snapshots, (\romannumeral2) prediction results of \textit{DDNE}, \textit{GCN-GAN}, and \textit{IDEA}, as well as (\romannumeral3) ground-truth of one example prediction operation.
To highlight the difference between small and zero weights, we set all $0$s to $-500$ in each visualized adjacency matrix, where dark blue, light blue, and yellow denote zero, small, and large weights. Similar to \textit{DDNE}, error minimization is also a classic objective of video prediction, which can only derive low-quality prediction results (e.g., the blurred image and dense adjacency matrix of \textit{DDNE} in Fig.~\ref{fig:HQ-TLP}). In contrast, the integration of GAN can effectively help derive high-quality results close to ground-truth (e.g., the clear image and sparse adjacency matrices of \textit{GCN-GAN} and \textit{IDEA} in Fig.~\ref{fig:HQ-TLP}).

However, existing methods that can support weighted TLP still rely on ESSD using adjacency matrices to describe weighted topology, which can only achieve coarse-grained representations of dynamic graphs compared with UESD.

\subsubsection{\textbf{Dealing with the Variation of Node Sets}}~

Some TLP methods (e.g., \textit{ctRBM}, \textit{GrNMF}, \textit{DDNE}, and \textit{CTDNE}) assume that the node set $V$ of a dynamic graph is known and fixed for all time steps. They rely on some techniques (e.g., matrix factorization and embedding lookup scheme introduced in Sections~\ref{Sec:SNAP-OTI} and \ref{Sec:Edge-OTI}), in which the dimensionality of model parameters is related to the number of nodes. As the dimensionality of model parameters should be fixed for all the time steps, these approaches can only support the TLP in level-1, failing to tackle the variation of node sets.

However, most real-world systems allow the addition and deletion of entities. To some extent, methods with ESSD can tackle the variation of node sets using `large' adjacency matrices $\{ {\bf{A}}_t \}$ induced by the union of all the associated nodes, where there may be isolated nodes without edges in some snapshots. Such a naive strategy may have unnecessary high space complexity. Moreover, it can only derive the prediction results w.r.t. the previously observed nodes but cannot be generalized to new unseen nodes in future time steps.

Inductive inference of dynamic graphs is an advanced topic in recent research. Typical inductive TLP methods (e.g., \textit{DyRep}, \textit{TGAT}, and \textit{IDEA}) use the available node attributes and the inductive nature of DL structures (e.g., GNNs and attention), in which the dimensionality of model parameters is not related to the number of nodes, to ensure a trained model can be directly generalized to new unseen nodes. Therefore, these inductive approaches can support the TLP in level-2, dealing with the variation of node sets. Nevertheless, most of them still rely on the availability of node attributes but lack discussions regarding how to handle the case without attributes. Only a few methods (e.g., \textit{CAW}) can extract features that are shared by all the possible nodes from the raw dynamic topology to support inductive inference.

\subsection{Future Research Directions}

\textbf{\textit{Simultaneous Prediction of Node and Edge Sets}}. As defined in Section~\ref{Sec:TLP-Setting}, node sets w.r.t. future links are usually assumed to be given in existing research. For instance, $V_{\tau}^{\tau+\Delta}$ and $V_{\Gamma(\tau,\tau+\Delta)}$ are inputs of the TLP in level-2 as described in (\ref{Eq:TLP-L2-1}) and (\ref{Eq:TLP-L2-2}). TLP aims to predict possible future edges induced by the given node sets. In addition to edge sets, the dynamics of a graph may also include the variation of node sets. In some real applications, the set of system entities in future time steps may also be unknown. For the TLP in level-2, simultaneously predicting the sets of future nodes and edges is more challenging and seldom studied in recent research. Moreover, the quality evaluation criteria of existing research are still based on the given future node sets (e.g., $V_{\tau}^{\tau+\Delta}$ or $V_{\Gamma(\tau,\tau+\Delta)}$). New quality metrics are also required to evaluate the prediction of both node and edge sets.

\textbf{\textit{Combining with Other Dynamic Inference Tasks}}. In addition to TLP, there are some other inference tasks on dynamic graphs (e.g., dynamic community detection \cite{enugala2015community,tamimi2015literature,rossetti2018community}, anomaly detection \cite{ranshous2015anomaly,ma2021comprehensive}, and traffic prediction \cite{tedjopurnomo2020survey,jiang2022graph}). In particular, the traffic prediction task is usually formulated as the dynamic node attribute prediction, while some TLP methods (e.g., \textit{EvolveGCN}, \textit{DySAT}, \textit{IDEA}, and \textit{TGAT}) assume that node attributes are available and only focus on the prediction of dynamic topology. Simultaneously optimizing a unified model to tackle multiple inference tasks (e.g., combining TLP with dynamic attribute prediction) is a promising future direction. It is usually expected that the optimization of multiple associated tasks can further improve the performance of each other \cite{zhang2021survey}.

For some settings considering both graph topology and attributes (e.g., prediction of future topology and attributes), recent studies \cite{qin2018adaptive,qin2021dual} have revealed complicated correlations between the two sources. On the one hand, attributes may carry complementary information beyond topology to support the better performance of a task. On the other hand, it may also result in unexpected performance decline due to the inconsistent noises hidden in attributes. Hence, it is promising to develop an advanced model that can adaptively adjust the contributions of different heterogeneous information sources (e.g., dynamic topology and attributes) according to the possible `mismatch' between them.

\textbf{\textit{Adaptive Selection of Sampling Rate and Execution Frequency}}. As introduced in Section~\ref{Sec:Data-Model}, we need to select a fixed time interval (or corresponding sampling rate) between successive snapshots when using ESSD to abstract a real-world system as a dynamic graph. Accordingly, the execution frequency of TLP can then be determined, where we execute one prediction operation once it comes to a new time step. Usually, the time interval (or sampling rate) is set according to the minimum duration of interactions in the system, which may result in high space complexities with many redundant descriptions of dynamic topology. In contrast, when using UESD, we sample a corresponding edge once there is a new interaction in the system without specifying the sampling rate, but still need to set a proper execution frequency of TLP. However, most related studies did not consider the selection of sampling rate and execution frequency, assuming that they are already determined by concrete applications or datasets. Only a few research considered the setting of sampling rate for ESSD (e.g., based on the autocorrelation between snapshots \cite{shao2022link}).

It is promising to explore an adaptive strategy for the selection of sampling rate and execution frequency according to the evolution and overhead of a system. For instance, we can set a high sampling rate and execution frequency in peak hours of a system to accurately capture its evolving patterns. When the system is not busy, we can set a relatively low sampling rate and execution frequency to minimize the system overhead.

\textbf{\textit{New Learning Paradigms}}. As introduced in Section~\ref{Sec:Learn-Parad}, OTI and OTOG are widely-used learning paradigms of the existing TLP techniques. On the one hand, OTI methods can effectively capture the latest variation of topology but may fail to satisfy the real-time constraints of applications due to their high complexities of online training.
On the other hand, OTOG approaches have the potential to satisfy the real-time constraints in their online generalization without additional model optimization, but may also fail to capture the latest characteristics as the topology evolves.
It is promising to develop new learning paradigms that integrate the advantages of both OTI and OTOG methods, where recent progress in online learning \cite{albers2003online,hoi2021online} and continual learning \cite{parisi2019continual,de2021continual} can be applied.

\textbf{\textit{Theoretical Analysis of Prediction Quality}}. As discussed in Section~\ref{Sec:Adv-Topic}, TLP can be used to pre-allocate key resources for better system performance. In particular, some methods (e.g., \textit{GCN-GAN} and \textit{IDEA}) focus on how to derive high-quality weighted links to support fine-grained resource allocation. However, some real-world systems may also have the reliability requirements \cite{rausand2003system,ahmad2017reliability} while a TLP model still has the risk of making wrong predictions. Therefore, theoretical analysis on the bound of prediction accuracy and error can help determine whether a TLP method can satisfy reliability requirements of a system.

\section{Conclusion}\label{Sec:Conc}
In this survey, we comprehensively reviewed existing representative TLP methods. We first gave the formal definitions regarding (\romannumeral1) data models of dynamic graphs (i.e., ESSD and UESD), (\romannumeral2) task settings of TLP (i.e., level-1 and -2), and (\romannumeral3) learning paradigms of related research (i.e., DI, OTI, and OTOG). Based on these definitions, we further introduced a fine-grained hierarchical taxonomy that categorizes existing methods in terms of (\romannumeral1) data models, (\romannumeral2) learning paradigms, and (\romannumeral3) techniques, covering multiple aspects. From a generic perspective, a unified encode-decoder framework was proposed to formulate all the methods reviewed in this survey. Each method can be described by an encoder, a decoder, and a loss function, where different methods only differ in terms of these components.
Based on this unified framework, we also refactored or implemented some TLP approaches and served the community with an open-source project \textit{OpenTLP}.
Finally, we summarized some advanced topics in recent research and future research directions. Due to space limit, we elaborate on (\romannumeral1) additional preliminaries (e.g., quality evaluation and classic techniques), (\romannumeral2) advanced applications, and (\romannumeral3) public datasets of TLP in supplementary materials.



\bibliographystyle{unsrt}

\appendix

\section{Summary of Notations \& Abbreviations}

\begin{table}[t]\small
\centering
\caption{Summary of major notations.}\label{Tab:Notations}
\vspace{-0.25cm}
\begin{tabular}{p{2.8cm}p{3.8cm}|p{2.8cm}p{4.2cm}}
\hline
\textbf{Notations} & \textbf{Definitions} & \textbf{Notations} & \textbf{Definitions} \\ \hline
$G = (G_1, \cdots, G_T)$ & ESSD-based dynamic graph & $G_t = (V_t, E_t)$ & snapshot at time step $t$ \\ \hline
$V_t = \{ v_1^t, \cdots, v_{N_t}^t\}$ & node set of $G_t$ & ${E_t} = \{ ((v_i^t,v_j^t),w)\}$ or $\{ (v_i^t,v_j^t)\}$ & edge set of $G_t$ for weighted or unweighted graphs\\ \hline
$\mathcal{A}_t = \{ \varphi (v_i^t) \}$ & attribute map of $G_t$ & $\varphi(v_i^t)$ & attribute map of node $v_i^t$\\ \hline
${\bf{A}}_t \in \Re^{N_t \times N_t}$ & adjacency matrix of $G_t$ & ${\bf{X}}_t ({\rm{or~}} {\bf{X}}) \in \Re^{N_t \times M}$ & attribute matrix of $G_t$\\\hline
$N_t$ (or $N$) & number of nodes & $M$ & dimensionality of attributes \\ \hline
$\Gamma = \{ t_1, t_2, \cdots \}$ & set of time steps for UESD & $G_\Gamma = (V_\Gamma, E_\Gamma, \Gamma)$ & UESD-based dynamic graph w.r.t. $\Gamma$ \\ \hline
$V_\Gamma = \{ v_1, \cdots, v_N \}$ & node set of $G_\Gamma$ w.r.t. $\Gamma$ & ${E_\Gamma } = \{ (({v_i},{v_j}),w,{t_e})\}$ or $\{ (({v_i},{v_j}),{t_e})\}$ & edge set of $G_\Gamma$ for weighted or unweighted graphs w.r.t. $\Gamma$\\ \hline
$\mathcal{A}_\Gamma = \{ \varphi ({v_i},t)\}$ & attribute map of $G_\Gamma$ w.r.t. $\Gamma$ & $\varphi ({v_i},t)$ & attribute map of node $v_i$ at time $t$\\ \hline
$\tau$ & index of current time step & $\mathcal{U}_d^s = ({\mathcal{U}_{s + 1}}, {\mathcal{U}_{s + 2}}, \cdots$ $,{\mathcal{U}_d})$ & sequence of an ESSD-based variable $\mathcal{U}$ w.r.t. indices $\{ s+1, s+2, \cdots, d\}$\\ \hline
$\Gamma(s, d) = \{ t | s < t \le d \}$ & set of sampling time steps between $(s, d]$ for UESD & $\mathcal{U}_{\Gamma(s, d)}$ & sequence of a UESD-based variable $\mathcal{U}$ w.r.t. $\Gamma(s, d)$\\ \hline
$L$ & number of historical time steps or historical time interval (i.e., window size) & $\Delta$ & number of future time steps or future time interval for prediction\\ \hline
${\tilde G}_{\tau}^{\tau+\Delta}$ & prediction result for ESSD & ${\tilde G}_{\Gamma(\tau, \tau+\Delta)}$ & prediction result for UESD\\ \hline
${\rm{Enc}}(\cdot)$ & \textit{encoder} in our unified framework & ${\rm{Dec}}(\cdot)$ & \textit{decoder} in our unified framework \\ \hline
${\mathcal{L}}(\cdot)$ & \textit{loss function} in our unified framework & $R$ & \textit{intermediate representation} given by ${\rm{Enc}}(\cdot)$ in our unified framework\\ \hline
$\delta$ & set of mode parameters to be optimized (learned) & ${\bf{z}}_i \in \Re^d$ & embedding of node $v_i$ \\ \hline
${\bf{Z}}_t ({\rm{or~}} {\bf{Z}}) \in \Re^{N \times d}$ & matrix form of embedding with the $i$-th row as embedding of $v_i$ & ${\bf{e}}_{ij} \in \Re^d ({\rm{or~}} \Re^{2d})$ & auxiliary edge embedding of node pair $(v_i, v_j)$\\ \hline
$d$ & embedding dimensionality & $\{ \alpha, \beta, \theta \}$ & tunable parameters in loss function of a method\\ \hline
$\{ {\bf{U}}, {\bf{V}}, {\bf{Y}}, {\bf{U}}_t, {\bf{V}}_t, {\bf{Y}}_t\}$ & latent matrices to be optimized in matrix factorization objectives & ${\bf{\tilde A}}_{\tau+r} \in \Re^{N_{\tau+r} \times N_{\tau+r}}$ & prediction result (in terms of an adjacency matrix) of a future snapshot $G_{\tau+r}$ for ESSD\\ \hline
$\mathcal{G}$ \& $\mathcal{D}$ & generator \& discriminator of GAN & $\omega = (v^{(0)}, v^{(1)}, \cdots,~$ $v^{(K)})$ & a TRW on UESD-based topology with $v^{(r)}$ as the $r$-th node\\ \hline
$\lambda (t)$ (or $\lambda_{ij} (t)$) & conditional intensity (w.r.t. node pair $(v_i, v_j)$) at time $t$ in Hawkes process & $\mu (t)$ (or $\mu_{ij} (t)$) & base intensity (w.r.t. $(v_i, v_j)$) at time $t$ in Hawkes process\\ \hline
$\kappa (t - s)$ & decaying kernel w.r.t. time interval $(t - s)$ in Hawkes process & $n(t)$ & number of events until $t$ in Hawkes process\\ \hline
$H_i$ & sequence of historical neighbors of $v_i$ for UESD-based topology & $H_i (t)$ & sequence of historical neighbors of $v_i$ before $t$ for UESD-based topology\\ \hline
$\Omega$ & set of time steps w.r.t. the training set for UESD-based methods & ${\bf{z}}_i (t)$ & embedding of node $v_i$ at time step $t$ for UESD-based OTOG methods\\ \hline
${\bf{e}}_{ij} (t)$ & auxiliary edge embedding of node pair $(v_i, v_j)$ at time step $t$ for UESD-based OTOG methods & ${\Phi }(\Delta t)$ & temporal encoding w.r.t. continuous time difference $\Delta t$\\ \hline
$\mathcal{\tilde E}_{ij}^{\tau+r}$ & probability that an edge $(v_i, v_j)$ appears at a future time step $(\tau + r)$ for UESD-based methods & ${\bar t}$ (or ${\bar t}_i$) & time step that an edge (or node $v_i$) is observed just before $t$ for UESD-based topology \\ \hline
\end{tabular}
\vspace{-0.35cm}
\end{table}

\begin{table}[t]\small
\centering
\caption{Summary of major abbreviations.}\label{Tab:Abbr}
\vspace{-0.25cm}
\begin{tabular}{ll|ll}
\hline
\textbf{Abbr.} & \textbf{Full Names} & \textbf{Abbr.} & \textbf{Full Names} \\ \hline
TLP & Temporal Link Prediction & DNE & Dynamic Network Embedding \\
ESSD & Evenly-Sampled Snapshot Sequence Description & UESD & Unevenly-Sampled Edge Sequence Description \\ 
DI & Direct Inference & OTI & Online Training \& Inference\\
OTOG & Offline Training \& Online Generalization & ROC & Receiver Operating Characteristic\\
AUC & Area under the ROC Curve & RMSE & Root-Mean-Square Error\\
MAE & Mean Absolute Error & MLSD & Mean Logarithmic Scale Difference\\
MR & Mismatch Rate & NMF & Non-negative Matrix Factorization\\
DL & Deep Learning & RBM & Restricted Boltzmann Machine\\
RNN & Recurrent Neural Network & LSTM & Long Short-Term Memory\\
GRU & Gated Recurrent Unit & MLP & Multi-Layer Perceptron\\
GAN & Generative Adversarial Networks & TRW & Temporal Random Walk\\
TPP & Temporal Point Process & NODE & Neural Ordinary Differential Equation\\
VAE & Variational Autoencoder & ELBO & Evidence Lower Bound\\
\hline
\end{tabular}
\vspace{-0.35cm}
\end{table}

Some major mathematical notations and abbreviations used in this survey are summarized in Tables~\ref{Tab:Notations} and \ref{Tab:Abbr}, respectively.

\section{Detailed Preliminaries}

In this section, we elaborate on some detailed preliminaries of this survey, including commonly-used quality evaluation criteria and basic techniques (e.g., NMF, MLP, RNN, attention, and GNN) of TLP.

\subsection{Quality Evaluation}

\subsubsection{\textbf{TLP on Unweighted Dynamic Graphs}}~

Existing TLP techniques usually consider the prediction on unweighted graphs, which can be treated as binary edge classification. Hence, metrics of binary classification can be used to measure the quality of prediction results.

For a given future time step $(\tau+r)$ with $1 \le r \le \Delta$, let ${\tilde E}_{\tau+r}$ be the set of predicted links. Accordingly, $E_{\tau+r}$ is the prediction ground-truth. To derive ${\tilde E}_{\tau+r}$, a TLP method first estimates the probability $p_{ij}^{\tau+r}$ that an edge $(v_i, v_j)$ appears at time step $(\tau+r)$. One can determine whether there is an edge between the node pair based on a threshold $\varepsilon$. Namely, if $p_{ij}^{\tau+r} \ge \varepsilon$ we let $(v_i, v_j) \in {\tilde E}_{\tau+r}$ and $(v_i, v_j) \notin {\tilde E}_{\tau+r}$ otherwise.
There exist four possible cases for each node pair $(v_i, v_j)$:
\begin{itemize}
    \item \textit{True Positive} (TP): $(v_i, v_j) \in E_{\tau+r}$ and $(v_i, v_j) \in {\tilde E}_{\tau+r}$;
    \item \textit{True Negative} (TN): $(v_i, v_j) \notin E_{\tau+r}$ and $(v_i, v_j) \notin {\tilde E}_{\tau+r}$;
    \item \textit{False Positive} (FP): $(v_i, v_j) \notin E_{\tau+r}$ but $(v_i, v_j) \in {\tilde E}_{\tau+r}$;
    \item \textit{False Negative} (FN): $(v_i, v_j) \in E_{\tau+k}$ but $(v_i, v_j) \notin {\tilde E}_{\tau+r}$.
\end{itemize}
\textit{Accuracy} and \textit{F1-score} are typical metrics defined based on statistics regarding the aforementioned cases:
\begin{equation}
    {\mathop{\rm Acc}\nolimits} ({E_{\tau  + r}},{{\tilde E}_{\tau  + r}}) \equiv \frac{{\# TP + \# TN}}{{\# TP + \# FP + \# TN + \# FN}},~
    {\mathop{\rm F1}\nolimits} ({E_{\tau  + r}},{{\tilde E}_{\tau  + r}}) \equiv \frac{{2Pre \cdot Rec}}{{Pre + Rec}} = \frac{{\# TP}}{{\# TP + (\# FP + \# FN)/2}},
\end{equation}
where F1-score is the harmonic mean of \textit{precision} $Pre \equiv \# TP/(\# TP + \# FP)$ and \textit{recall} $Rec \equiv \# TP/(\# TP + \# FN)$.

Given the ${\tilde E}_{\tau+r}$ w.r.t. a value $\varepsilon$, we can also compute the \textit{true positive rate} (i.e., recall) $TPR \equiv \# TP/(\# TP + \# FN)$ and \textit{false positive rate} $FPR \equiv \# FP/(\# FP + \# TN)$. By respectively letting $\varepsilon$ be the probability value $p_{ij}^{\tau+r}$ w.r.t. all the possible node pairs (i.e., $\varepsilon \in \{ p_{ij}^{\tau+r} | v_i, v_j \in V_{\tau+r}\}$), one can draw a \textit{receiver operating characteristics} (ROC) curve \cite{bradley1997use}, where the result w.r.t. each value of $\varepsilon$ is plotted to a 2D space with $FPR$ and $TPR$ as the $x$- and $y$-axis. The metric of \textit{area under the ROC curve} (AUC) \cite{bradley1997use} is defined as the area covered by the ROC curve, whose value is within the range $[0.5, 1]$.

Usually, higher accuracy, F1-score, and AUC indicate better prediction quality of ${\tilde E}_{\tau+r}$. For the case with a large number of nodes, some studies \cite{nguyen2018continuous,xu2020inductive} adopt a sampling strategy to compute the aforementioned metrics, where only a small ratio of the positive and negative node pairs (s.t. $\{ (v_i, v_j) \in E_{\tau+r}\}$ and $\{ (v_i, v_j) \notin E_{\tau+r} \}$) are used for the evaluation, instead of considering all the $N^2$ node pairs.

\subsubsection{\textbf{TLP on Weighted Dynamic Graphs}}~

As discussed in Section 4.1 of the main paper, TLP on weighted dynamic graphs is an advanced topic seldom considered in recent research. Typical metrics for unweighted graphs, which are based on binary edge classification, cannot be used to evaluate the quality of a weighted prediction result.

To the best of our knowledge, most existing approaches that can handle the TLP on weighted dynamic graphs use the data model of ESSD and describe the topology of each snapshot $G_t$ using an adjacency matrix ${\bf{A}}_t \in \Re^{N_t \times N_t}$. \textit{Root-mean-square error} (RMSE) and \textit{mean absolute error} (MAE) are widely-used metrics for weighted TLP, which measure the error between prediction result ${\bf{\tilde A}}_{\tau+r}$ and corresponding ground-truth ${\bf{A}}_{\tau+r}$ based on the F-norm and $l_1$-norm, respectively. Given ${\bf{A}}_{\tau+r}$ and ${\bf{\tilde A}}_{\tau+r}$ w.r.t. the snapshot $G_{\tau+r}$ to be predicted, RMSE and MAE are defined as
\begin{equation}\label{Eq:RMSE}
    {\mathop{\rm RMSE}\nolimits} ({{\bf{A}}_{\tau  + r}},{{\bf{\tilde A}}_{\tau  + r}}) \equiv \sqrt {\left\| {{{\bf{A}}_{\tau  + r}} - {{\bf{\tilde A}}_{\tau  + r}}} \right\|_F^2/N_{\tau  + r}^2},
\end{equation}
\begin{equation}\label{Eq:MAE}
    {\mathop{\rm MAE}\nolimits} ({{\bf{A}}_{\tau  + r}},{{{\bf{\tilde A}}}_{\tau  + r}}) \equiv \frac{1}{{N_{\tau  + r}^2}}\sum\limits_{i,j = 1}^N {|{{({{\bf{A}}_{\tau  + r}})}_{ij}} - {{({{{\bf{\tilde A}}}_{\tau  + r}})}_{ij}}|}.
\end{equation}
Qin et al. \cite{qin2023high} argued that \textit{conventional metrics of RMSE and MAE cannot measure the ability of a method to derive high-quality prediction results} (i.e., the ability to handle the \textit{wide-value-range} and \textit{sparsity} issues as described in Section 4.1 of the main paper) and introduced new metrics of \textit{mean logarithmic scale difference} (MLSD) and \textit{mismatch rate} (MR).

For instance, the scale difference between $(1, 2)$ is much larger than that $(1990, 2000)$ but the latter case has larger reconstruction errors with $(1-2)^2 < (1990-2000)^2$ and $|1-2| < |1990-2000|$. Hence, RMSE and MAE cannot measure the scale difference between ${\bf{A}}_{\tau+r}$ and ${\bf{\tilde A}}_{\tau+r}$ for the \textit{wide-value-range} issue.
In contrast, MLSD uses the logarithmic function $\log_{10}(\cdot)$ to measure such scale difference, where we have $|\log_{10}(1/2)| > |\log_{10}(1990/2000)|$. To compute MLSD, two auxiliary matrices ${\bf{\hat U}}_{\tau + r} \in \Re^{N_{\tau+r} \times N_{\tau+r}}$ and ${\bf{\hat V}}_{\tau+r} \in \Re^{N_{\tau+r} \times N_{\tau+r}}$ are used to avoid the zero-exception of ${{{\log }_{10}}} (\cdot)$ (i.e., ${\log _{10}}(0) = {\rm{nan}}$), where ${({{\bf{\hat U}}_{\tau  + r}})_{ij}} \equiv \max \{ {({{\bf{A}}_{\tau  + r}})_{ij}},\epsilon \} $ and ${({{\bf{\hat V}}_{\tau  + r}})_{ij}} \equiv \max \{ {({{{\bf{\tilde A}}}_{\tau  + r}})_{ij}},\epsilon \} $, with $\epsilon$ as a small threshold (e.g., $\epsilon = 10^{-5}$) to clip zero elements. MLSD is then defined as
\begin{equation}
    {\mathop{\rm MLSD}\nolimits} ({{\bf{A}}_{\tau  + r}},{{{\bf{\tilde A}}}_{\tau  + r}}) \equiv \frac{1}{{N_{\tau  + r}^2}}\sum\limits_{i,j = 1}^N {|{{\log }_{10}}\frac{{{{({{\bf{\hat U}}_{\tau  + r}})}_{ij}}}}{{{{({{\bf{\hat V}}_{\tau  + r}})}_{ij}}}}|}.
\end{equation}
Focusing on the \textit{sparsity} issue, MR first defines that a given node pair $(v_i^{\tau+r}, v_j^{\tau+r})$ is \textit{mismatched} if (\romannumeral1) $({\bf{A}}_{\tau+r})_{ij} = 0$ but $({\bf{\tilde A}}_{\tau+r})_{ij} > 0$ or (\romannumeral2) $({\bf{A}}_{\tau+r})_{ij} > 0$ but $({\bf{\tilde A}}_{\tau + r})_{ij} = 0$, which corresponds to the two exceptions of TLP that conventional RMSE and MAE metrics cannot measure. Accordingly, MR is defined as
\begin{equation}
    {\mathop{\rm MR}\nolimits} ({{\bf{A}}_{\tau  + r}},{{{\bf{\tilde A}}}_{\tau  + r}}) \equiv {{\mathop{\rm C}\nolimits} _{mis}}({{\bf{A}}_{\tau  + r}},{{{\bf{\tilde A}}}_{\tau  + r}})/N_{\tau  + r}^2,
\end{equation}
where ${{\mathop{\rm C}\nolimits} _{mis}}({{\bf{A}}_{\tau  + r}},{{{\bf{\tilde A}}}_{\tau  + r}})$ denotes the number of \textit{mismatched} node pair in the prediction result. In this setting, $1 - {\mathop{\rm MR}\nolimits} ({\bf{A}}_{\tau+r}, {\bf{\tilde A}}_{\tau+r})$ is the accuracy of successfully matching zero and non-zero elements between ${\bf{A}}_{\tau+r}$ and ${\bf{\tilde A}}_{\tau+r}$.

Usually, smaller RMSE, MAE, MLSD, and MR indicate better quality of a weighted prediction result ${\bf{\tilde A}}_{\tau+r}$.

\subsection{Non-negative Matrix Factorization (NMF)}
For a non-negative data matrix ${\bf{M}} \in \Re^{N \times N}$ (e.g., the adjacency matrix of snapshot $G_t$ with ${\bf{M}} = {\bf{A}}_t$), the standard NMF problem \cite{lee1999learning,huang2012non} can be formulated as the following optimization objective:
\begin{equation}
    \mathop {\arg \min }\limits_{{\bf{U}} \ge 0,{\bf{V}} \ge 0} {O_{{\rm{NMF}}}}\left\| {{\bf{M}} - {\bf{U}}{{\bf{V}}^T}} \right\|_F^2,
\end{equation}
where ${\bf{U}} \in \Re^{N \times d}$ (a.k.a. the \textit{basis matrix}) and ${\bf{V}} \in \Re^{N \times d}$ (a.k.a. the \textit{feature matrix}) are model parameters to be learned with non-negative constraint (i.e., elements in ${\bf{U}}$ and ${\bf{V}}$ must be non-negative). In the rest of this subsection, we elaborate on the model optimization strategy of NMF using the aforementioned objective as an example. All the NMF-based TLP methods reviewed in this survey (e.g., \textit{CRJMF}, \textit{TLSI}, \textit{MLjFE}, \textit{GrNMF}, and \textit{DeepEye}) can be optimized in a similar way.

In general, a matrix factorization problem can be solved via the block coordinate descent algorithm, where we properly initialize model parameters $\{ {\bf{U}}, {\bf{V}} \}$ (i.e., latent matrices to be learned) and in terms update their values using certain updating rules until converge. For NMF, $\{ {\bf{U}}, {\bf{V}} \}$ should be initialized with non-negative values.

Note that we have $\left\| {\bf{X}} \right\|_F^2 = {\mathop{\rm tr}\nolimits} ({\bf{X}}{{\bf{X}}^T})$. To obtain the updating rules, we first derive the partial derivative of $O_{\rm NMF}$ w.r.t. each latent matrix to be learned (i.e., ${\bf{U}}$ and ${\bf{V}}$):
\begin{equation}
    \frac{{\partial {O_{{\rm{NMF}}}}}}{{\partial {\bf{U}}}} = 2({\bf{U}}{{\bf{V}}^T}{\bf{V}} - {\bf{MV}}) {~\rm{and}~} \frac{{\partial {O_{{\rm{NMF}}}}}}{{\partial {\bf{V}}}} = 2({\bf{V}}{{\bf{U}}^T}{\bf{V}} - {{\bf{M}}^T}{\bf{U}}).
\end{equation}
According to the gradient descent algorithm, we have the following \textit{addictive updating rules} for ${\bf{U}}$ and ${\bf{V}}$:
\begin{equation}
   {{\bf{U}}_{ir}} \leftarrow {{\bf{U}}_{ir}} - {\gamma _{ir}}{({[ \cdot ]_ + } - {[ \cdot ]_ - })_{ir}} {~\rm{and}~} {{\bf{V}}_{ir}} \leftarrow {{\bf{V}}_{ir}} - {\gamma _{ir}}{({[ \cdot ]_ + } - {[ \cdot ]_ - })_{ir}},
\end{equation}
where $\gamma_{ir}$ is a pre-set learning rate; ${[ \cdot ]_ + }$ and ${[ \cdot ]_ - }$ are simplified notations to represent terms in the partial derivative with positive and negative coefficients (e.g., ${[ \cdot ]_ + } = 2{\bf{U}}{{\bf{V}}^T}{\bf{V}}$ and ${[ \cdot ]_ - } = 2{\bf{MV}}$ for ${\bf{U}}$). By setting ${\gamma _{ir}} = {{\bf{U}}_{ir}}/{({[ \cdot ]_ + })_{ir}}$ and ${\gamma _{ir}} = {{\bf{V}}_{ir}}/{({[ \cdot ]_ + })_{ir}}$, we can obtain the following \textit{multiplicative updating rules}:
\begin{equation}
    {{\bf{U}}_{ir}} \leftarrow {{\bf{U}}_{ir}}\frac{{{{({{[ \cdot ]}_ - })}_{ir}}}}{{{{({{[ \cdot ]}_ + })}_{ir}}}} = {{\bf{U}}_{ir}}\frac{{{{({\bf{MV}})}_{ir}}}}{{{{({\bf{U}}{{\bf{V}}^T}{\bf{V}})}_{ir}}}} {~\rm{and}~} {{\bf{V}}_{ir}} \leftarrow {{\bf{V}}_{ir}}\frac{{{{({{[ \cdot ]}_ - })}_{ir}}}}{{{{({{[ \cdot ]}_ + })}_{ir}}}} = {{\bf{V}}_{ir}}\frac{{{{({{\bf{M}}^T}{\bf{U}})}_{ir}}}}{{{{({\bf{V}}{{\bf{U}}^T}{\bf{U}})}_{ir}}}},
\end{equation}
which can be considered as the adaptive adjustment of the learning rate in gradient descent. If all the variables (e.g., ${\bf{U}}$ and ${\bf{V}}$) are initialized with non-negative values, the aforementioned \textit{multiplicative updating rules} will not change their signs, thus ensuring the non-negative constraints (i.e., ${\bf{U}} \ge 0$ and ${\bf{V}} \ge 0$).

\subsection{Multi-Layer Perceptron (MLP)}
MLP is the basic building block of many DL-based models. It usually follows a multi-layer structure. In this survey, we use ${\bf{\bar Z}} = {\mathop{\rm MLP}\nolimits} ({\bf{Z}})$ to denote an MLP with ${\bf{Z}}$ and ${\bf{\bar Z}}$ as its input and output. Let ${\bf{Z}}^{(k-1)}$ and ${\bf{Z}}^{(k)}$ be the input and output of the $k$-th layer. The general form of the $k$-th layer in an MLP can be represented as
\begin{equation}
    {{\bf{Z}}^{(k)}} = {\mathop{\rm MLP}\nolimits}_k ({{\bf{Z}}^{(k - 1)}}) \equiv {f_{{\rm{act}}}}({{\bf{Z}}^{(k - 1)}}{{\bf{W}}^{(k)}} + {{\bf{b}}^{(k)}}),
\end{equation}
where ${f_{{\rm{act}}}} (\cdot)$ is an activation function (e.g., sigmoid, tanh, ReLU, ELU, and LeakyReLU) to be specified; ${\bf{W}}^{(k)}$ and ${\bf{b}}^{(k)}$ are learnable weight matrix and bias vector of the $k$-th layer. Accordingly, ${\bf{\bar Z}}$ is the output of the last layer.

\subsection{Recurrent Neural Network (RNN)}

Some TLP methods (e.g., \textit{dyngraph2vec}, \textit{DDNE}, and \textit{GCN-GAN}) use RNNs to capture the evolving patterns of dynamic graphs. Typical RNN structures include the vanilla RNN, LSTM \cite{gers2000learning}, and GRU \cite{chung2014empirical}. In this survey, we describe the three RNN structures as $[{{\bf{H}}_1},{{\bf{H}}_2}, \cdots ,{{\bf{H}}_L}] = {\mathop{\rm VanRNN}\nolimits} ([{{\bf{Z}}_1},{{\bf{Z}}_2}, \cdots ,{{\bf{Z}}_L}])$, $[{{\bf{H}}_1},{{\bf{H}}_2}, \cdots ,{{\bf{H}}_L}] = {\mathop{\rm LSTM}\nolimits} ([{{\bf{Z}}_1},{{\bf{Z}}_2}, \cdots ,{{\bf{Z}}_L}])$, and $[{{\bf{H}}_1},{{\bf{H}}_2}, \cdots ,{{\bf{H}}_L}] = {\mathop{\rm GRU}\nolimits} ([{{\bf{Z}}_1},{{\bf{Z}}_2}, \cdots ,{{\bf{Z}}_L}])$. Namely, given an input sequence ${\bf{Z}}_1^{L} = [{\bf{Z}}_1, \cdots, {\bf{Z}}_L]$ with length $L$, the RNN structure successively derives a hidden state ${\bf{H}}_t$ ($1 \le t \le L$), forming an output sequence ${\bf{H}}_1^L = [{\bf{H}}_1, \cdots, {\bf{H}}_L]$ that can preserve the evolving patterns of input sequence ${\bf{Z}}_1^L$.

For each time step $t$, RNN outputs the corresponding hidden state ${\bf{H}}_{t}$ based on the joint inputs of current feature ${\bf{Z}}_t$ and hidden state ${\bf{H}}_{t-1}$ of the previous time step. For simplicity, we denote this procedure as ${{\bf{H}}_t} = {\mathop{\rm VanRNN}\nolimits} ({{\bf{Z}}_t},{{\bf{H}}_{t - 1}})$, ${{\bf{H}}_t} = {\mathop{\rm LSTM}\nolimits} ({{\bf{Z}}_t},{{\bf{H}}_{t - 1}})$, and ${{\bf{H}}_t} = {\mathop{\rm GRU}\nolimits} ({{\bf{Z}}_t},{{\bf{H}}_{t - 1}})$ for vanilla RNN, LSTM, and GRU, respectively.

Concretely, ${{\bf{H}}_t} = {\mathop{\rm VanRNN}\nolimits} ({{\bf{Z}}_t},{{\bf{H}}_{t - 1}})$ is usually defined as
\begin{equation}
    {{\bf{H}}_t} = {\mathop{\rm VanRNN}\nolimits} ({{\bf{Z}}_t},{{\bf{H}}_{t - 1}}) \equiv {f_{{\rm{act}}}}({{\bf{Z}}_t}{{\bf{W}}_Z} + {{\bf{H}}_{t - 1}}{{\bf{W}}_H}),
\end{equation}
where ${f_{{\rm{act}}}} (\cdot)$ represents an activation function to be specified; $\{ {\bf{W}}_Z, {\bf{W}}_H \}$ are trainable model parameters.

Compared with vanilla RNN, LSTM is a more sophisticated structure (with the designs of input gate, forget gate, output gate, and memory cell) that can effectively capture the long-term dependencies of sequential data. Details of ${{\bf{H}}_t} = {\mathop{\rm LSTM}\nolimits} ({{\bf{Z}}_t},{{\bf{H}}_{t - 1}})$ are described as follows:
\begin{equation}
    {{\bf{I}}_t} = \sigma ({{\bf{Z}}_t}{\bf{W}}_Z^I + {{\bf{H}}_{t - 1}}{\bf{W}}_H^I + {{\bf{b}}^I}),
\end{equation}
\begin{equation}
    {{\bf{F}}_t} = \sigma ({{\bf{Z}}_t}{\bf{W}}_Z^F + {{\bf{H}}_{t - 1}}{\bf{W}}_H^F + {{\bf{b}}^F}),
\end{equation}
\begin{equation}
    {{\bf{O}}_t} = \sigma ({{\bf{Z}}_t}{\bf{W}}_Z^O + {{\bf{H}}_{t - 1}}{\bf{W}}_H^O + {{\bf{b}}^O}),
\end{equation}
\begin{equation}
     {{\bf{C}}_t} = {{\bf{F}}_t} \odot {{\bf{C}}_{t - 1}} + {{\bf{I}}_t} \odot {{{\bf{\tilde C}}}_t},
\end{equation}
\begin{equation}
    {{{\bf{\tilde C}}}_t} = \sigma ({{\bf{Z}}_t}{\bf{W}}_Z^C + {{\bf{H}}_{t - 1}}{\bf{W}}_H^C + {{\bf{b}}^C}),
\end{equation}
\begin{equation}
    {{\bf{H}}_t} = {{\bf{O}}_t} \odot \tanh ({{\bf{C}}_t}),
\end{equation}
where ${\bf{I}}_t$, ${\bf{F}}_t$, ${\bf{O}}_t$, and ${\bf{C}}_t$ are intermediate states given by the input gate, forget gate, output gate, and memory cell, respectively; $\{ {\bf{W}}_Z^I,  {\bf{W}}_H^I, {\bf{b}}^I, {\bf{W}}_Z^F,  {\bf{W}}_H^F, {\bf{b}}^F, {\bf{W}}_Z^O,  {\bf{W}}_H^O, {\bf{b}}^O, {\bf{W}}_Z^C,  {\bf{W}}_H^C, {\bf{b}}^C \}$ are model parameters to be optimized; $\sigma (\cdot)$ denotes the sigmoid function; $\odot$ is the element-wise multiplication.

Also aiming to capture the long-term dependencies of sequential data, GRU is a simplified structure (with the designs of update gate and reset gate) compared with LSTM. Details of ${{\bf{H}}_t} = {\mathop{\rm GRU}\nolimits} ({{\bf{Z}}_t},{{\bf{H}}_{t - 1}})$ are described as follows:
\begin{equation}
    {{\bf{U}}_t} = \sigma ({{\bf{Z}}_t}{\bf{W}}_Z^U + {{\bf{H}}_{t - 1}}{\bf{W}}_H^U),
\end{equation}
\begin{equation}
    {{\bf{R}}_t} = \sigma ({{\bf{Z}}_t}{\bf{W}}_Z^R + {{\bf{H}}_{t - 1}}{\bf{W}}_H^R),
\end{equation}
\begin{equation}
    {{{\bf{\hat H}}}_t} = \tanh ({{\bf{Z}}_t}{{{\bf{\hat W}}}_Z} + ({{\bf{R}}_t} \odot {{\bf{H}}_{t - 1}}){{{\bf{\hat W}}}_H}),
\end{equation}
\begin{equation}
    {{\bf{H}}_t} = (1 - {{\bf{U}}_t}) \odot {{\bf{H}}_{t - 1}} + {{\bf{U}}_t} \odot {{{\bf{\hat H}}}_t},
\end{equation}
where ${\bf{U}}_t$ and ${\bf{R}}_t$ are intermediate states given by the update gate and reset gate; $\{ {\bf{W}}_Z^U, {\bf{W}}_H^U, {\bf{W}}_Z^R, {\bf{W}}_H^R, {\bf{\hat W}}_Z, {\bf{\hat W}}_H\}$ are trainable model parameters.

\subsection{Attention Mechanism}

Attention \cite{vaswani2017attention,niu2021review} is another type of DL structures designed for sequential data. For TLP, it is adopted by some methods (e.g., \textit{STGSN} and \textit{DySAT}) to capture the evolving patterns of dynamic graphs.
The inputs of a typical attention unit include a query, a key, and a value described by matrices ${\bf{Q}} \in \Re^{m \times d}$, ${\bf{K}} \in \Re^{n \times d}$, and ${\bf{V}} \in \Re^{n \times d}$, respectively. A commonly used design of attention, which follows an advanced multi-head setting, can be described as
\begin{equation}\label{Eq:Attention}
\begin{array}{l}
{\bf{Z}} = {\mathop{\rm Att}\nolimits} ({\bf{Q}},{\bf{K}},{\bf{V}}) \equiv [{{\bf{Z}}^{(1)}}|| \cdots ||{{\bf{Z}}^{(h)}}],\\
{{\bf{Z}}^{(r)}} = {{\mathop{\rm Att}\nolimits} _r}({\bf{Q}},{\bf{K}},{\bf{V}}) \equiv {f_{{\rm{act}}}}{\rm{(}}{\mathop{\rm softmax}\nolimits} ({{{\bf{\hat Q}}}^{(r)}}{{{\bf{\hat K}}}^{(r)T}}/\sqrt {\hat d} ){{{\bf{\hat V}}}^{(r)}}),\\
{\bf{\hat Q}}^{(r)} = {\bf{Q}} {\bf{W}}_Q^{(r)}, {\bf{\hat K}}^{(r)} = {\bf{K}} {\bf{W}}_K^{(r)}, {\bf{\hat V}}^{(r)} = {\bf{V}} {\bf{W}}_V^{(r)},
\end{array}
\end{equation}
where $h$ is the number of heads; $\{ {\bf{\hat Q}}^{(r)} \in \Re^{m \times {\bar d}},  {\bf{\hat K}}^{(r)} \in \Re^{n \times {\bar d}}, {\bf{\hat V}}^{(r)} \in \Re^{n \times {\bar d}} \}$ are the linear mapping of $\{ {\bf{Q}}, {\bf{K}}, {\bf{V}} \}$ with $\{ {\bf{W}}_Q^{(r)} \in \Re^{d \times {\bar d}}, {\bf{W}}_K^{(r)} \in \Re^{d \times {\bar d}}, {\bf{W}}_V^{(r)} \in \Re^{d \times {\bar d}}\} $ as model parameters to be optimized and ${\bar d} = d/h$;
${f_{{\rm{act}}}} (\cdot)$ is an activation function to be specified. The $r$-th attention head outputs a matrix ${\bf{Z}}^{(r)} \in \Re^{m \times {\bar d}}$, where each row of ${\bf{Z}}^{(r)}$ is the linear combination of rows in ${\bf{\hat V}}^{(r)}$, with the combination weights determined by the row-wise softmax w.r.t. the inner product between $\{ {\bf{\hat Q}}^{(r)}, {\bf{\hat K}}^{(r)} \}$. The attention unit derives its final output ${\bf{Z}} \in \Re^{m \times d}$ by concatenating the outputs of all the heads (i.e., ${\bf{Z}} = [{{\bf{Z}}^{(1)}} || \cdots || {{\bf{Z}}^{(h)}}]$).
Some methods also adopt the following design to derive the $i$-th row of ${\bf{Z}}^{(r)}$ in the $r$-th attention head:
\begin{equation}
    {\bf{Z}}_{i,:}^{(r)} = {{\mathop{\rm Att}\nolimits} _r}({\bf{Q}},{\bf{K}},{\bf{V}}) \equiv {f_{\rm act}}(\sum\nolimits_j {{a_{ij}}{\bf{V}}_j^{(r)}} ){\rm{~with~}}{a_{ij}} \equiv \frac{{{g_{{\rm{act}}}}([{\bf{\hat Q}}_{i,:}^{(r)}||{\bf{\hat K}}_{j,:}^{(r)}]{{\bf{a}}})}}{{\sum\nolimits_s {{g_{{\rm{act}}}}([{\bf{\hat Q}}_{i,:}^{(r)}||{\bf{\hat K}}_{s,:}^{(r)}]{{\bf{a}}})} }},
\end{equation}
where $f_{\rm act} (\cdot)$ and $g_{\rm act} (\cdot)$ are activation functions to be specified; ${\bf{a}} \in \Re^{2 {\bar d}}$ is a learnable parameter; $a_{ij}$ is the attentive weight determined by the concatenation of corresponding rows of ${\bf{\hat Q}}^{(r)}$ and ${\bf{\hat K}}^{(r)}$ (i.e., $[{\bf{\hat Q}}_{i,:}^{(r)}||{\bf{\hat K}}_{j,:}^{(r)}]$).

\subsection{Graph Neural Network (GNN)}

Some TLP approaches (e.g., \textit{EvolveGCN}, \textit{GCN-GAN}, \textit{IDEA}, and \textit{DySAT}) use GNNs to explore the structural characteristics of graph topology at a specific time step. Most GNNs were originally designed for attributed graphs, which aggregate node attributes (or latent embedding) according to graph topology and derive another latent representation (i.e., embedding) for each node. GCN \cite{kipf2016semi} and GAT \cite{velivckovic2017graph} are widely-used GNN structures.

Let ${\bf{A}}$ and ${\bf{Z}}$ be the adjacency matrix and feature matrix that describe the topology and node attributes (or latent embedding) of a static graph, where the $i$-th row of ${\bf{Z}}$ (i.e., ${\bf{Z}}_{i,:}$) describes the features (or embedding) of node $v_i$. The operation of one GCN layer can be described as
\begin{equation}
    {\bf{\bar Z}} = {\mathop{\rm GCN}\nolimits} ({\bf{A}},{\bf{Z}}) \equiv {f_{\rm act}}({{{\bf{\hat D}}}^{ - 0.5}}{\bf{\hat A}}{{\bf{\hat D}}^{ - 0.5}}{\bf{Z}}{{\bf{W}}_{{\rm{GCN}}}}),
\end{equation}
where $f_{\rm{act}}(\cdot)$ is an activation function to be specified; ${\bf{\hat A}} = {\bf{A}} + {\bf{I}}$ denotes the adjacency matrix with self-connected edges; ${\bf{\hat D}}$ is the diagonal degree matrix w.r.t. ${\bf{\hat A}}$; ${\bf{W}}_{\rm GCN}$ is the model parameter to be learned. The $i$-th row of the output (i.e., ${\bf{\bar Z}}_{i,:}$) corresponds to the latent representation of node $v_i$, which is the nonlinear weighted mean aggregation of features from $v_i \cup {\mathop{\rm Nei}\nolimits}(v_i)$, with ${\mathop{\rm Nei}\nolimits}(v_i)$ as the set of neighbors of $v_i$.

Different from GCN, GAT applies the attention mechanism to adaptively adjust the feature aggregation of each node according to the feature inputs of neighbors. Existing methods usually adopt the advanced multi-head setting of GAT. Let $h$ and ${\bf{\bar Z}}^{(r)}$ be the number of heads and output (in terms of latent representations) of the $r$-th head. The final output ${\bf{\bar Z}}$ of a GAT layer can be the concatenation of the outputs of multiple heads (i.e., ${\bf{\bar Z}} = [{{{\bf{\bar Z}}}^{(1)}}|| \cdots ||{{{\bf{\bar Z}}}^{(h)}}]$). To derive the latent representation of node $v_i$ in the $r$-th head, the GAT layer can be described as
\begin{equation}
    {\bf{\bar Z}}_{i,:}^{(r)} = {\mathop{\rm GAT}\nolimits} ({\bf{A}},{\bf{Z}}) \equiv {f_{{\rm{act}}}}(\sum\limits_{{v_j} \in {\mathop{\rm Nei}\nolimits} ({v_i})} {a_{ij}^{(r)}{{\bf{Z}}_{j,:}}{{\bf{W}}_h}^{(r)}} ),
\end{equation}
\begin{equation}\label{Eq:GAT-Wei}
    a_{ij}^{(r)} = \frac{{\exp \{ {f_{{\rm{act}}}}([{{\bf{Z}}_{i,:}}{\bf{W}}_a^{(r)}||{{\bf{Z}}_{j,:}}{\bf{W}}_a^{(r)}]{{\bf{a}}^{(r)}})\} }}{{\sum\nolimits_{{v_k} \in {\mathop{\rm Nei}\nolimits} ({v_i})} {\{ {f_{{\rm{act}}}}([{{\bf{Z}}_{i,:}}{\bf{W}}_a^{(r)}||{{\bf{Z}}_{k,:}}{\bf{W}}_a^{(r)}]{{\bf{a}}^{(r)}})\} } }},
\end{equation}
where $a_{ij}^{(r)}$ is the attention weight w.r.t. an edge $(v_i, v_j)$ in the $r$-th head determined by the softmax w.r.t. ${\mathop{\rm Nei}\nolimits} ({v_i})$; $\{ {\bf{W}}_h^{(r)}, {\bf{W}}_a^{(r)}, {\bf{a}}^{(r)} \}$ are trainable model parameters of the $r$-th head. Note that (\ref{Eq:GAT-Wei}) is only designed for unweighted graphs. It can be further extended to explore the weighted topology based on the following form:
\begin{equation}
    a_{ij}^{(r)} = \frac{{\exp \{ {f_{{\rm{act}}}}( {\bf{A}}_{ij} [{{\bf{Z}}_{i,:}}{\bf{W}}_a^{(r)}||{{\bf{Z}}_{j,:}}{\bf{W}}_a^{(r)}]{{\bf{a}}^{(r)}})\} }}{{\sum\nolimits_{{v_k} \in {\mathop{\rm Nei}\nolimits} ({v_i})} {\{ {f_{{\rm{act}}}}( {\bf{A}}_{ik} [{{\bf{Z}}_{i,:}}{\bf{W}}_a^{(r)}||{{\bf{Z}}_{k,:}}{\bf{W}}_a^{(r)}]{{\bf{a}}^{(r)}})\} } }},
\end{equation}
where the adjacency matrix ${\bf{A}}$, which describes the weighted topology, is integrated into the computation of $a_{ij}^{(r)}$.

\section{Advanced Applications Supported by TLP}

In this section, we introduce some advanced applications that can be supported by TLP in various scenarios.

\subsection{Friend Recommendation \& Next Item Recommendation in Online Social Networks \& Media}

\begin{figure}[t]
\begin{center}
 \begin{minipage}{0.57\linewidth}
 \subfigure[Friend recommendation]{
  \frame{
\includegraphics[width=\textwidth,trim=18 18 18 18,clip]{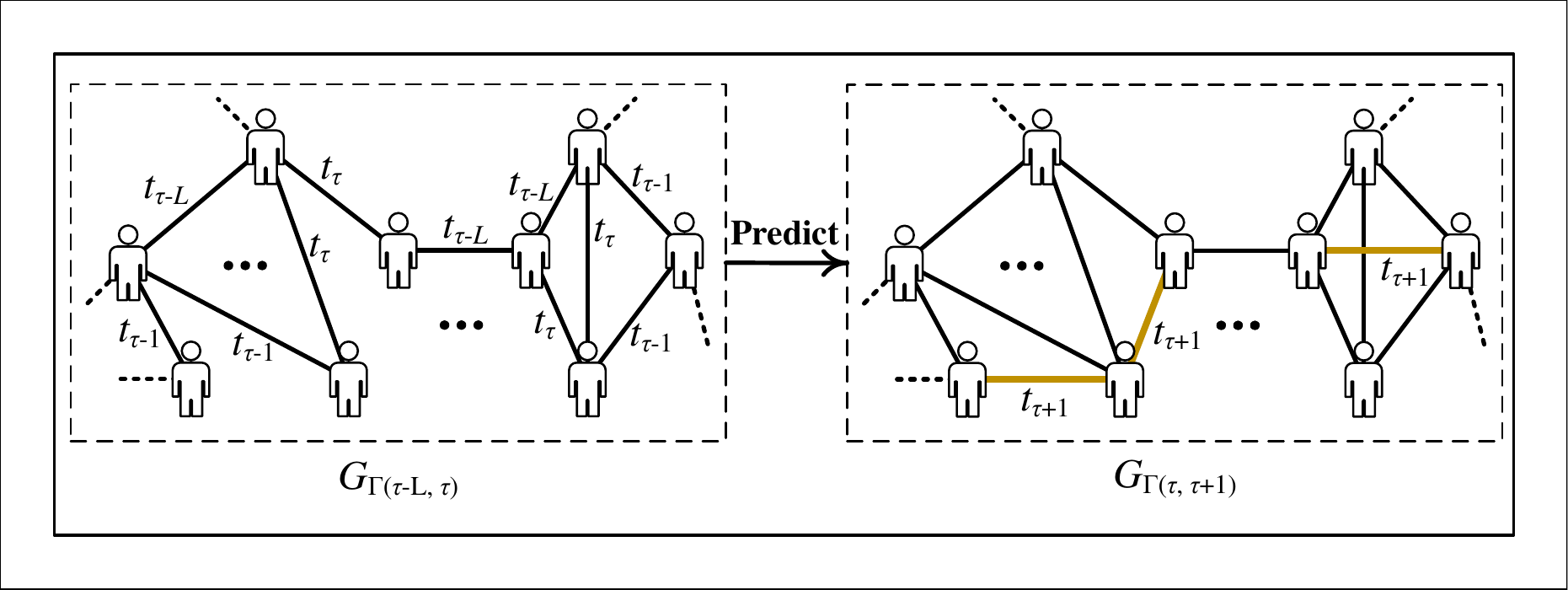}}
  }
 \end{minipage}
 \begin{minipage}{0.32\linewidth}
 \subfigure[Next item recommendation]{
  \frame{
\includegraphics[width=\textwidth,trim=18 18 18 18,clip]{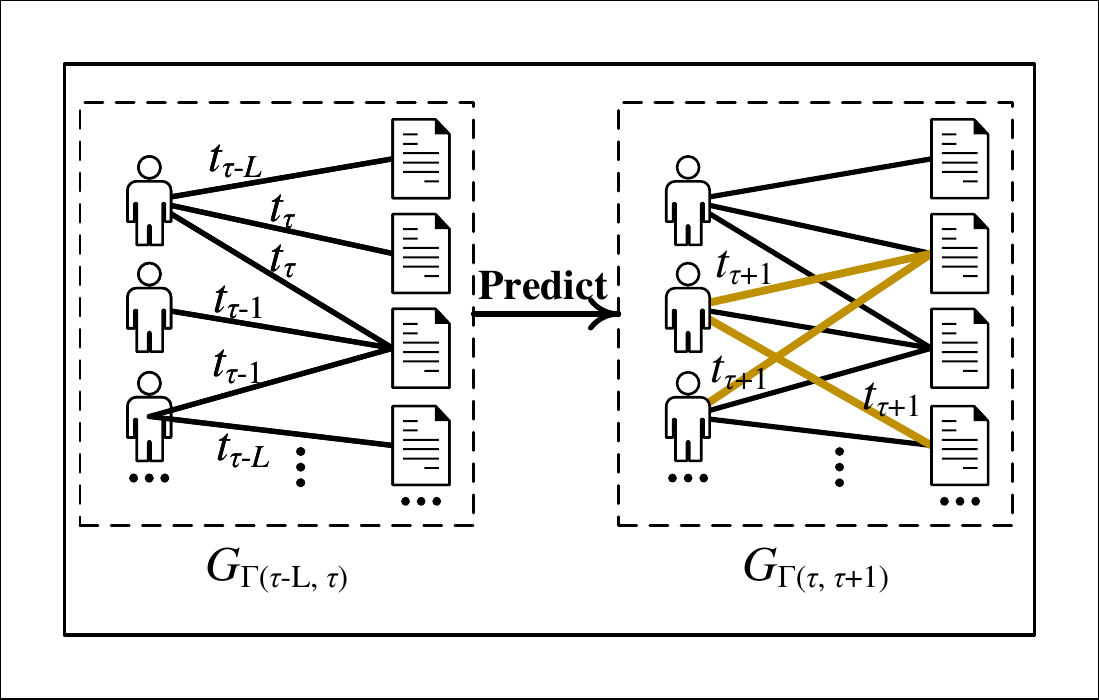}}
  }
 \end{minipage}
\end{center}
\vspace{-0.25cm}
\caption{Dynamic graph abstraction of online social networks for friend recommendation and next item recommendation.}\label{fig:APP-Social}
\vspace{-0.35cm}
\end{figure}

Recommendation is a straightforward application that can be supported by TLP. In online social networks and media, typical recommendation tasks include the friend recommendation \cite{campana2017recommender} and next item recommendation \cite{wang2020next}.

For instance, in Fig.~\ref{fig:APP-Social} (a), we can describe the evolution of friend relations using a UESD-based dynamic graph, where each node represents a unique \textit{user}; each unweighted edge associated with a time step $t$ indicates that the corresponding users establish a \textit{friend relation} at time step $t$. Given the abstracted historical topology $G_{\Gamma(\tau-L, \tau)}$, TLP aims to predict new edges that appear in the future time period $\Gamma(\tau, \tau+\Delta)$ denoted as ${\tilde E}_{\Gamma(\tau, \tau+\Delta)}$. One can directly recommend a new friend $v_j$ for user $v_i$ according to each edge $(v_i, v_j) \in {\tilde E}_{\Gamma(\tau, \tau+\Delta)}$.

Furthermore, we can also utilize the UESD-based dynamic bipartite graph to describe interactions between users and items. For instance, in Fig.~\ref{fig:APP-Social} (b), we represent each \textit{user} or \textit{item} (e.g., a product or an article) as a node. When \textit{a user $v_i$ interacts with an item $v_j$} (e.g., buying a product or clicking on an article) at time step $t$, we sample a new edge $((v_i, v_j), t)$. Similar to friend recommendation, the predicted future linkage ${\tilde E}_{\Gamma(\tau, \tau+\Delta)}$ can be directly utilized to recommend next items for each user (e.g., recommending item $v_j$ for user $v_i$ if $(v_i, v_j) \in {\tilde E}_{\Gamma(\tau, \tau+\Delta)}$).

\subsection{Intrusion Detection in Enterprise Internet}

\begin{figure}[t]
\begin{center}
 \includegraphics[width=0.8\linewidth, trim=18 18 18 18,clip]{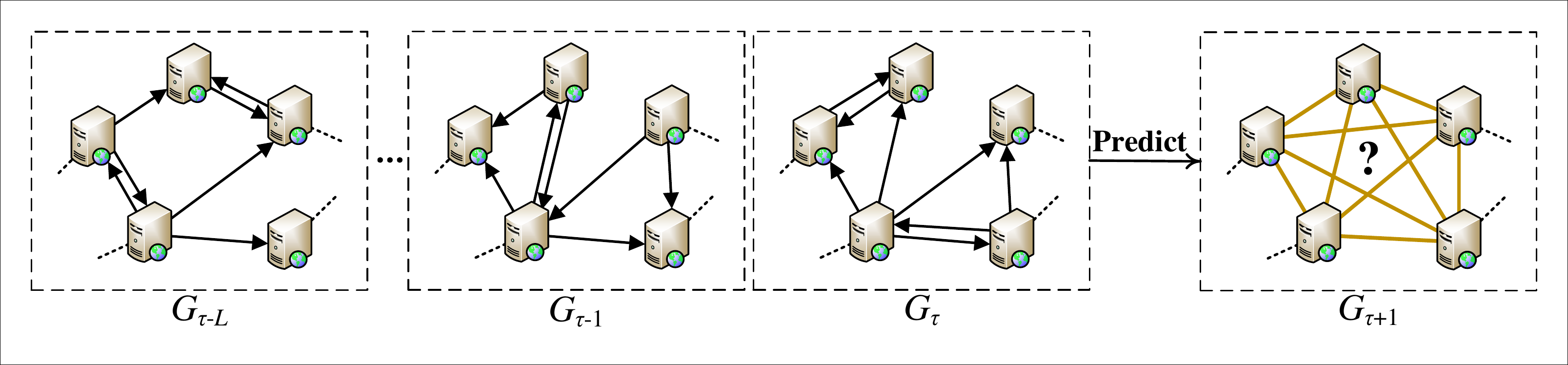}
\end{center}
\vspace{-0.25cm}
\caption{Dynamic graph abstraction of enterprise Internets for intrusion detection.}\label{fig:APP-Instrusion}
\vspace{-0.35cm}
\end{figure}

In \cite{king2023euler}, TLP was applied to detect intrusions in enterprise Internet. As in Fig.~\ref{fig:APP-Instrusion}, the behavior of an enterprise Internet can be described by a dynamic graph with ESSD. Concretely, \textit{host servers} and \textit{interactions between these servers} (e.g., transmitting data from a source server to a destination server) at a time step $t$ are represented as nodes and directed unweighted edges in the corresponding snapshot $G_t$.

For each node pair $(v_i, v_j)$, the TLP module outputs the probability $p_{ij}^{\tau+1}$ that there is an edge between $(v_i, v_j)$ in the next snapshot $G_{\tau+1}$, based on the historical topology $G_{\tau-L}^{\tau}$ of the abstracted graph. 
The node pair $(v_i, v_j)$ with a probability below a pre-set threshold (i.e., $p_{ij}^{\tau+1} < \varepsilon$) is defined to be anomalous. The detected anomalous links are usually believed to be indicative of \textit{lateral movement}, an important stage in a cyber-attack where the attacker attempts to find high-value hosts and spread malware from a compromised node throughout the network.

\subsection{Channel Allocation in Wireless Internet-of-Things (IoT) Networks}

\begin{figure}[t]
\begin{center}
 \includegraphics[width=0.8\linewidth, trim=18 18 18 18,clip]{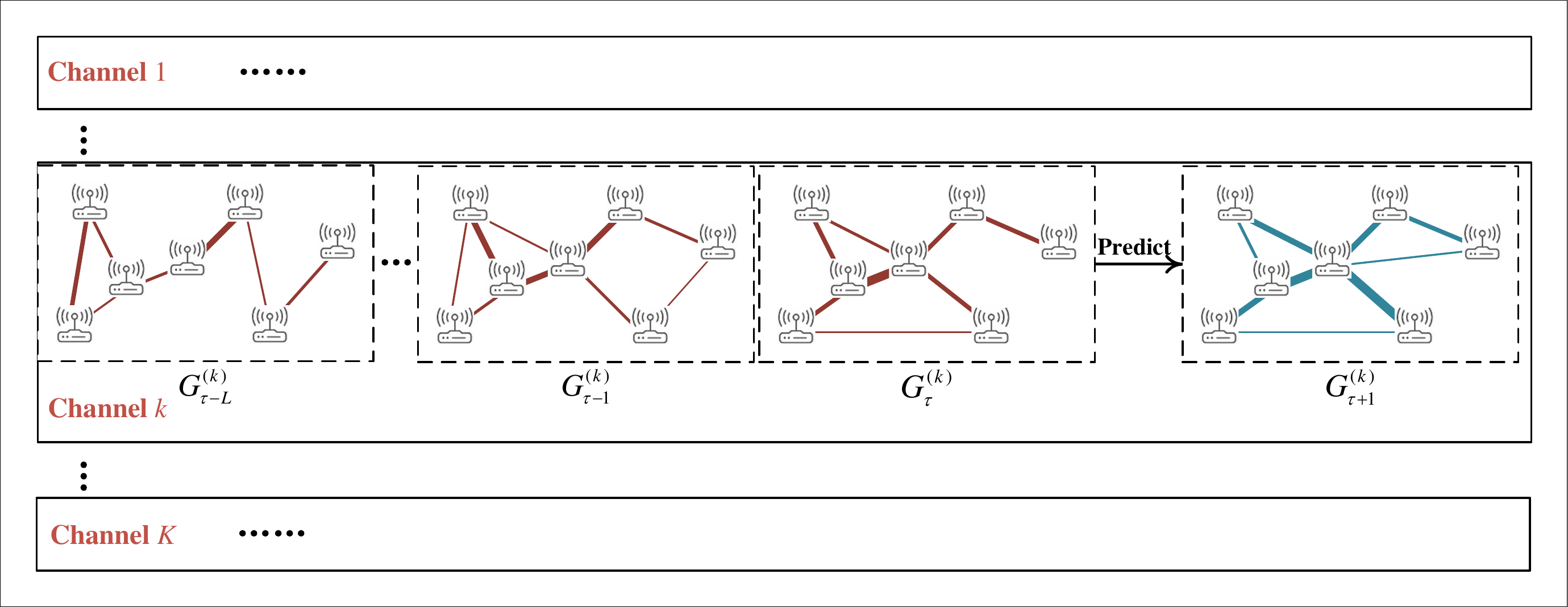}
\end{center}
\vspace{-0.25cm}
\caption{Dynamic graph abstract of wireless IoT networks for channel allocation.}\label{fig:APP-IoT}
\vspace{-0.35cm}
\end{figure}

Gao et al. \cite{gao2020edge} proposed a prediction-based channel allocation strategy for wireless IoT networks. As illustrated in Fig.~\ref{fig:APP-IoT}, a wireless IoT network with $K$ available channels can be abstracted as a set of ESSD-based dynamic graphs $\{ G^{(1)}, \cdots, G^{(K)} \}$, where $G^{(k)} = (G_1^{(k)}, G_2^{(k)}, \cdots)$ describes the dynamic states of the $k$-th channel.
All the graphs share a common node set, with each node corresponding to a unique \textit{wireless IoT device} (e.g., monitor and sensor). The \textit{link quality} (in terms of packet delivery ratio) between a pair of devices in the $k$-th channel at time slot $t$ is then represented as a weighted edge between the corresponding nodes in $G_t^{(k)}$, where each \textit{time slot} $t$ in the IoT network is mapped to a time step $t$ in the abstracted graph.

Given a sequence of data transmission requests (e.g., sending a package from a source device $v_i$ to a destination device $v_j$), the \textit{channel allocation task} assigns each request with (\romannumeral1) an available channel and (\romannumeral2) a time slot to conduct the corresponding data transmission while ensuring that there are no conflicts on each channel. Before formally allocating the channels, one can use a TLP module to predict future link quality states using historical system snapshots $\{ (G^{(1)})_{\tau-L}^{\tau}, \cdots, (G^{(K)})_{\tau-L}^{\tau}\}$. Gao et al. \cite{gao2020edge} have demonstrated that better system reliability can be achieved by greedily allocating channels according to the predicted link quality states (e.g., in terms of adjacency matrices $\{ {\bf{\tilde A}}_{\tau+1}^{(1)}, \cdots, {\bf{\tilde A}}_{\tau+1}^{(K)} \}$).

\subsection{Burst Traffic Detection \& Dynamic Routing in Optical Networks}

\begin{figure}[t]
\begin{center}
 \includegraphics[width=0.9\linewidth, trim=18 18 18 18,clip]{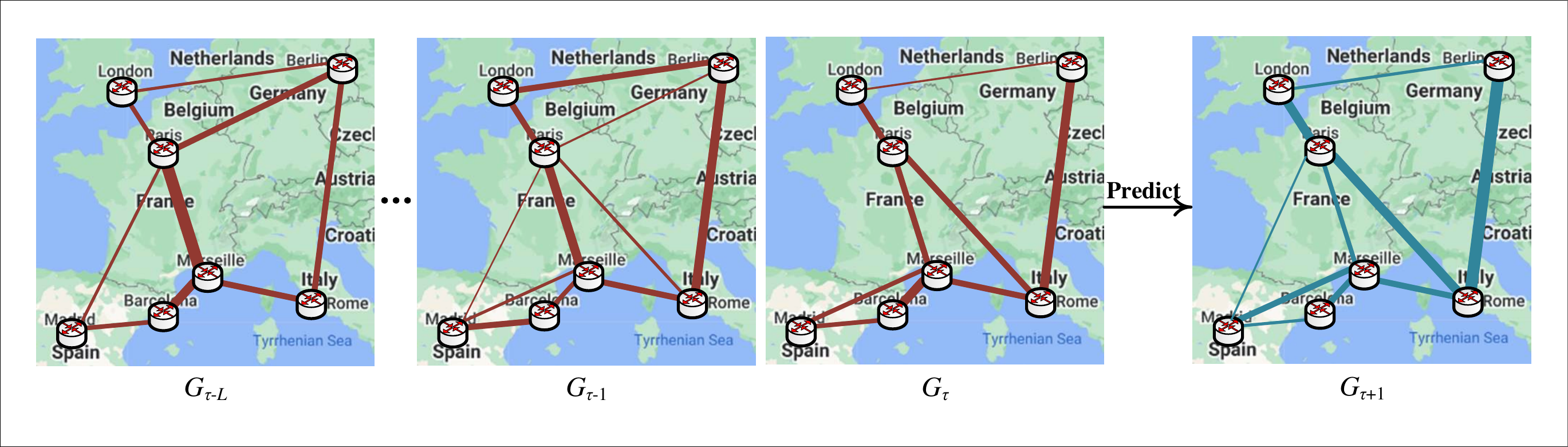}
\end{center}
\vspace{-0.25cm}
\caption{Dynamic graph abstraction of optical networks for burst traffic detection and dynamic routing.}\label{fig:APP-Opt-Net}
\vspace{-0.35cm}
\end{figure}

Aibin et al. \cite{vinchoff2020traffic,aibin2021short} focused on burst traffic detection in optical networks and analyzed its effects to the performance of dynamic routing. As shown in Fig.~\ref{fig:APP-Opt-Net}, we can describe the behavior of an optical network using an ESSD-based dynamic graph, where each \textit{routing node} (i.e., a router associated with a city) is abstracted as a unique node; the \textit{traffic intensity} (in terms of Erlang) on a fiber link between a pair of routers is represented as a weighted edge between the corresponding nodes. In this setting, each snapshot $G_t$ describes the network traffic at time step $t$. Given the historical network traffic described by adjacency matrices ${\bf{A}}_{\tau-L}^{\tau}$ w.r.t. snapshots $G_{\tau-L}^{\tau}$, one can predict the next snapshot $G_{\tau+1}$ (in terms of an adjacency matrix ${\bf{\tilde A}}_{\tau+1}$) that describes the possible future traffic by applying a TLP method.

In \cite{vinchoff2020traffic,aibin2021short}, the authors defined three types of burst traffic (i.e., single-burst, double-burst, and plateau-burst traffic) with different variation patterns and detected each type of traffic on the predicted snapshot ${\bf{\tilde A}}_{\tau + 1}$ instead of current snapshot ${\bf{A}}_\tau$. They further demonstrated that better routing performance (i.e., lower request blocking percentage) can be achieved if one makes routing decisions according to the (future) burst traffic detected on ${\bf{\tilde A}}_{\tau + 1}$.

\subsection{Dynamics Simulation \& Conformational Analysis of Molecules}

\begin{figure}[t]
\begin{center}
 \includegraphics[width=0.65\linewidth, trim=18 18 18 18,clip]{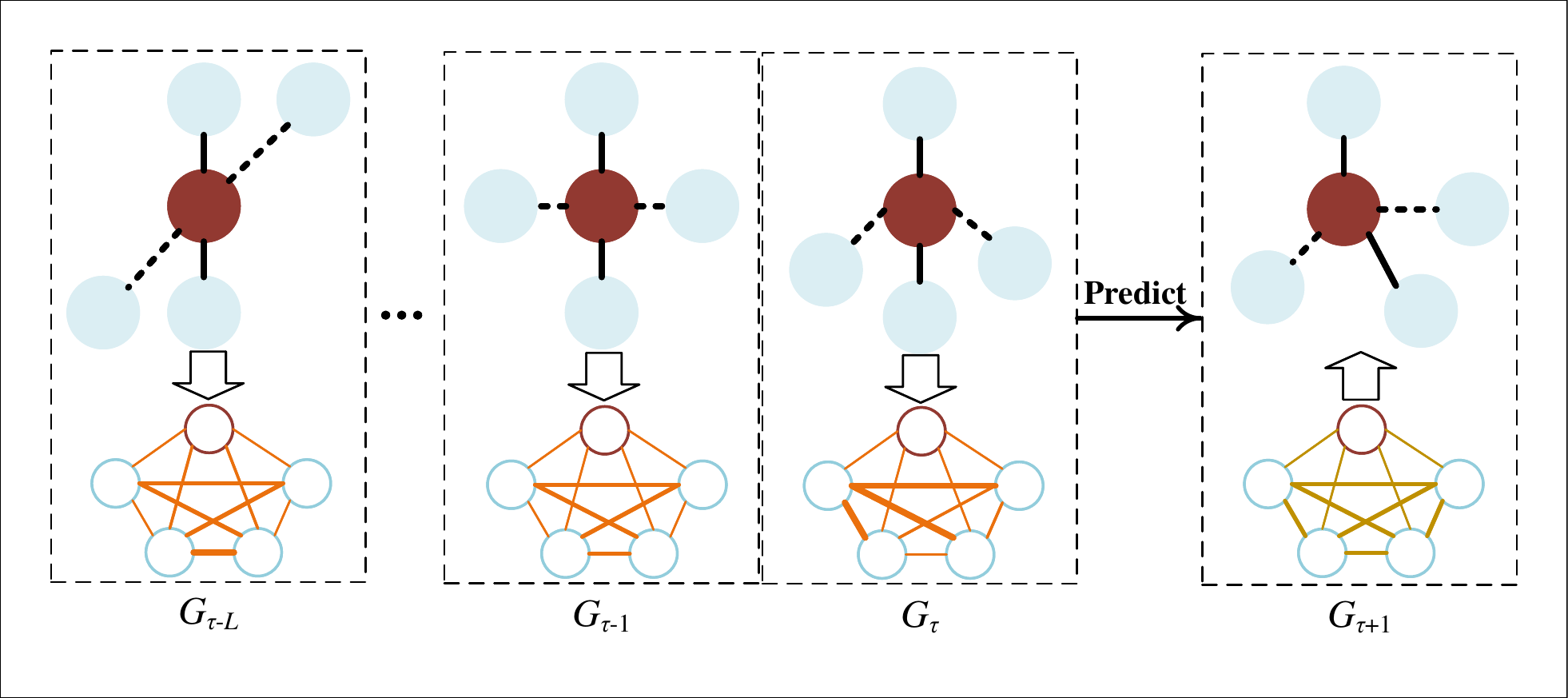}
\end{center}
\vspace{-0.25cm}
\caption{Dynamic graph abstraction of molecular structures for dynamic simulation and conformational analysis.}\label{fig:APP-Molecule}
\vspace{-0.35cm}
\end{figure}

Ashby et al. \cite{ashby2021geometric} encoded the geometric properties of molecules using a graph-based representation and explored the potential of TLP to support conformational analysis. As illustrated in Fig.~\ref{fig:APP-Molecule}, given a molecule with a static structure, we can represent each \textit{atom} as a unique node and abstract the \textit{Euclidean distance between each pair of atoms} as a weighted edge between corresponding nodes. In this setting, each molecule can be represented as a weighted complete graph.

Since the molecule structure may vary over time, one can describe the geometric variation using an ESSD-based dynamic graph, where snapshot $G_t$ encodes the geometric structure at time step $t$ in terms of a weighted complete graph. Given the trajectory of geometric variation described by adjacency matrices ${\bf{A}}_{\tau-L}^{\tau}$ w.r.t. snapshots $G_{\tau-L}^{\tau}$, Ashby et al. tried to predict the next snapshot $G_{\tau+1}$ (in terms of an adjacency matrix ${\bf{\tilde A}}_{\tau+1}$) regarding the possible future structure for molecular simulation. Based on the predicted structure, one can further analyze the molecular energy and forces on atoms, supporting conformational analysis.

\section{Public Datasets}

\begin{table}[t]\footnotesize
\centering
\caption{Summary of public datasets for TLP.}\label{Tab:Data}
\vspace{-0.25cm}
\begin{tabular}{l|p{2.5cm}|p{1.5cm}p{3.0cm}l|l|l|l}
\hline
\textbf{Datasets} & \textbf{Scenarios} & \textbf{Nodes} & \textbf{Edges} & \textbf{\begin{tabular}[c]{@{}l@{}}\tiny{Weighted}\\ \tiny{Links}\end{tabular}} & \textbf{\begin{tabular}[c]{@{}l@{}}\tiny{Min Time}\\ \tiny{Granularity}\end{tabular}} & \textbf{\begin{tabular}[c]{@{}l@{}}\tiny{Data}\\ \tiny{Models}\end{tabular}} & \textbf{\tiny{Level}} \\ \hline
\textit{Social Evolution} & Social Network & Cell phones & Bluetooth signal, calls, or messages between cell phones & No & 1 min & UESD & 2 \\ \hline
\textit{CollegeMsg} & Online social network & App users & Messages sent from a source user to a destination user & No & 1 sec & UESD & 2 \\ \hline
\textit{Wiki-Talk} & Online social network & Wikipedia users & Relations that a user edits another user's talk page & No & 1 sec & UESD & 2 \\ \hline
\textit{Enron} & Email network & Email users & Emails from a source user to a destination user & No & 1 sec & UESD & 2 \\ \hline
\textit{Reddit-Hyperlink} & Hyperlink network & Subreddits & Hyperlinks from one subreddit to another & No & 1 sec & UESD & 2 \\ \hline
\textit{DBLP} & Paper collaboration network & Paper authors & Collaboration relations between authors & No & 1 day & UESD & 2 \\ \hline
\textit{AS-733} & BGP autonomous systems of Internet & BGP routers & Who-talks-to-whom communication between routers & No & 1 day & ESSD & 2 \\ \hline
\textit{Bitcoin-Alpha} & Bitcoin transaction network & Bitcoin users & Trust scores between users & Yes & 1 sec & UESD & 2 \\ \hline
\textit{Bitcoin-OTC} & Bitcoin transaction network & Bitcoin users & Trust scores between users & Yes & 1 sec & UESD & 2 \\ \hline
\textit{UCSB-Mesh} & Wireless mesh network & Wireless routers & Link quality (in terms of expected transmission time) between routers & Yes & 1 min & ESSD & 1 \\ \hline
\textit{NumFabric} & Data center network & Host servers & Traffic (in terms of KB) between host servers & Yes & 1e-6 sec & UESD & 1 \\ \hline
\textit{UCSD-WTD} & WiFi mobility network & Access points/PDA devices & Signal strength (in terms of dBm) between access points and PAD devices & Yes & 1 sec & UESD & 2 \\ \hline
\textit{UNSW-IoT} & IoT network & IoT devices & Traffic (in terms of KB) between IoT devices & Yes & 1e-6 sec & UESD & 2 \\ \hline
\textit{WIDE} & Internet backbone & Host servers/user devices & Traffic (in terms of KB) between servers/devices & Yes & 1e-6 sec & UESD & 2 \\ \hline
\end{tabular}
\vspace{-0.35cm}
\end{table}

In addition, we also summarize some public datasets that can be used to evaluate different settings of TLP, which include 
(\romannumeral1) \textit{Social Evolution}\footnote{http://realitycommons.media.mit.edu/socialevolution.html} \cite{madan2011sensing},
(\romannumeral2) \textit{CollegeMsg}\footnote{https://snap.stanford.edu/data/CollegeMsg.html} \cite{paranjape2017motifs},
(\romannumeral3) \textit{Wiki-Talk}\footnote{https://snap.stanford.edu/data/wiki-talk-temporal.html} \cite{paranjape2017motifs},
(\romannumeral4) \textit{Enron}\footnote{http://konect.cc/networks/enron/} \cite{klimt2004enron},
(\romannumeral5) \textit{Reddit-Hyperlink}\footnote{https://snap.stanford.edu/data/soc-RedditHyperlinks.html} \cite{kumar2018community},
(\romannumeral6) \textit{DBLP}\footnote{https://dblp.uni-trier.de/xml/} \cite{yang2012defining},
(\romannumeral7) \textit{AS-733}\footnote{https://snap.stanford.edu/data/as-733.html} \cite{leskovec2005graphs}, 
(\romannumeral8) \textit{Bitcoin-Alpha}\footnote{https://snap.stanford.edu/data/soc-sign-bitcoin-alpha.html} \cite{kumar2016edge},
(\romannumeral9) \textit{Bitcoin-OTC}\footnote{https://snap.stanford.edu/data/soc-sign-bitcoin-otc.html} \cite{kumar2016edge},
(\romannumeral10) \textit{UCSB-Mesh}\footnote{https://ieee-dataport.org/open-access/crawdad-ucsbmeshnet} \cite{ramachandran2007routing},
(\romannumeral11)\textit{NumFabric}\footnote{https://github.com/shouxi/numfabric} \cite{nagaraj2016numfabric},
(\romannumeral12) \textit{UCSD-WTD}\footnote{http://www.sysnet.ucsd.edu/wtd} \cite{mcnett2005access},
(\romannumeral13) \textit{UNSW-IoT}\footnote{https://iotanalytics.unsw.edu.au/iottraces.html} \cite{sivanathan2018classifying},
and (\romannumeral14) \textit{WIDE}\footnote{https://mawi.wide.ad.jp/mawi} \cite{borgnat2009seven}.

Most existing survey papers merely focus on some statistics of these public datasets (e.g., numbers of nodes, edges, snapshots, and time steps).
Instead of focusing on these statistics, which may be different according to data pre-processing (e.g., different selections of sampling rate may result in different numbers of snapshots for ESSD), we summarize some properties of the original versions of these datasets in Table~\ref{Tab:Data}, including the (\romannumeral1) scenario, (\romannumeral2) abstraction of nodes, (\romannumeral3) abstraction of edges, (\romannumeral4) applicability to weighted links, (\romannumeral5) minimum time granularity, (\romannumeral6) data model (i.e., ESSD or UESD used by the original data files), and (\romannumeral7) variation of node sets (i.e., level-1 or -2 for whether the node set is assumed to be known for all the time steps in the original data files).

In particular, one can further change some of the original properties during pre-processing. For instance, we can convert UESD to ESSD by manually selecting a proper sampling rate to extract the corresponding topology in each snapshot. Some transductive TLP methods also covert level-2 to level-1 using the union of node sets w.r.t. all the time steps, even though there may exist many isolated nodes in some time steps. A weighted dynamic graph can also be converted to an unweighted version by just removing the edge weights.

\end{document}